\documentclass[journal]{IEEEtran}
\usepackage[T1]{fontenc}
\usepackage{color}
\usepackage[tbtags]{amsmath}
\usepackage{url}
\usepackage{array}
\usepackage{fancyhdr}
\usepackage{latexsym}
\usepackage{diagbox}
\usepackage{epsfig}
\usepackage{amssymb}
\usepackage{amsfonts}
\usepackage{pifont}
\usepackage{booktabs}
\usepackage{amsxtra}
\usepackage{xspace}
\usepackage{makeidx}
\usepackage{graphics}
\usepackage{pstool}
\usepackage{theorem}
\usepackage{cuted}
\usepackage[caption=false]{subfig}
\usepackage{caption2}
\usepackage{algorithm}
\usepackage{hyperref}
\usepackage{cleveref}
\usepackage{algorithmic}
\usepackage{multirow}
\usepackage{diagbox}
\usepackage{amsmath}
\usepackage{cite}
\usepackage{setspace,graphics,rotating}
\usepackage{framed}
\usepackage{epstopdf}
\usepackage{bm}
\usepackage{makecell}
\usepackage{stfloats}
\usepackage{enumerate}
\graphicspath{{Figure/}}

\newtheorem{lemma}{\textbf{Lemma}}

\newtheorem{proposition}{\textbf{Proposition}}

\IEEEoverridecommandlockouts

\def\gap{0.85ex}
\abovedisplayskip\gap
\belowdisplayskip\gap
\abovedisplayshortskip\gap
\belowdisplayshortskip\gap

\begin{document}

\renewcommand{\algorithmicrequire}{\textbf{Input:}}  
\renewcommand{\algorithmicensure}{\textbf{Output:}}

\title{Sensing-Then-Beamforming: Robust Transmission Design for RIS-Empowered Integrated Sensing and Covert Communication}

\author{Xingyu~Zhao,~\IEEEmembership{Graduate Student Member,~IEEE,} Min~Li,~\IEEEmembership{Member,~IEEE,} Ming-Min~Zhao,~\IEEEmembership{Senior Member,~IEEE,} Shihao~Yan,~\IEEEmembership{Senior Member,~IEEE,} and Min-Jian~Zhao,~\IEEEmembership{Member,~IEEE}
\thanks{X. Zhao, M. Li, M. M. Zhao and M. J. Zhao are with the College of Information Science and Electronic Engineering and the Zhejiang Provincial Key Laboratory of Multi-Modal Communication Networks and Intelligent Information, Zhejiang University, Hangzhou 310027, China (e-mail: zhaoxingyu23@zju.edu.cn; min.li@zju.edu.cn; zmmblack@zju.edu.cn; mjzhao@zju.edu.cn). (\emph{Corresponding authors: Min Li; Min-Jian Zhao}.)

S. Yan is with the School of Science, Edith Cowan University, Perth,
WA 6027, Australia (e-mail: s.yan@ecu.edu.au).}

\thanks{The work of Min Li was supported in part by National Natural Science Foundation of China under Grant 62271440.}

}

\maketitle

\begin{abstract}
Traditional covert communication often relies on the knowledge of the warden's channel state information, which is inherently challenging to obtain due to the non-cooperative nature and potential mobility of the warden. The integration of sensing and communication technology provides a promising solution by enabling the legitimate transmitter to sense and track the warden, thereby enhancing transmission covertness. In this paper, we develop a framework for sensing-then-beamforming in reconfigurable intelligent surface (RIS)-empowered integrated sensing and covert communication (ISACC) systems, where the transmitter (Alice) estimates and tracks the mobile aerial warden's channel using sensing echo signals while simultaneously sending covert information to multiple legitimate users (Bobs) with the assistance of RIS, under the surveillance of the warden (Willie). Considering channel estimation errors, we formulate a robust non-convex optimization problem that jointly designs the communication beamformers, the sensing signal covariance matrix at Alice, and the phase shifts at the RIS to maximize the covert sum rate of Bobs while satisfying the constraints related to covert communication, sensing, transmitter power, and the unit modulus of the RIS elements. To solve this complex problem, we develop an efficient algorithm using alternating optimization, successive convex approximation, S-procedure, sequential rank-one constraint relaxation, and semidefinite relaxation techniques. Numerical results confirm the convergence of the proposed algorithm and demonstrate its effectiveness in tracking the warden's channel while ensuring robust covert transmission. Furthermore, the results highlight the advantages of using RIS to enhance the covert transmission rate compared to baseline schemes, and also illustrate the intricate trade-off between communication and sensing in ISACC systems.

\end{abstract}

\IEEEpeerreviewmaketitle

\section{Introduction}\label{sec:introduction}
Covert communication has gained significant attention due to its ability to provide stronger protection than traditional physical layer security (PLS), aiming to prevent wardens from detecting legitimate transmission from their received signals \cite{chen2023covert}. Meanwhile, reconfigurable intelligent surface (RIS) has emerged  as a promising technique for enhancing  spectral and energy efficiency in future wireless systems by dynamically reconfiguring wireless channels through precise control of passive reflecting elements its passive reflecting elements \cite{yuan2021reconfigurable,xiu2021reconfigurable}. Owing to these capabilities, RIS can also substantially improve covert communication performance \cite{chen2023covert}. 

In recent years, considerable efforts have been devoted to combining covert communication with RIS from various perspectives. In particular,  \cite{ma2021robust} developed both covert and low-complexity zero-forcing beamformers to maximize the covert rate based on perfect channel state information (CSI) of the warden, and further extended the design to account for imperfect CSI. Expanding to  RIS-empowered Internet of Things (IoT) networks,  \cite{ma2022covert} jointly optimized transmit power and RIS reflection coefficients to strengthen the legitimate signal while degrading warden detection. Considering both the warden's detection error probability constraint and the legitimate user's communication outage probability, \cite{wang2021intelligent} developed a RIS-empowered multi-antenna covert communication system to improve covert rate. Additionally, \cite{si2021covert} studied a RIS-assisted covert transmission scheme with either instantaneous or partial CSI of the warden's link to maximize the covert rate. However, these studies \cite{ma2021robust,ma2022covert,zhou2021intelligent,wang2021intelligent,si2021covert} generally assume prior knowledge of the warden’s CSI—an unrealistic assumption due to the non-cooperative nature of the warden, especially when passive RIS is deployed. This highlights the critical need for effective approaches to acquire warden-related CSI for covert communication design.

Fortunately, integrated sensing and communications (ISAC) provides a promising solution to this challenge  by enabling legitimate transmitters  to sense and track the warden \cite{chen2023covert}. Unlike separate independent sensing and communication systems, ISAC systems can improve spectrum utilization and reduce hardware cost by sharing both spectrum and hardware for both functionalities \cite{liu2022}. Furthermore, the robustness-oriented ISAC designs—such as those accounting for imperfect CSI and time synchronization error—has advanced its practical feasibility \cite{xiu2025movable2}. Combining covert communication, RIS, and ISAC therefore positions RIS-empowered integrated sensing and covert communication (ISACC) as a compelling technology for civil and military applications. In such systems, the sensing waveform itself embeds communication information, which, with RIS assistance, remains concealed from wardens during sensing operations \cite{chen2023covert,du2022reconfigurable}. 

There has been a growing body of recent research on ISACC. In \cite{shuai:2023}, robust beamformers for joint  target detection and single-user communication were designed to maximize the covert rate while guaranteeing the covertness requirement for legitimate communication and radar detection mutual information. In \cite{hu2024covert}, a robust beamforming scheme was proposed to maximize the covert throughput subject to both radar detection and covertness constraints, revealing a trade-off between the two functionalities. The work \cite{Hu:2024} studied a RIS-assisted ISACC system to jointly optimize communications rate and radar probing power. More recently,   \cite{zhang2024robust} and \cite{zhao2024robust} investigated robust beamforming design under imperfect CSI of the warden to balance radar, communication, and covertness requirements. Despite these contributions, most works assume known perfect or imperfect CSI of the warden and do not fully address the channel acquisition of warden. In addition, \cite{shuai:2023,hu2024covert,Hu:2024} are restricted to single-user setups.

Leveraging sensing-assisted communication, recent studies such as  \cite{Wei:2023} and \cite{wang2024sensing} explored radar sensing to track and jam aerial adversaries while safeguarding legitimate communication. Specifically, \cite{Wei:2023} employed extended Kalman filtering (EKF) to predict adversary CSI from backscattered echoes and proposed a robust resource allocation framework for multi-user secure communication, considering prediction errors. However, its focus was physical-layer security rather than covertness, making the results less directly applicable to covert communication. In \cite{wang2024sensing}, sensing information was used to construct the adversary target’s CSI, and joint beamforming and radar waveform design were proposed to maximize covert rate under a single-user setup with tracking accuracy constraints. Yet, this work did not consider scenarios with multiple legitimate receivers or explore the trade-off between covert communication and sensing. Moreover, neither  \cite{Wei:2023} nor \cite{wang2024sensing} incorporated RIS, leaving its potential for enhancing covert ISACC systems largely unexplored.

\begin{table*}[htbp!]
\small
\centering
{
\caption{A Brief Comparison of the Related Works}
\label{TAB1}
\begin{tabular}{!{\vrule width 1pt}c||c|c|c|c|c|c|c|c|c|c|c!{\vrule width 1pt}}
\Xhline{1pt}
\diagbox{Content}{Literature}                    & This work & \cite{ma2021robust}& \cite{ma2022covert}&\cite{wang2021intelligent}&\cite{si2021covert}&\cite{zhou2021intelligent} &\cite{shuai:2023,hu2024covert}&\cite{Hu:2024} & \cite{zhang2024robust,zhao2024robust}&\cite{Wei:2023}&\cite{wang2024sensing}\\ \Xhline{1pt}
ISAC                                   & $\checkmark$  & \ding{55}    & \ding{55} & \ding{55}& \ding{55}& \ding{55} & $\checkmark$  & $\checkmark$     &$\checkmark$ &$\checkmark$&$\checkmark$ \\ \hline
Covert communication                   & $\checkmark$ & $\checkmark$  &$\checkmark$ &$\checkmark$&$\checkmark$&$\checkmark$& $\checkmark$   &$\checkmark$     &$\checkmark$& \ding{55} &$\checkmark$  \\ \hline
RIS-empowered                            & $\checkmark$ & \ding{55} & $\checkmark$ & $\checkmark$ & $\checkmark$ & $\checkmark$    &\ding{55}         & $\checkmark$        &  \ding{55} &\ding{55} & \ding{55}  \\ \hline
Acquisition of warden's CSI     & $\checkmark$ &\ding{55} &\ding{55} &\ding{55}&\ding{55}&\ding{55}     &  \ding{55}   &\ding{55}         &  \ding{55}  & $\checkmark$& $\checkmark$\\ \hline
Imperfect warden's CSI                             & $\checkmark$  &$\checkmark$ &$\checkmark$&  \ding{55}  & $\checkmark$ & \ding{55}   &$\checkmark$ & $\checkmark$&$\checkmark$&$\checkmark$&$\checkmark$ \\ \hline
Multi-user                            & $\checkmark$  &\ding{55}  &\ding{55}  & \ding{55}  & \ding{55} & \ding{55} & \ding{55}   &\ding{55}             & $\checkmark$& $\checkmark$&\ding{55}    \\ \Xhline{1pt}
\end{tabular}}
\end{table*}

In this paper, we investigate the robust transmission design for RIS-empowered ISACC systems in the presence of an aerial adversary warden. Due to its exceptional maneuverability and mobility, the warden can pose a significant threat to terrestrial communications. To address this issue, we propose a sensing-then-beamforming framework to enhance the system's robustness and covertness against such aerial adversaries. On the one hand, Alice with dual-functional signals can construct the relevant CSI of the warden  by actively sensing and tracking in real-time. On the other hand, considering the channel estimation errors of the warden, a robust transmission design related to the joint optimization of the communication beamformers, the sensing signal covariance matrix at Alice, and the phase shifts at the RIS is proposed to further improve the covert sum rate of Bobs. The main contributions of this paper are as follows:
 \begin{itemize}
     \item We propose a novel sensing-then-beamforming framework for  RIS-Empowered ISACC systems. In this framework,  Alice leverages sensing capabilities and the EKF technique to track Willie’s trajectory and construct  the corresponding CSI for both Alice-Willie and RIS-Willie links so as to achieve covert communication with legitimate Bobs.  
     \item Considering imperfect CSI prediction of Willie, we formulate a robust optimization problem to maximize the covert sum rate for multiple legitimate Bobs through jointly designing Alice’s communication beamformers, the sensing signal covariance matrix, and the RIS phase shifts, subject to covert communication, sensing accuracy, power, and unit-modulus constraints. To tackle the highly non-convex and multivariate problem,    
     we  develop  an effective alternating optimization algorithm that integrates successive convex approximation (SCA), the S-procedure, semidefinite relaxation (SDR) and sequential rank-one constraint relaxation (SROCR).
     \item Simulation results show that the proposed framework  effectively tracks Willie's channel while ensuring robust covert transmission, demonstrating the value of leveraging sensing information to enhance covert communication. Additionally, compared with baseline schemes, our results further confirm that RIS deployment significantly boosts the covert sum rate for Bobs and reveal the fundamental trade-off between covert communication and sensing.
 \end{itemize}
 
Table \ref{TAB1} presents a brief comparison of research contents of this work and the most relevant articles in the literature.

The rest of this paper is organized as follows. Section~\ref{sys-model} presents the RIS-empowered ISACC system model. Section~\ref{alg-design} first elaborates the transmission protocol, formulates the robust transmission design problem, and then proposes an efficient algorithm for solving it. Numerical results are provided in Section~\ref{sim-result}, while the conclusions are drawn in Section~\ref{conclu}.

\begin{table}[!t]
\small
{
\caption{Key Mathematical Notations}
\label{TAB-symbol}
\centering
\resizebox{.98\columnwidth}{!}{
\begin{tabular}{!{\vrule width 1pt}c||c!{\vrule width 1pt}}
\Xhline{1pt}
\textbf{Notation} & \textbf{Description}\\ \Xhline{1pt}
$K$  & Number of legitimate users (Bobs) \\ \hline  $N_{\mathrm{A}}^{\mathrm{t}}$ $(N_{\mathrm{A}}^{\mathrm{r}})$  & Number of transmit (receive) antennas at Alice   \\ \hline  $s_k$  & Transmitted information
signal for Bob $k$   \\ \hline  $\mathbf{s}_{\mathrm{r}}$  & Sensing
signal  \\ \hline  $\mathbf{R}_{\mathrm{s}}$  & Covariance matrix of the sensing
signal  \\ \hline  $\mathbf{h}_{\mathrm{d}, k}$  & Channel vector of the Alice-Bob $k$ link  \\ \hline   $\mathbf{h}_{\mathrm{r}, k}$  & Channel vector of the RIS-Bob $k$ link   \\ \hline $\mathbf{G}$  & Channel vector of the Alice-RIS link  \\ \hline $\mathbf{h}_{\mathrm{aw}}$ & Channel vector of the Alice-Willie link\\ \hline $\mathbf{h}_{\mathrm{rw}}$ & Channel vector of the RIS-Willie link\\ \hline $\mathbf{w}_k$ & Beamforming at Alice for Bob $k$ \\ \hline  $\boldsymbol{\Theta}$  & RIS phase shift matrix  \\ \hline $\hat{\boldsymbol{\chi}}_{\mathrm{w}}$ & Estimated state of Willie \\ \hline $\hat{\mathbf{Z}}_{\mathrm{w}}$ & Measurements of Willie\\ \hline $\widehat{\mathbf{h}}_{\mathrm{aw}}$ ($\widehat{\mathbf{h}}_{\mathrm{rw}}$)  & Estimated channel vector of $\mathbf{h}_{\mathrm{aw}}$ ($\mathbf{h}_{\mathrm{rw}}$) \\ \hline $\boldsymbol{\Delta}_{\mathrm{aw}}$ ($\boldsymbol{\Delta}_{\mathrm{rw}}$) & Channel estimation error of $\mathbf{h}_{\mathrm{aw}}$ ($\mathbf{h}_{\mathrm{rw}}$)  \\  \Xhline{1pt} 
\end{tabular}}}
\end{table}

\textit{Notations:} $a$ (or $A$), ${\mathbf{a}}$ and ${\mathbf{A}}$ stand for a scalar, a column vector and a matrix, respectively. The imaginary unit of a complex number is denoted by $j=\sqrt{-1}$. $\mathbb{C}^{m \times n}$ and $\mathbb{H}^m$ denote the sets of all $m \times n$ complex-valued matrices and all $m \times m$ Hermitian matrices, respectively. ${\left(  \cdot  \right)^ * }$, ${\left(  \cdot  \right)^T}$ and ${\left(  \cdot  \right)^H}$ respectively represent the conjugate, transpose and conjugate transpose operations of ${\mathbf{a}}$ or ${\mathbf{A}}$. $\left|  \cdot  \right|$, ${\left\|  \cdot  \right\|}$,  ${\left\|  \cdot  \right\|_F}$ and $\Re\{\cdot\}$  denote the determinant (or module for a complex variable), $l_2$ norm,  Frobenius norm, and real part, respectively.  ${\mathrm{tr}}\left(\cdot  \right)$, $\operatorname{diag}\{\cdot\}$ and ${\left(  \cdot  \right)^{ - 1}}$ denote the trace, the diagonal and the matrix inversion operators, respectively. $\mathrm{blkdiag}\left(\mathbf{X},\mathbf{Y}\right)$ is a block diagonal matrix whose diagonal components are matrices $\mathbf{X}$ and $\mathbf{Y}$.  $\otimes$ is the Kronecker product and $\mathbf{A} \succeq \mathbf{0}$ means that $\mathbf{A}$ is a symmetric and positive semi-definite (PSD) matrix. ${{\bf{I}}_N}$ denotes the $N \times N$ identity matrix, while ${\mathbf{0}}_{N\times M}$ denotes an $N \times M$ all-zero matrix. ${\mathbf{x}} \sim \mathcal{C}\mathcal{N}( {{\boldsymbol{\mu}},{\mathbf{K}}} )$ means that ${\mathbf{x}}$ is a circularly symmetric complex Gaussian vector whose mean and covariance matrix are  $\boldsymbol{\mu}$ and ${\mathbf{K}}$, respectively. The maximum eigenvalue of matrix $\mathbf{A}$ and the corresponding eigenvector are represented by $\lambda_{\max }(\mathbf{A})$ and $\mathbf{u}_{\max }(\mathbf{A})$. $\nabla_{\mathbf{A}} f$ denotes the gradient of $f$ with respect to matrix $\mathbf{A}$. Furthermore, $\mathcal{O}$ represents the big-O notation. Finally, $\mathcal{K} \triangleq\{1, \ldots, K\}$ and $\mathcal{K}_k \triangleq \mathcal{K} \backslash\{k\}$. The key mathematical notations are summarized in Table~\ref{TAB-symbol}.

\section{System Model}\label{sys-model}
\begin{figure}[!t]
\vspace{0cm}
\begin{center}
\centering
\includegraphics[width=8cm,height=5cm]{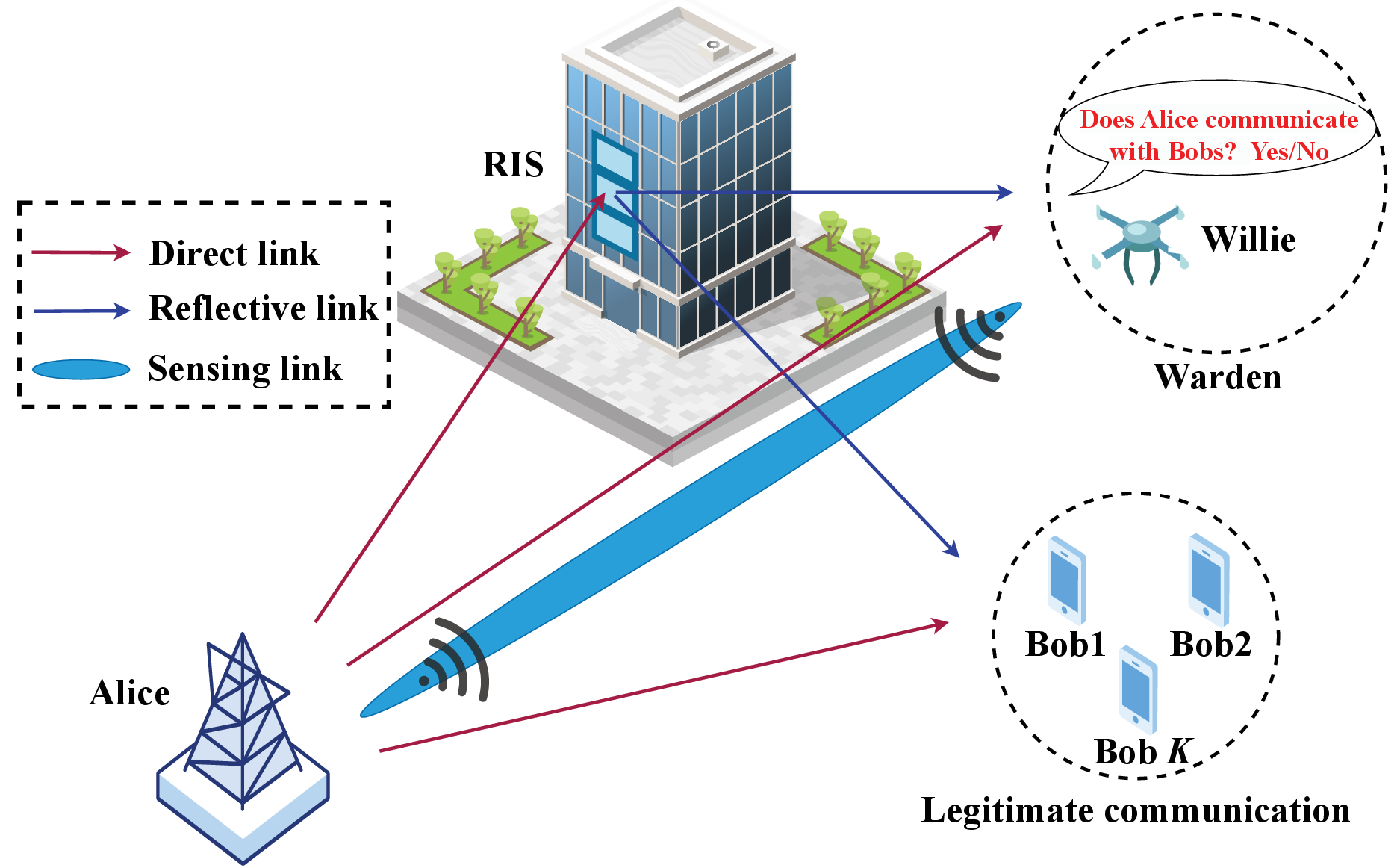}
\captionsetup{justification=raggedright,singlelinecheck=false}
\caption{ Illustration of the RIS-empowered ISACC system.}
\label{fig:system-model}
\end{center}
\vspace{-0.8cm} 
\end{figure}
As shown in Fig.~\ref{fig:system-model}, we consider a downlink narrowband RIS-empowered ISACC system, which consists of a dual-functional radar-communication Alice, $K$ single-antenna legitimate users (Bobs), a mobile illegal warden (Willie) with a single antenna, and one RIS. Alice equipped with two rectangular uniform planar arrays (UPAs) performs multi-user covert communication and related sensing tasks to track the trajectory of Willie, such as an adversary unmanned aerial vehicle (UAV), simultaneously. Without loss of generality, the numbers of transmit and receive antennas, denoted by $N^{\mathrm{t}}_{\mathrm{A}}$ and $N^{\mathrm{r}}_{\mathrm{A}}$, are assumed to be identical, i.e., $N^{\mathrm{t}}_{\mathrm{A}}=N^{\mathrm{r}}_{\mathrm{A}}=N_HN_V=N_{\mathrm{A}}$, where $N_H$ and $N_V$ denote the numbers of antennas in the horizontal and vertical dimensions, respectively. RIS with $N_\mathrm{R}$ passive elements is deployed to improve the system performance. Meanwhile, Willie aims to determine whether or not  Alice is transmitting signals to Bobs.

\subsection{Signal Model}
\subsubsection{Communication Signal}
Consider a frame consisting of $N$ time slots. Let $T$ be the frame duration, and thus each time slot  occupies $\delta = T/N$ unit time. Then, the received communication signal at Bob  $k$ in time slot $n$ is given by
\begin{equation}\label{comm-signal}
\begin{aligned}
y_k(n, t)= & \underbrace{\mathbf{h}_k^H[n] \mathbf{w}_k[n] s_k(n, t)}_{\text {desired signal }}+\underbrace{\sum_{j \in \mathcal{K}_k} \mathbf{h}_k^H[n] \mathbf{w}_j[n] s_j(n, t)}_{\text {inter-user interference }} \\
& +\underbrace{\mathbf{h}_k^H[n] \mathbf{s}_{\mathrm{r}}(n, t)}_{\text {sensing signal }}+\underbrace{n_k(n, t)}_{\text {noise }}, k \in \mathcal{K},
\end{aligned}
\end{equation}
where $t \in(0, \delta)$ is a time instant within time slot $n \in \{1,..,N\}$. $s_k(n, t) \sim {\mathcal{CN}}(0, 1)$ represents the transmitted information signal to Bob $k$ with the corresponding beamformer $\mathbf{w}_{k}[n] \in \mathbb{C}^{N_{\mathrm{A}} \times 1}$. The term $\mathbf{s}_{\mathrm{r}}(n,t)$ represents the sensing signal with the same deterministic covariance matrix $\mathbf{R}_{\mathrm{s}}[n] \in \mathbb{C}^{N_{\mathrm{A}} \times N_{\mathrm{A}}}$ for each signal sample in time slot $n$. In addition, the equivalent channel vector $\mathbf{h}_k^H[n]\triangleq\mathbf{h}_{\mathrm{d}, k}^H+\mathbf{h}_{\mathrm{r}, k}^H \boldsymbol{\Theta}[n] \mathbf{G}$ is comprised of the Alice-RIS link $\mathbf{G} \in \mathbb{C}^{N_{\mathrm{R}} \times N_{\mathrm{A}}}$, the RIS-Bob $k$ link $\mathbf{h}_{\mathrm{r}, k}^H \in \mathbb{C}^{1 \times N_{\mathrm{R}}}$, and the Alice-Bob $k$ link $\mathbf{h}_{\mathrm{d}, k}^H \in \mathbb{C}^{1 \times N_{\mathrm{A}}}$. $\boldsymbol{\Theta}[n] \triangleq \operatorname{diag}\left\{\boldsymbol{\theta}[n]\right\}$ with $\boldsymbol{\theta}[n]=\left[e^{j \theta_1[n]}, \ldots, e^{j \theta_{N_{\mathrm{R}}}[n]}\right]^T$ denotes the RIS phase shift matrix  \cite{Hu:2024, ma2022covert}. Considering the coordination nature among legitimate parties and with the help of some advanced channel estimation methods\cite{ruan2022approximate,zhou2022channel}, we assume the availability of perfect CSI for all links related to legitimate communication during the whole transmission period. Finally, $n_k(n, t) \sim \mathcal{C N}(0, \sigma_{\mathrm{b},k}^2)$  is
the additive white Gaussian noise (AWGN) at Bob $k$ with $\sigma_{b,k}^2$ denoting the noise variance.

\subsubsection{Sensing Signal}
In order to transmit and receive signals simultaneously, the full-duplex mode is considered at the transmitter side, so that the received echo signal at Alice reflected by Willie can be expressed as\footnote{Note that we omit the sensing signals reflected by RIS, i.e., the Alice-RIS-Willie-RIS-Alice link signals,  which suffer severe product-distance round-trip path-loss over the multiple signal reflections compared to the direct reflected link \cite{Ming:2023}.}\cite{Wei:2023,LiuP:2023}
\begin{equation}\label{echo}
\begin{split}
\mathbf{r}_{\mathrm{w}}(n, t)= & \underbrace{e^{j 2 \pi v_{\mathrm{w}}[n] t} \mathbf{H}_{\mathrm{rw}}[n] \sum_{k \in \mathcal{K}} \mathbf{w}_k[n] s_k\left(n, t-\tau_{\mathrm{w}}[n]\right)}_{\text {echo of communication signal }} \\
& +\underbrace{e^{j 2 \pi v_{\mathrm{w}}[n] t} \mathbf{H}_{\mathrm{rw}}[n] \mathbf{s}_{\mathrm{r}}\left(n, t-\tau_{\mathrm{w}}[n]\right)}_{\text {echo of sensing signal }}+\mathbf{n}_{\mathrm{r}}(n, t),
\end{split}
\end{equation}
where $v_{\mathrm{w}}[n]$ and $\tau_{\mathrm{w}}[n]$ denote the Doppler shift and round-trip time delay, respectively. $\mathbf{n}_{\mathrm{r}}(n, t)$ is the background AWGN with each entry obeying $\mathcal{C} \mathcal{N}\left(0, \sigma_\mathrm{r}^2\right)$. Due to the strong line-of-sight (LoS) nature for an aerial adversary Willie \cite{zeng2019accessing}, the target response channel matrix  $\mathbf{H}_{\mathrm{rw}}[n]$ is given by

\begin{equation}\label{echo_channel}
\mathbf{H}_{\mathrm{rw}}[n]=\sqrt{\frac{\rho_0 \varsigma}{d_{\mathrm{aw}}^4[n]}} \mathbf{a}\left(\theta_{\mathrm{aw}}[n], \phi_{\mathrm{aw}}[n]\right) \mathbf{a}^H\left(\theta_{\mathrm{aw}}[n], \phi_{\mathrm{aw}}[n]\right),
\end{equation}
where $d_{\mathrm{aw}}[n]$ is the distance between Alice and Willie. $\rho_0$ and $\varsigma$ represent the channel power at the reference distance of one meter and the radar cross-section (RCS) of Willie, respectively. The steering vector $\mathbf{a}\left(\theta, \phi\right) \in \mathbb{C}^{N_\mathrm{A} \times 1}$ is given by
\begin{equation}
\begin{aligned}
\mathbf{a}(\theta, \phi) & =\left[1, e^{j  \pi  \sin (\phi) \sin (\theta)}, \cdots, e^{j \pi \left(N_H-1\right)\sin (\phi) \sin (\theta)}\right]^T \\
& \quad \otimes\left[1, e^{j  \pi  \cos (\phi)}, \cdots, e^{j  \pi\left(N_V-1\right) \cos (\phi)}\right]^T,
\end{aligned}
\end{equation}
where we assume half-wavelength antenna spacing for UPA, while $\theta$ and $\phi$ represent the azimuth angle of departure (AoD) and the elevation AoD from Alice to Willie, respectively.

\subsection{Tracking Model}\label{Trackig-Mo}
Unlike the assumptions in \cite{ma2022covert,shuai:2023,Hu:2024} that the perfect/statistical CSI of the Alice-Willie link can be obtained, we consider a more practical scenario where Alice must actively estimate and track the channel of the mobile Willie. Thus, Alice is required to estimate the location of Willie first and thereby construct the channels for the Alice-Willie link and the RIS-Willie link to achieve effective covert transmission. To this end, we employ the EKF technique to track and predict the location of Willie based on the estimated velocity information \cite{Liu:2020,Wei:2023}. In particular, the known and fixed locations of Alice and RIS are denoted as $\mathbf{q}_{\mathrm{a}}=\left[x_{\mathrm{a}}, y_{\mathrm{a}}, z_{\mathrm{a}}\right]^T$ and $\mathbf{q}_{\mathrm{r}}=\left[x_{\mathrm{r}}, y_{\mathrm{r}}, z_{\mathrm{r}}\right]^T$, respectively. In time slot $n$, the location of Willie is denoted as $\mathbf{q}_{\mathrm{w}}[n]=\left[x_{\mathrm{w}}[n], y_{\mathrm{w}}[n], z_{\mathrm{w}}[n]\right]^T$ and the corresponding speed is $\dot{\mathbf{q}}_{\mathrm{w}}[n]=\left[\dot{x}_{\mathrm{w}}[n], \dot{y}_{\mathrm{w}}[n], \dot{z}_{\mathrm{w}}[n]\right]^T$. The state of Willie, i.e., $\boldsymbol{\chi}_{\mathrm{w}}[n]=\left[\mathbf{q}_{\mathrm{w}}^T[n], \dot{\mathbf{q}}_{\mathrm{w}}^T[n]\right]^T\in \mathbb{R}^{6 \times 1}$, is unknown to Alice and required to be estimated. The entire modeling and execution process of EKF are elaborated as follows:

\subsubsection{State evolution model of Willie}
Considering the non-cooperativity between Alice and Willie, we adopt a simple yet practical constant velocity movement model where the velocity of Willie is assumed to remain unchanged within two adjacent slots. The state evolution model of Willie is then given by
\begin{equation}\label{state-evo}
\boldsymbol{\chi}_{\mathrm{w}}[n]=\mathbf{F} \boldsymbol{\chi}_{\mathrm{w}}[n-1]+\mathbf{n}_{\boldsymbol{\chi}_{\mathrm{w}}},\mathbf{F}=\left[\begin{array}{cc}\mathbf{I}_3 & \delta \mathbf{I}_3 \\ \mathbf{0}_3 & \mathbf{I}_3\end{array}\right],
\end{equation}
where $\mathbf{F} \in \mathbb{R}^{6 \times 6}$ is the state transition matrix, $\mathbf{n}_{\boldsymbol{\chi}_{\mathrm{w}}}\sim \mathcal{N}\left(\mathbf{0}, \mathbf{Q}_{\boldsymbol{\chi}_{\mathrm{w}}}\right)$ denotes the state evolution noise vector with $\mathbf{Q}_{\boldsymbol{\chi}_{\mathrm{w}}}=\operatorname{diag}\left\{\sigma_{x_{\mathrm{w}}}^2, \sigma_{y_{\mathrm{w}}}^2, \sigma_{z_{\mathrm{w}}}^2, \sigma_{\dot{x}_{\mathrm{w}}}^2, \sigma_{\dot{y}_{\mathrm{w}}}^2, \sigma_{\dot{z}_{\mathrm{w}}}^2\right\}$ obtained based on the long-term measurements at Alice \cite{Wei:2023,wang2024sensing}.

\subsubsection{Measurement model of Willie}
The measurement model associated with $v_{\mathrm{w}}[n], \tau_{\mathrm{w}}[n], \theta_{\mathrm{aw}}[n]$, and $\phi_{\mathrm{aw}}[n]$ can be established as
\begin{equation}\label{mes-model}
\mathbf{Z}_{\mathrm{w}}[n]=\mathbf{g}\left(\boldsymbol{\chi}_{\mathrm{w}}[n]\right)+\mathbf{n}_{\mathbf{Z}_{\mathrm{w}}[n]},
\end{equation}
 where 
 \begin{equation*}
 \begin{aligned}
     &\mathbf{Z}_{\mathrm{w}}[n]=\left[\hat{\tau}_{\mathrm{w}}[n], \hat{v}_{\mathrm{w}}[n], \sin \hat{\theta}_{\mathrm{aw}}[n], \cos \hat{\theta}_{\mathrm{aw}}[n], \sin \hat{\phi}_{\mathrm{aw}}[n]\right]^T,\\ &\mathbf{n}_{\mathbf{Z}_{\mathrm{w}}}=\left[n_{\tau_{\mathrm{w}}[n]}, n_{v_{\mathrm{w}}[n]}, n_{\sin{\theta}_{\mathrm{w}}[n]}, n_{\cos{\theta}_{\mathrm{w}}[n]}, n_{\sin{\phi}_w[n]}\right]^T,
 \end{aligned}
 \end{equation*}
and the non-linear function $\mathbf{g}: \mathbb{R}^{6 \times 1} \rightarrow \mathbb{R}^{5 \times 1}$ represents the mapping from $\boldsymbol{\chi}_{\mathrm{w}}[n]$ to $\mathbf{Z}_{\mathrm{w}}[n]$. $\mathbf{n}_{\mathbf{Z}_{\mathrm{w}}[n]}$ denotes the zero-mean Gaussian noise vector with covariance matrix $\widehat{\mathbf{Q}}_{\mathbf{Z}_{\mathrm{W}[n]}}$.

According to the EKF method, we can linearize the nonlinear measurement model in \eqref{mes-model} as
\begin{equation}\label{mes-line}
\begin{aligned}
    &  \hat{\mathbf{Z}}_{\mathrm{w}}[n] \\&\approx\mathbf{g}\left(\hat{\boldsymbol{\chi}}_{\mathrm{w}}[n|n-1]\right)+\mathbf{J}_n\left(\boldsymbol{\chi}_{\mathrm{w}}[n]-\hat{\boldsymbol{\chi}}_{\mathrm{w}}[n|n-1]\right)+\mathbf{n}_{\mathbf{Z}_{\mathrm{w}}[n]},
\end{aligned}
\end{equation}
where $\hat{\boldsymbol{\chi}}_{\mathrm{w}}[n|n-1]$ is the predicted state of Willie in time slot $n$.
$\mathbf{J}_n=\left.\frac{\partial \mathbf{g}}{\partial \boldsymbol{\chi}_{\mathrm{w}}}\right|_{\boldsymbol{\chi}_{\mathrm{w}}=\hat{\boldsymbol{\chi}}_{\mathrm{w}}[n|n-1]}$ is the Jacobian matrix of $\mathbf{g}$.

The detailed derivations of \eqref{mes-model} and \eqref{mes-line} is presented in Appendix D.

\subsubsection{EKF process}\label{EKF-process}
Based on  \eqref{state-evo} and \eqref{mes-line}, the whole process of EKF can be specifically presented as below:
\begin{itemize}
    \item \textit{Predict the state of Willie $\hat{\boldsymbol{\chi}}_{\mathrm{w}}[n|n-1]$ }: 
    \begin{equation}\label{pre-state}
      \hat{\boldsymbol{\chi}}_{\mathrm{w}}[n|n-1]=\mathbf{F}\hat{\boldsymbol{\chi}}_{\mathrm{w}}[n-1],  
    \end{equation}
    where $\hat{\boldsymbol{\chi}}_{\mathrm{w}}[n-1] $ is the estimated state of Willie in time slot $n-1$.
    
    \item \textit{Predict the covariance matrix $\mathbf{C}[n|n-1]$}: 
    \begin{equation}\label{PRCM}
        \mathbf{C}[n|n-1]=\mathbf{F} \mathbf{C}[n-1] \mathbf{F}^T+\mathbf{Q}_{\boldsymbol{\chi}_{\mathrm{w}}}.
    \end{equation}

    \item \textit{Calculate the Kalman gain matrix $\mathbf{K}_n$}:
    \begin{equation}\label{KGM}
        \mathbf{K}_n=\mathbf{C}[n| n-1] \mathbf{J}_n^{T}\left(\mathbf{J}_n \mathbf{C}[n|n-1] \mathbf{J}_n^{T}+\widehat{\mathbf{Q}}_{\mathbf{Z}_{\mathrm{w}}[n]}\right)^{-1}.
    \end{equation}
    \item \textit{Update the corresponding posterior covariance matrix $\mathbf{C}[n]$}: 
\begin{equation}\label{PSCM}
\mathbf{C}[n]=\left(\mathbf{I}_6-\mathbf{K}_n \mathbf{J}_n\right) \mathbf{C}[n|n-1].
\end{equation}
\item  \textit{Estimate the state of Willie $\hat{\boldsymbol{\chi}}_{\mathrm{w}}[n]$}:
\begin{equation}\label{estimated-state}
\hat{\boldsymbol{\chi}}_{\mathrm{w}}[n]=\hat{\boldsymbol{\chi}}_{\mathrm{w}}[n|n-1]+\mathbf{K}_n\left(\hat{\mathbf{Z}}_{\mathrm{w}}[n]-\mathbf{J}_n\left(\hat{\boldsymbol{\chi}}_{\mathrm{w}}[n|n-1]\right)\right).
\end{equation}
\end{itemize}

\subsection{Performance Metrics}
Based on  the aforementioned signal and tracking models, we define the following performance metrics for the considered RIS-empowered ISACC system.
\subsubsection{Communication}
According to \eqref{comm-signal}, the covert sum rate of Bobs in time slot $n$ can be expressed as
\footnote{In the current work, the covert sum rate is formulated based on the asymptotic Shannon rate, which has been widely adopted in prior studies \cite{Hu:2024, Zhang:2022,ma2022covert,wang2024sensing,ma2021robust}.  Optimizing system design with this objective also leads to improved performance when finite blocklength effects are considered. While exact expressions for the sum rate under finite blocklength are not available, approximate formulations involving the asymptotic term and a channel-dispersion-related correction term have been proposed \cite{polyanskiy2010channel,nasir2020resource}. However, adopting such formulations would alter the structure of the objective function and make the optimization problem significantly less tractable, which requires solving techniques different from the current framework. Exploring this direction would indeed be valuable, and we consider it an interesting avenue for future work.}
\begin{equation}
\sum_{k\in \mathcal{K}} R_k[n]=\sum_{k\in \mathcal{K}} \log _2\left(1+\mathrm{SINR}_k\right),
\end{equation}
where the signal-to-interference-plus-noise ratio (SINR) of Bob $k$ is 
\begin{equation}
    \mathrm{SINR}_k=\frac{\left|\mathbf{h}_k^H[n] \mathbf{w}_k[n]\right|^2}{\sum_{j\in \mathcal{K}_k}\left|\mathbf{h}_k^H[n] \mathbf{w}_j[n]\right|^2+\mathbf{h}_k^H[n] \mathbf{R}_{\mathrm{s}}[n] \mathbf{h}_k[n]+\sigma_{\mathrm{b}, k}^2}.
\end{equation}

\subsubsection{Sensing}
By substituting \eqref{KGM} into \eqref{PSCM}, we have
\begin{equation}\label{PCSM=new}
    \mathbf{C}[n]=\left(\mathbf{C}^{-1}[n|n-1]+\mathbf{J}_n^{\mathrm{T}} \widehat{\mathbf{Q}}_{\mathbf{Z}_{\mathrm{w}}[n]}^{-1} \mathbf{J}_n\right)^{-1}.
\end{equation}
Note that $\operatorname{tr}(\mathbf{C}[n])$ characterizes the posterior mean square error (MSE) for tracking the state of Willie, which means that limiting the value of $\operatorname{tr}(\mathbf{C}[n])$ to a target threshold, e.g.,  $\mathrm{MSE}_{\max }$, can effectively ensure certain sensing accuracy, i.e., 
\begin{equation}
    \operatorname{tr}(\mathbf{C}[n]) \leq \mathrm{MSE}_{\max }.
\end{equation}

\begin{figure*}[!t]
\vspace{-0.3cm}
\begin{center}
\centering
\includegraphics[width=0.8\textwidth]{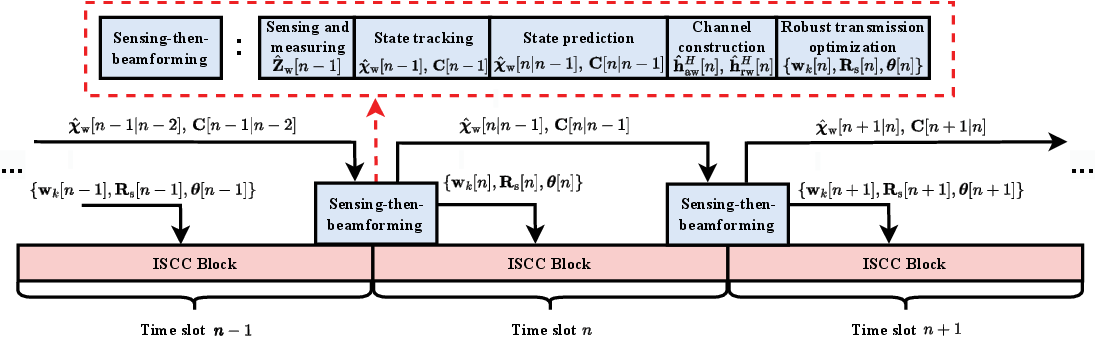}
\captionsetup{justification=centering}
\caption{ Diagram of the transmission  protocol of RIS-empowered  ISACC systems.}
\label{fig:protocl}
\end{center}
\vspace{-0.5cm} 
\end{figure*}
\subsubsection{Covertness}

In our considered system, the focus is on protecting the intention behind transmitting information signals. Therefore, Willie conducts a binary hypothesis test on his received signals in each time slot to determine whether or not Alice is communicating with Bobs. In this setting, the sensing signal serves a dual purpose: it enables active sensing and facilitates the tracking of Willie’s position, while simultaneously acting as artificial noise that helps obscure the presence of useful information. Specifically, considering that each time slot contains $L$ symbols, we denote the received signal of Willie in time slot $n$ as $\mathbf{y}_\mathrm{w}[n]=\left[y_\mathrm{w}[n, 1], y_\mathrm{w}[n, 2], \ldots y_\mathrm{w}[n, L]\right]\in \mathbb{C}^{1 \times L}$  and then the hypothesis test on $\mathbf{y}_\mathrm{w}[n]$ can be written as

\begin{equation}\label{hyp}
\mathbf{y}_\mathrm{w}[n]= \begin{cases}\mathbf{h}_{\mathrm{w}}^H[n]\mathbf{S}_{\mathrm{r}}[n]+\mathbf{n}_{\mathrm{w}}[n], & \mathcal{H}_0, \\ \mathbf{h}_{\mathrm{w}}^H[n]\left( \mathbf{W}_{\mathrm{c}}[n] \mathbf{S}_{\mathrm{c}}[n]+\mathbf{S}_{\mathrm{r}}[n]\right)+\mathbf{n}_{\mathrm{w}}[n], & \mathcal{H}_1,\end{cases}
\end{equation}
Here, $\mathbf{h}_{\mathrm{w}}^H[n] \triangleq \mathbf{h}_{\mathrm{aw}}^H[n]+\mathbf{h}_{\mathrm{rw}}^H[n] \boldsymbol{\Theta}[n] \mathbf{G}$, the Alice-Willie link $\mathbf{h}_{\mathrm{aw}}$ and RIS-Willie $\mathbf{h}_{\mathrm{rw}}$ link can be modeled as \footnote{Usually,  LoS propagation dominates the air-ground links according to the air-ground channel properties. Thus, similar to \cite{Wei:2023,bai2019energy,han2022joint,li2019secure}, adopting a LoS channel model for the link between the Alice/RIS deployed on the ground and Willie in the air  is a reasonable approximation. This is particularly valid in rural areas where there is minimal blockage and scattering or when Willie operates at a sufficiently high altitude \cite{zeng2019accessing}.} \begin{equation}\label{Willie-real-channel}
\mathbf{h}_{\mathrm{iw}}^H[n]=\sqrt{\frac{\rho_0}{d_{\mathrm{iw}}^2[n]}} \mathbf{a}^H\left(\theta_{\mathrm{iw}}[n], \phi_{\mathrm{iw}}[n]\right), \quad \forall \mathrm{i} \in\{\mathrm{a}, \mathrm{r}\}
\end{equation}
where $d_\mathrm{iw}[n]$, ${\theta}_{\mathrm{iw}}[n]$ , and $\phi_{\mathrm{iw}}$, $\forall \mathrm{i} \in\{\mathrm{a}, \mathrm{r}\}$ denote the distance between Alice/RIS and Willie, the azimuth AoD, and the elevation AoD from Alice/RIS to Willie, respectively. $\mathbf{W}_{\mathrm{c}}[n]\triangleq\left[\mathbf{w}_1[n], \ldots, \mathbf{w}_K[n]\right]$. $\mathbf{S}_{\mathrm{c}}[n] \triangleq\left[\mathbf{s}_{\mathrm{c}}[n, 1], \ldots, \mathbf{s}_{\mathrm{c}}[n, L]\right] \in \mathbb{C}^{K \times L}$ denotes the communication symbol matrix with i.i.d. elements $\sim \mathcal{C N}(0,1)$, and 
each column $\mathbf{s}_{\mathrm{c}}[n, l] =\left[s_1[n, l], \ldots, s_K[n, l]\right]^T$ with $l \in\{1, \ldots, L\}$ is the concatenation of each legitimate Bob’s $l$-th symbol. 
$\mathbf{S}_{\mathrm{r}}[n] \triangleq\left[\mathbf{s}_{\mathrm{r}}[n, 1], \ldots, \mathbf{s}_{\mathrm{r}}[n, L]\right] \in \mathbb{C}^{N_{\mathrm{A}} \times L}$ represents the sensing symbol matrix, and each symbol satisfies
\begin{equation}\label{sensing-cov}
   \mathbb{E}\left\{\mathbf{s}_{\mathrm{r}}[n, l] \mathbf{s}_{\mathrm{r}}[n, l]^H\right\}=\mathbf{R}_{\mathrm{s}}[n] \succeq \mathbf{0}, \forall l.
\end{equation}
Finally, $\mathbf{n}_{\mathrm{w}}[n] \triangleq[n_{\mathrm{w}}[n, l], \ldots, n_{\mathrm{w}}[n, l]]$ (with i.i.d. elements $\sim \mathcal{C N}(0, \sigma_{\mathrm{w}}^2)$) denotes the noise at Willie. 

Given $\mathbf{y}_\mathrm{w}[n]$, Willie makes a binary decision ($\mathcal{D}_1$ or $\mathcal{D}_0$) that infers whether or not the communication between Alice and Bobs is present in form of \eqref{hyp}. Assuming an equal prior probability for $\mathcal{H}_0$ and  $\mathcal{H}_1$ as in \cite{Zhang:2022}, the sum of detection error probabilities can be expressed as $\xi=\operatorname{Pr}\left(\mathcal{D}_1 |\mathcal{H}_0\right)+\operatorname{Pr}\left(\mathcal{D}_0| \mathcal{H}_1\right)$, where $ \operatorname{Pr}\left(\mathcal{D}_1 | \mathcal{H}_0\right)$ and $\operatorname{Pr}\left(\mathcal{D}_0 | \mathcal{H}_1\right)$ denote the false alarm
probability and the missed detection probability, respectively.  By using an optimal detector with the minimum detection error probability $\xi^{\star}$ at Willie,  the covertness constraint can be written as $\xi^{\star} \geq 1-\varepsilon$ for a required covertness level $\varepsilon>0$. However, due to the intractability of  $\xi^{\star}$, we instead lower-bound it via the Pinsker’s inequality \cite{shuai:2023,Zhang:2022} as $\xi^{\star} \geq 1-\sqrt{\frac{D\left(\mathbb{P}_0 \| \mathbb{P}_1\right)}{2}}$, where $\mathbb{P}_0$ and $\mathbb{P}_1$ are the probability distributions of $\mathbf{y}_{\mathrm{w}}[n]$ under  $\mathcal{H}_0$ and $\mathcal{H}_1$, respectively. 

Based on the construction of $\mathbf{y}_\mathrm{w}[n]$ in \eqref{hyp}, and the assumptions that $\mathbf{W}_{\mathrm{c}}[n]$ is fixed for the $n$-th slot and each sensing symbol satisfies \eqref{sensing-cov}, ${y_w[n,l], \forall l=1,..., L}$ are independently identical distributed. Therefore, we have 
\begin{equation}\label{P}
\begin{aligned}
&\mathbb{P}_0=\mathbb{P}\left(\mathbf{y}_{\mathrm{w}}[n] | \mathcal{H}_0\right)=\frac{1}{\left(\pi \lambda_0\right)^L} \exp \left(-\frac{\left|\mathbf{y}_{\mathrm{w}}[n]\right|^2}{\lambda_0}\right), \\
& \mathbb{P}_1=\mathbb{P}\left(\mathbf{y}_{\mathrm{w}}[n] | \mathcal{H}_1\right)=\frac{1}{\left(\pi \lambda_1\right)^L} \exp \left(-\frac{\left|\mathbf{y}_{\mathrm{w}}[n]\right|^2}{\lambda_1}\right),
\end{aligned}
\end{equation}
where $\lambda_0 \triangleq\mathbf{h}_{\mathrm{w}}^H[n] \mathbf{R}_{\mathrm{s}}[n]\mathbf{h}_{\mathrm{w}}[n]+\sigma_{\mathrm{w}}^2$ and $\lambda_1 \triangleq \mathbf{h}_{\mathrm{w}}^H[n]\left(\mathbf{R}_{\mathrm{s}}[n]+\mathbf{W}_{\mathrm{c}}[n] \mathbf{W}_{\mathrm{c}}^H[n]\right) \mathbf{h}_{\mathrm{w}}[n]+\sigma_{\mathrm{w}}^2$.

Then, the relative entropy $D\left(\mathbb{P}_0 \| \mathbb{P}_1\right)$ of $\mathbb{P}_0$ to $ \mathbb{P}_1$ can be quantified as\cite{cover1999elements,Zhang:2022}
\begin{equation}\label{covert-equ}
\begin{aligned}
D\left(\mathbb{P}_0 \| \mathbb{P}_1\right) & =\int_{-\infty}^{+\infty} \mathbb{P}_0 \ln \frac{\mathbb{P}_0}{\mathbb{P}_1} d y= L\left(\ln \frac{\lambda_1}{\lambda_0}+\frac{\lambda_0}{\lambda_1}-1\right),
\end{aligned}
\end{equation}
and the covertness constraint $D\left(\mathbb{P}_0 \| \mathbb{P}_1\right)\leq 2 \varepsilon^2$ is thus enforced in order to achieve covert communication at any given covertness level $\varepsilon>0$. Note that perfect knowledge of $\mathbf{h}_{\mathrm{aw}}$ and $\mathbf{h}_{\mathrm{rw}}$ in \eqref{covert-equ} is not available in practice. To address this issue, Alice will actively sense and track Willie’s location, as detailed in Section II-B, to obtain the estimates of $\mathbf{h}_{\mathrm{aw}}$ and $\mathbf{h}_{\mathrm{rw}}$. In the next section, we will formulate a robust covert transmission design that accounts for the estimation errors.

\section{Robust Transmission Design based on the Sensing-Then-Beamforming Framework}\label{alg-design}
In this section, we first elaborate the transmission  protocol of the considered  RIS-empowered ISACC system, followed by the sensing-then-beamforming framework, and then proceed to design the robust transmission scheme.

\subsection{Transmission Protocol}\label{Protocol}
As shown in Fig.~\ref{fig:protocl}, Alice can first send the sensing signal $\mathrm{s}_{\mathrm{r}}(n, t)$ and then obtain the measurement values $\hat{\mathbf{Z}}_\mathrm{w}[n-1]$ based on the echo signal in \eqref{echo}. Combining $\hat{\mathbf{Z}}_\mathrm{w}[n-1]$ with the prior obtained $\hat{\boldsymbol{\chi}}_{\mathrm{w}}[n-1|n-2]$ and $ \mathbf{C}[n-1 |n-2]$, we can correct the final estimated state $\hat{\boldsymbol{\chi}}_{\mathrm{w}}[n-1]$ and posterior covariance matrix $\mathbf{C}[n-1]$ according to \eqref{estimated-state} in time slot $n-1$. 

Then, based on the first-order Markov process for modeling the movement of Willie in \eqref{pre-state}, Alice can obtain the predicted state $\hat{\boldsymbol{\chi}}_{\mathrm{w}}[n|n-1]$ and construct the Alice-Willie link $\widehat{\mathbf{h}}_{\mathrm{aw}}^H[n]$ and RIS-Willie link $\widehat{\mathbf{h}}_{\mathrm{rw}}^H[n]$ in time slot $n$ by replacing $d_\mathrm{iw}[n]$, $\theta_{\mathrm{iw}}[n]$, and $\phi_{\mathrm{iw}}[n]$ in \eqref{Willie-real-channel} with the corresponding predicted values $\hat{d}_\mathrm{iw}[n|n-1]$, $\hat{\theta}_{\mathrm{iw}}[n|n-1]$, and $\hat{\phi}_{\mathrm{iw}}[n|n-1]$, respectively, which are given by 
\begin{equation}\label{channl-par}
\begin{aligned}
&\hat{d}_{\mathrm{iw}}[n|n-1]=\left\|\hat{\mathbf{q}}_{\mathrm{w}}[n |n-1]-\mathbf{q}_{\mathrm{i}}\right\|,\\&\hat{\phi}_{\mathrm{iw}}[n|n-1]=\sin ^{-1} \left(\frac{\hat{z}_{\mathrm{w}}[n|n-1]-z_{\mathrm{i}}}{\hat{d}_{\mathrm{iw}}[n|n-1]}\right),\\ &\hat{\theta}_{\mathrm{iw}}[n|n-1]=\\
&\sin ^{-1}\left(\frac{\hat{y}_{\mathrm{w}}[n| n-1]-y_{\mathrm{i}}}{\sqrt{\left|\hat{x}_{\mathrm{w}}[n | n-1]-x_{\mathrm{i}}\right|^2+\left|\hat{y}_{\mathrm{w}}[n | n-1]-y_{\mathrm{i}}\right|^2}}\right), \\ & \quad \quad\quad\quad\quad\quad\quad\quad\quad\quad\quad\quad\quad\quad\quad\quad\quad\quad\quad\quad \forall \mathrm{i} \in\{\mathrm{a}, \mathrm{r}\}.
\end{aligned}
\end{equation}

However, due to the inability to accurately determine the true position of Willie, the EKF method itself will introduce estimation errors and the real channels $\mathbf{h}_{\mathrm{aw}}^H[n]$ and  $\mathbf{h}_{\mathrm{rw}}^H[n]$ can be modeled as \cite{ma2022covert,Wei:2023}
\begin{equation}\label{channel-willie}
\mathbf{h}_{\mathrm{aw}}^H[n]=\widehat{\mathbf{h}}_{\mathrm{aw}}^H[n]+\boldsymbol{\Delta}_{\mathrm{aw}}[n], \mathbf{h}_{\mathrm{rw}}^H[n]=\widehat{\mathbf{h}}_{\mathrm{rw}}^H[n]+\boldsymbol{\Delta}_{\mathrm{rw}}[n],
\end{equation}
where we characterize the channel estimation errors as follows
\begin{equation}\label{channel-err_set}
\begin{aligned}
    &\mathcal{U}_{\mathrm{aw}}[n] \triangleq\left\{\boldsymbol{\Delta}_{\mathrm{aw}}[n]\left\|\boldsymbol{\Delta}_{\mathrm{aw}}[n]\right\|_2^2 \leq \delta_{\mathrm{aw}}^2[n]\right\},\\
    &\mathcal{U}_{\mathrm{rw}}[n] \triangleq\left\{\boldsymbol{\Delta}_{\mathrm{rw}}[n]\left\|\boldsymbol{\Delta}_{\mathrm{rw}}[n]\right\|_2^2 \leq \delta_{\mathrm{rw}}^2[n]\right\}.
\end{aligned}
\end{equation}
Note that $\delta_{\mathrm{aw}}[n]$ and $\delta_{\mathrm{rw}}[n]$ represent the upper bounds of the corresponding CSI errors related to the tracking MSE of Willie and is given by 
\begin{equation} \label{error-chan}  
\delta_{\mathrm{aw}}^2[n]=\left\|\boldsymbol{\gamma}_{\mathrm{aw}}^{\mathrm{H}} \mathbf{C}[n |n-1]\right\|^2,\delta_{\mathrm{rw}}^2[n]=\left\|\boldsymbol{\gamma}_{\mathrm{rw}}^{\mathrm{H}} \mathbf{C}[n |n-1]\right\|^2
\end{equation}
where parameters $\boldsymbol{\gamma}_{\mathrm{aw}}\in \mathbb{R}^{6 \times 1}$ and  $\boldsymbol{\gamma}_{\mathrm{rw}}\in \mathbb{R}^{6 \times 1}$ can be obtained via
Monte Carlo simulation as in \cite{wei2022safeguarding}. 

Subsequently, by considering the channel estimation errors in \eqref{channel-willie}, Alice can optimize the robust transmission strategy to simultaneously execute integrated sensing and covert communication operations within the ISACC architecture.

Finally, by exploiting the echo signal in time slot $n$, the new measurements  $\hat{\mathbf{Z}}_{\mathrm{w}}[n]$ are taken and the estimated state $\hat{\boldsymbol{\chi}}_{\mathrm{w}}[n]$ is calculated, which also serve as the input in the next time slot to predict the new state of Willie.

\textit{Remark 1}: : Alice can employ advanced detection instruments, such as the “Ghostbuster” introduced in \cite{chaman2018detecting,chaman2018ghostbuster}, to identify the passive warden by detecting its signal leakage. The initial location of the warden can then be estimated based on this leakage, as described in \cite{chaman2018detecting,chaman2018ghostbuster,mukherjee2012detecting}. Additionally, dual-functional Alice can emit sensing signal to detect Willie before sending any information to Bobs. By analyzing the reflected echoes and removing signals corresponding to known legitimate Bobs, Alice can determine Willie's initial position \cite{su2024sensing}. It is noted that this initial detection and localization process typically precedes the tracking of mobile warden Willie, which is the primary focus of our study.

\subsection{Problem Formulation}
Based on the predicted channels and considering the channel estimation errors in \eqref{channel-willie}, we focus on a communication-centric robust transmission design, aiming to enhance the covert sum rate for Bobs. This is achieved by jointly optimizing the communication beamformers, the sensing signal covariance matrix at Alice, and the phase shifts at the RIS, while adhering to constraints on covert communication, sensing performance, transmit power, and the unit modulus of the RIS reflection coefficients. Specifically, for covert communication provisioning, we make a worst-case assumption regarding the covertness requirement of Willie and therefore  the robust optimization problem in time slot $n$ can be formulated as
\begin{equation}\label{P1}
\begin{array}{rl}
\mathcal{P}_1: & \underset{\boldsymbol{\theta}[n], \mathbf{R}_{\mathrm{s}}[n],\mathbf{w}_k[n]}{\max}\sum_{k\in \mathcal{K}} R_k[n]\\
\text { s.t. } & \mathrm{C1}: \operatorname{tr}\left(\sum_{k\in \mathcal{K}} \mathbf{w}_k[n] \mathbf{w}_k^H[n]+\mathbf{R}_{\mathrm{s}}[n]\right) \leq P_{\mathrm{t}}, \\
 &\mathrm{C2}:\underset{\substack{\boldsymbol{\Delta}_{\mathrm{aw}}[n] \in \mathcal{U}_{\mathrm{aw}}[n] \\ \boldsymbol{\Delta}_{\mathrm{rw}}[n] \in \mathcal{U}_{\mathrm{rw}}[n]}}{\mathrm{max}}  D\left(\mathbb{P}_0 \| \mathbb{P}_1\right) \leq 2 \varepsilon^2 , \\
& \mathrm{C3}:\operatorname{tr}(\mathbf{C}[n]) \leq \mathrm{MSE}_{\max}, \\
&\mathrm{C4}: |\boldsymbol{\theta}_i[n]|^2=1, \quad i \in\left[1: N_{\mathrm{R}}\right], \\&\mathrm{C}5:\mathbf{R}_{\mathrm{s}}[n] \succeq \mathbf{0}.
\end{array}
\end{equation}
Herein, $\mathrm{C1}$ is imposed to ensure that the transmit power of Alice is limited to her maximum power $P_\mathrm{t}$. Based on the channel estimation errors in (28),  $\mathrm{C2}$ can guarantee the minimum covertness requirement, where $D\left(\mathbb{P}_0 \| \mathbb{P}_1\right)$ is a function of $\boldsymbol{\theta}[n]$, $\mathbf{R}_\mathrm{s}[n]$, and $\mathbf{w}_k[n]$ according to \eqref{covert-equ}. $\mathrm{C3}$ depends on $\mathbf{R}_\mathrm{s}[n]$, and it ensures Alice's effective sensing ability for tracking Willie's motion trajectory, which, in turn, affects the channel estimation errors in the next time slot, as detailed in \eqref{PCSM=new} and \eqref{error-chan}. It is worth noting that the coupling between $\mathrm{C2}$ and $\mathrm{C3}$ is established through $\mathbf{R}_\mathrm{s}$, and this coupling affects the optimal power allocation between sensing and communication. Finally, $\mathrm{C4}$ reflects the unit modulus of the passive elements at the  RIS, and $\mathrm{C5}$ is imposed as the covariance matrix $\mathbf{R}_{\mathrm{s}}[n]$ of the sensing signal is a Hermitian PSD matrix.

However, solving $\mathcal{P}_1$ is quite challenging for the following reasons: $\text{1)}$ the optimization variables are strongly coupled, $\text {2)}$ the continuity of the CSI uncertainty sets $\mathcal{U}_{\text {aw }}[n]$ and $\mathcal{U}_{\text {rw }}[n]$ in $\mathrm{C}2$ leads to infinitely many non-convex constraints, $\text {3)}$ the sensing constraint in $\mathrm{C}3$ and unit modulus constraints in  $\mathrm{C}4$ are non-convex.
Additionally, compared with the traditional ISAC works \cite{xu2024robust,Ming:2023,liu2022}, it is very challenging to construct and deal with the covert constraint $\mathrm{C}2$. On the one hand, due to the non-cooperative mechanism between Alice and Willie, Alice needs to leverage sensing function to predict CSI about Willie to construct $\mathrm{C}2$. On the other hand, considering the complex form of $\mathrm{C}2$ including double CSI error characteristics, we need to further carefully analyze and convert $\mathrm{C}2$ to facilitate algorithm design. In the following
subsections, we develop an efficient AO-based algorithm to solve $\mathcal{P}_1$  effectively. For notational simplicity, we omit the time slot index $n$ in the following subsections. The key step of the AO-based algorithm is to solve each subproblem given the fixed solutions of the other subproblems, thus we first decompose the complicated $\mathcal{P}_1$  into two subproblems with regard to $\left\{\mathbf{R}_{\mathrm{s}}, \mathbf{w}_k\right\}$ and $\left\{{\boldsymbol{\theta}}\right\}$ in what follows. 
\subsection{Subproblem With Respect to \texorpdfstring{$\left\{\mathbf{R}_{\mathrm{s}}, \mathbf{w}_k\right\}$}{Problem-{Rs,Wk}}}\label{AO-1}
By fixing ${\boldsymbol{\theta}}$, the subproblem with respect to $\left\{\mathbf{R}_{\mathrm{s}}, \mathbf{w}_k\right\}$ can be expressed as

\begin{equation}\label{P2-1}
\begin{array}{rl}
\mathcal{P}_2: & \underset{ \mathbf{R}_{\mathrm{s}},\mathbf{w}_k}{\max}\quad \sum_{k \in \mathcal{K}} R_k\\
\text { s.t } & \mathrm{C1}: \operatorname{tr}\left(\sum_{k \in \mathcal{K}} \mathbf{w}_k \mathbf{w}_k^H+\mathbf{R}_{\mathrm{s}}\right) \leq P_{\mathrm{t}},\\
&\mathrm{C2}:\underset{\substack{\boldsymbol{\Delta}_{\mathrm{aw}} \in \mathcal{U}_{\mathrm{aw}} \\ \boldsymbol{\Delta}_{\mathrm{rw}} \in \mathcal{U}_{\mathrm{rw}}}}{\mathrm{max}}  D\left(\mathbb{P}_0 \| \mathbb{P}_1\right) \leq 2 \varepsilon^2,  \\
& \mathrm{C3}:\operatorname{tr}(\mathbf{C}) \leq \mathrm{MSE}_{\max}, \\&\mathrm{C}5:\mathbf{R}_{\mathrm{s}} \succeq \mathbf{0}.
\end{array}
\end{equation}
However, problem $\mathcal{P}_2$ is still non-convex due to the non-convex constraints $\mathrm{C} 2$ and $\mathrm{C} 3$. To address this challenge, we first define $\mathbf{W}_k \triangleq \mathbf{w}_k \mathbf{w}_k^H$ that satisfies $\operatorname{rank}\left(\mathbf{W}_k\right)=1 $ and $\mathbf{W}_k \succeq \mathbf{0}, \forall k$, then the objective function and constraints of $\mathcal{P}_2$ can be translated into more tractable forms as follows.
\subsubsection{Objective function of \texorpdfstring{$\mathcal{P}_2$}{Obj-func-P2}}
We can rewrite the objective function of $\mathcal{P}_2$ in the form of the difference of two concave functions, i.e., 
    \begin{equation}\label{C1-NEW}
    \begin{aligned}
        \sum_{k \in \mathcal{K}} R_k&=\sum_{k \in \mathcal{K}} \log _2\left(\frac{\sum_{i\in \mathcal{K}} \mathbf{h}_k^H \mathbf{W}_i^H \mathbf{h}_k+\mathbf{h}_k^H \mathbf{R}_{\mathrm{s}} \mathbf{h}_k+\sigma_{\mathrm{b}, k}^2}{\sum_{i\in \mathcal{K}_k} \mathbf{h}_k^H \mathbf{W}_i^H \mathbf{h}_k+\mathbf{h}_k^H \mathbf{R}_{\mathrm{s}} \mathbf{h}_k+\sigma_{\mathrm{b}, k}^2}\right)\\
        &=Q_1-P_1,
    \end{aligned}
    \end{equation}
where 
\begin{equation}\label{Q1P1}
\begin{aligned}
& Q_1 \triangleq\sum_{k \in \mathcal{K}} \log _2\left(\sum_{i \in \mathcal{K}} \mathbf{h}_k^H \mathbf{W}_i\mathbf{h}_k+\mathbf{h}_k^H \mathbf{R}_{\mathrm{s}} \mathbf{h}_k+\sigma_{\mathrm{b}, k}^2\right), \\
& P_1 \triangleq\sum_{k \in \mathcal{K}} \log _2\left(\sum_{i \in \mathcal{K}_k} \mathbf{h}_k^H \mathbf{W}_i \mathbf{h}_k+\mathbf{h}_k^H \mathbf{R}_{\mathrm{s}} \mathbf{h}_k+\sigma_{\mathrm{b}, k}^2\right).
\end{aligned}
\end{equation}

Unfortunately, \eqref{C1-NEW} is still a non-convex objective function. Thus, we adopt the SCA method to address this difficulty . For any feasible $\mathbf{W}^{(m)}=\{\mathbf{W}_k^{(m)}\}_{k \in \mathcal{K}}$ and $\mathbf{R}_{\mathrm{s}}^{(m)}$, where $m$ denotes the iteration index, $P_1$  could be upper bounded as 
\begin{equation}\label{P1-UPPER}
\begin{aligned}
& P_1 \leq P_1^{(m)}+\operatorname{tr}\left(\nabla_{\mathbf{R}_{\mathrm{s}}}^H P_1^{(m)}\left(\mathbf{R}_{\mathrm{s}}-\mathbf{R}_{\mathrm{s}}^{(m)}\right)\right)\\& \quad\quad+\sum_{k \in \mathcal{K}} \operatorname{tr}\left(\nabla_{\mathbf{w}_k}^H P_1^{(m)}\left(\mathbf{W}_k-\mathbf{W}_k^{(m)}\right)\right),
\end{aligned}
\end{equation}
where $P_1^{(m)}=P_1\left(\mathbf{W}^{(m)}, \mathbf{R}_{\mathrm{s}}^{(m)}\right)$ and 
\begin{equation*}
\begin{aligned}
& \nabla_{\mathbf{R}_{\mathrm{s}}}^H P_1\left(\mathbf{W}^{(m)}, \mathbf{R}_{\mathrm{s}}^{(m)}\right)= \\
& \frac{1}{\ln 2} \sum_{j \in \mathcal{K}}\frac{\mathbf{h}_j \mathbf{h}_j^H}{\operatorname{tr}\left(\mathbf{h}_j \mathbf{h}_j^H \mathbf{R}_{\mathrm{s}}^{(m)}\right)+\sum_{i \in \mathcal{K}_j} \operatorname{tr}\left(\mathbf{h}_j \mathbf{h}_j^H \mathbf{W}_i^{(m)}\right)+\sigma_{\mathrm{b}, j}^2}, \\
& \nabla_{\mathbf{w}_k}^H P_1\left(\mathbf{W}^{(m)}, \mathbf{R}_{\mathrm{s}}^{(m)}\right)= \\
& \frac{1}{\ln 2} \sum_{j\in \mathcal{K}_k} \frac{\mathbf{h}_j \mathbf{h}_j^H}{\operatorname{tr}\left(\mathbf{h}_j \mathbf{h}_j^H \mathbf{R}_{\mathrm{s}}^{(m)}\right)+\sum_{i\in \mathcal{K}_j} \operatorname{tr}\left(\mathbf{h}_j \mathbf{h}_j^H \mathbf{W}_i^{(m)}\right)+\sigma_{\mathrm{b}, j}^2}.
\end{aligned}
\end{equation*}

Therefore, the original objective function in $\mathcal{P}_2$ can be transformed into a concave function based on the approximation in \ $\mathcal{P}_2$ to facilitate the design of an effective algorithm.

\subsubsection{Covertness constraint in \texorpdfstring{$\mathrm{C}2$}{Covert-C2} } 
We first introduce the following proposition to translate $\mathrm{C}2$ into a tractable form.
    \begin{proposition}\label{pro-covert-trans}
     Considering the channel estimation errors in \eqref{channel-willie}, the covertness constraint in $\mathrm{C}2$ can be equivalently transformed into 
    \begin{equation}\label{C2-NEW1}
    \begin{aligned}
         &\mathrm{C} 2 &\Leftrightarrow&\underset{\substack{\boldsymbol{\Delta}_{\mathrm{aw}} \in \mathcal{U}_{\mathrm{aw}} \\ \boldsymbol{\Delta}_{\mathrm{rw}} \in \mathcal{U}_{\mathrm{rw}}}}{\mathrm{max}} \left(\mathbf{h}_{\mathrm{aw}}^H+\mathbf{h}_{\mathrm{rw}}^H \boldsymbol{\Theta} \mathbf{G}\right) \mathbf{R}\left(\mathbf{h}_{\mathrm{aw}}^H+\mathbf{h}_{\mathrm{rw}}^H \boldsymbol{\Theta} \mathbf{G}\right)^H\\ & & &\leq(\eta_2-1) \sigma_{\mathrm{w}}^2,
    \end{aligned}
    \end{equation}
    where $\mathbf{R}=\mathbf{W}_{\mathrm{c}} \mathbf{W}_{\mathrm{c}}^H-(\eta_2-1) \mathbf{R}_{\mathrm{s}}=\sum_{k \in \mathcal{K}} \mathbf{W}_k-(\eta_2-1) \mathbf{R}_{\mathrm{s}}$, and $\eta_2 \geq1$ is the solution to the equation $\ln x+\frac{1}{x}-1-\frac{2 \varepsilon^2}{L}=0$.
    \end{proposition}

    \textit{Proof}: 
Let  $h(x)=\ln x+\frac{1}{x}-1-\frac{2 \varepsilon^2}{L}$. Then, it can be readily seen that there exist two roots $\eta_1$ and $\eta_2$ (assuming that $\eta_1<1<\eta_2$) that satisfy  $h(x)=0$. Consequently, $\mathrm{C}2$  can be equivalently converted to
\begin{equation}
    \underset{\substack{\boldsymbol{\Delta}_{\mathrm{aw}} \in \mathcal{U}_{\mathrm{aw}} \\ \boldsymbol{\Delta}_{\mathrm{rw}} \in \mathcal{U}_{\mathrm{rw}}}}{\mathrm{max}}  D\left(\mathbb{P}_0 \| \mathbb{P}_1\right) \leq 2 \varepsilon^2  \stackrel{(a)} {\Leftrightarrow} \underset{\substack{\boldsymbol{\Delta}_{\mathrm{aw}} \in \mathcal{U}_{\mathrm{aw}} \\ \boldsymbol{\Delta}_{\mathrm{rw}} \in \mathcal{U}_{\mathrm{rw}}}}{\mathrm{max}}  \frac{\lambda_1}{\lambda_0} \leq \eta_2,
\end{equation}
where ${(a)}$ holds because $\frac{\lambda_1}{\lambda_0} \geq 1$ and the fact that $h(x)$ is monotonically increasing when $x \geq 1$. Furthermore, based on the expression of
$\mathbf{h}_{\mathrm{w}}^H$ defined in \eqref{hyp}, we have 
\begin{equation}
\begin{aligned}
& \max _{\substack{\Delta_{\mathrm{aw}} \in \mathcal{U}_{\mathrm{aw}} \\
\Delta_{\mathrm{rw}} \in \mathcal{U}_{\mathrm{rw}}}} \frac{\lambda_1}{\lambda_0} \leq \eta_2 \Leftrightarrow \max _{\substack{\Delta_{\mathrm{aw}} \in \mathcal{U}_{\mathrm{aw}} \\
\Delta_{\mathrm{rw}} \in \mathcal{U}_{\mathrm{nv}}}} \frac{\mathbf{h}_{\mathrm{w}}^H \mathbf{W}_{\mathrm{c}} \mathbf{W}_{\mathrm{c}}^H \mathbf{h}_{\mathrm{w}}}{\mathbf{h}_{\mathrm{w}}^H \mathbf{R}_{\mathrm{s}} \mathbf{h}_{\mathrm{w}}+\sigma_{\mathrm{w}}^2} \leq \eta_2-1 \\
& \Leftrightarrow \max _{\substack{\Delta_{\mathrm{aw}} \in \mathcal{U}_{\mathrm{aw}} \\
\Delta_{\mathrm{rw}} \in \mathcal{U}_{\mathrm{rw}}}}\left(\mathbf{h}_{\mathrm{aw}}^H+\mathbf{h}_{\mathrm{rw}}^H \boldsymbol{\Theta} \mathbf{G}\right) \mathbf{R}\left(\mathbf{h}_{\mathrm{aw}}^H+\mathbf{h}_{\mathrm{rw}}^H \boldsymbol{\Theta} \mathbf{G}\right)^H \\ & \quad \thinspace \thinspace\leq(\eta_2-1) \sigma_{\mathrm{w}}^2.
\end{aligned}
\end{equation}
The proof is thus completed.   \hfill $\blacksquare$  
    
Although the transformed form in \eqref{C2-NEW1} is more tractable than the original form in $\mathrm{C}2$, it is still non-convex due to the continuity of the CSI uncertainty sets $\mathcal{U}_{\mathrm{aw}}$ and $\mathcal{U}_{\mathrm{rw}}$. To tackle this issue, we introduce the following Lemma and transform the non-convex constraint in \eqref{C2-NEW1} into a more tractable form in terms of linear matrix inequality (LMI).

\begin{lemma} (S-Procedure\cite{1994Linear}): Let
\begin{equation}\label{S-Pro}
f_i\left(\mathbf{x}\right)=\mathbf{x}^H \mathbf{P}_i \mathbf{x}+2 \Re\left\{\mathbf{q}_i^H \mathbf{x}\right\}+q_i,\quad i \in[0,P],
\end{equation}
where $\mathbf{P}_i\in \mathbb{H}^N$,$\mathbf{q}_i \in \mathbb{C}^N$ and $q_i\in \mathbb{R}$. Then, the implication relation  $f_0(\mathbf{x}) \leq 0$ for all $\mathbf{x}$ with $\left\{f_i(\mathbf{x}) \leq 0\right\}_{i=1}^P$ holds if and only if there exists a variable $\mu_i \geqslant0$ such that
\begin{equation}
 \sum_{i=1}^P \mu_i\left[\begin{array}{cc}
\mathbf{P}_i & \mathbf{q}_i \\
\mathbf{q}_i^{H} & q_i
\end{array}\right]\succeq \left[\begin{array}{cc}
\mathbf{P}_0 & \mathbf{q}_0 \\
\mathbf{q}_0^{H} & q_0
\end{array}\right].
\end{equation}
\end{lemma}

Based on Lemma 1, the following Proposition~\ref{pro-covert-w} that can transform \eqref{C2-NEW1} into an LMI form is presented.
\begin{proposition}\label{pro-covert-w}
   By inserting \eqref{channel-willie} into the covertness constraint in \eqref{C2-NEW1}, $\mathrm{C2}$ can be equivalently transformed into the following form\footnote{Different from \cite{Ming:2023}, the LMI form obtained based on Proposition~\ref{pro-covert-w} is more concise. Moreover, the dimension of the LMI in \eqref{C2-NEW2} is $N_\mathrm{A}+N_\mathrm{R}+1$, which entails much lower computational complexity as compared to the LMI with dimension $N_\mathrm{A}+N_\mathrm{A}N_\mathrm{R}+1$ in \cite{Ming:2023}.} 
\begin{equation}\label{C2-NEW2}
\mathrm{C}2\Leftrightarrow\overline{\mathrm{C2a} }:\mathbf{\Pi}\triangleq\left[\begin{array}{cc}
\mathbf{P}+\boldsymbol{\mu} & \mathbf{q} \\
\mathbf{q}^H & q
\end{array}\right] \succeq \mathbf{0}\thinspace \mathrm{and}\thinspace\overline{\mathrm{C2b} }: \mu_1, \mu_2 \geq0, 
\end{equation}

\end{proposition}
where $\mathbf{P}$, $\boldsymbol{\mu}$, and $q$ are given in Appendix A, $\mu_1\geq0$ and $\mu_2\geq0$ are slack variables.

\textit{Proof}: Please refer to  Appendix A. \hfill $\blacksquare$

\subsubsection{Sensing MSE constraint in \texorpdfstring{$\mathrm{C}3$}{C3}} To deal with the non-convex constraint $\mathrm{C}3$, we first introduce auxiliary variables $c_i \geq 0$, $i \in\{1, \ldots, 6\}$ to bound the diagonal entries of the
posterior covariance matrix $\mathbf{C}$, i.e., $\left\{\mathbf{C}\right\}_{i i} \leq c_i$. Then, $\mathrm{C}3$ in (\ref{P2-1}) can be equivalently rewritten as
\begin{equation}\label{C3-NEW}
\begin{aligned}
&\mathrm{C} 3 \Leftrightarrow \\
&\overline{\mathrm{C3a} }:\sum_{i=1}^6 m_i \leq \mathrm{MSE}_{\max }\thinspace \mathrm{and} \thinspace\overline{\mathrm{C3b} }:\left[\begin{array}{cc}
\mathbf{C}^{-1} & \mathbf{e}_i \\
\mathbf{e}_i^T & c_i
\end{array}\right] \succeq \mathbf{0}, \forall i,
\end{aligned}
\end{equation}
where $\mathbf{e}_i \in \mathbb{R}^{6 \times 1}$ is the $i$-th column of $\mathbf{I}_6$.

Based on the derivation above, problem $\mathcal{P}2$ can be rewritten as
\begin{equation}\label{P2-2}
\begin{array}{ll}
\bar{\mathcal{P}}_2:&\underset{\mathbf{R}_{\mathrm{s}},\mathbf{W}_k,\lambda_1,\lambda_2,c_i}{\max} Q_1-\tilde{P}_1 \\
\text { s.t } & \mathrm{C1}: \operatorname{tr}\left(\sum_{k \in \mathcal{K}} \mathbf{W}_k+\mathbf{R}_{\mathrm{s}}\right) \leq P_\mathrm{t},\\&\overline{\mathrm{C2a}}-\overline{\mathrm{C2b}}: \eqref{C2-NEW2},\thinspace \overline{\mathrm{C3a}}-\overline{\mathrm{C3b}}: \eqref{C3-NEW},\\
& \mathrm{C}5:\mathbf{R}_{\mathrm{s}} \succeq \mathbf{0}, \quad\mathrm{C6}: \mathbf{W}_k \succeq \mathbf{0}, \forall k,\\&\mathrm{C7}: \mathrm{rank}\left(\mathbf{W}_k\right)=1, \forall k,
\end{array}
\end{equation}
where $\tilde{P}_1=\sum_{k\in \mathcal{K}} \operatorname{tr}(\nabla_{\mathbf{w}_k}^H P_1(\mathbf{W}^{(m)}, \mathbf{R}_{\mathrm{s}}^{(m)}) \mathbf{W}_k)+\operatorname{tr}(\nabla_{\mathbf{R}_{\mathrm{s}}}^H P_1(\mathbf{W}^{(m)}, \mathbf{R}_{\mathrm{s}}^{(m)}) \mathbf{R}_{\mathrm{s}})$.

Omitting the rank-one constraint $\mathrm{C7}$, $\bar{\mathcal{P}}_2$ is jointly convex with respect to  $\mathbf{R}_{\mathrm{s}}$, $\mathbf{W}_k$, $\lambda_1$, $\lambda_2$, and $c_i$, which can be solved by off-the-shelf convex optimization solvers, such as CVX\cite{grant2009cvx}. In fact, if the relaxed problem is feasible, the rank-one constraints $\mathrm{C7}$ can always be satisfied, as we will prove in Proposition~\ref{pro-rank}.

\begin{proposition}\label{pro-rank}
    If the relaxed problem $\bar{\mathcal{P}}_2$ is feasible, the optimal solution $\mathbf{W}_k^{{\star}}$ always satisfies the rank-one condition $\mathrm{C7}$.
\end{proposition}

\textit{Proof}: Please refer to  Appendix B. \hfill $\blacksquare$

\subsection{Subproblem With Respect to \texorpdfstring{$\{\boldsymbol{\theta}\}$}{Problem-theta}}
Then, fixing $\left\{\mathbf{R}_{\mathrm{s}}, \mathbf{w}_k\right\}$, the RIS phase-shift  design subproblem is given by 
\begin{equation}\label{P3-1}
\begin{array}{rl}
\mathcal{P}_3: & \underset{\boldsymbol{\theta}}{\max}\quad\sum_{k \in \mathcal{K}} R_k\\
\text { s.t } & \mathrm{C2}:\underset{\substack{\boldsymbol{\Delta}_{\mathrm{aw}} \in \mathcal{U}_{\mathrm{aw}} \\ \boldsymbol{\Delta}_{\mathrm{rw}}\in \mathcal{U}_{\mathrm{rw}}}}{\mathrm{max}}  D\left(\mathbb{P}_0 \| \mathbb{P}_1\right) \leq 2 \varepsilon^2,  \\
& \mathrm{C4}: |\boldsymbol{\theta}_i|^2=1, \quad i \in\left[1: N_{\mathrm{R}}\right],
\end{array}
\end{equation}
where the remaining non-convexity is caused by $\mathrm{C2}$ and $\mathrm{C4}$. Similar to Section~\ref{AO-1}, we will transform the constraints and objective function in $\mathcal{P}_3$ into tractable forms. Before handling this challenging issue, we first introduce an auxiliary variable $\beta \in \mathbb{C}$ with $|\beta|=1$ and then define
\begin{equation}\label{V-define}
    \mathbf{v}\triangleq\left[\beta\boldsymbol{\theta}^H, \beta\right]^T,\quad\mathbf{V}\triangleq\mathbf{v} \mathbf{v}^H=\left[\begin{array}{cc}\boldsymbol{\theta}^* \boldsymbol{\theta}^T & \boldsymbol{\theta}^* \\ \boldsymbol{\theta}^T & 1\end{array}\right].
\end{equation}

\subsubsection{Objective function of \texorpdfstring{$\mathcal{P}_3$}{Obj-fun-P3}}
We first define  $\mathbf{G}_k\triangleq\operatorname{diag}\left(\mathbf{h}_{\mathrm{r}, k}^H\right) \mathbf{G}$, $\mathbf{T}_k \triangleq\left[\mathbf{G}_k^H, \mathbf{h}_{\mathrm{d}, k}\right]$, $\mathbf{\Phi}_{k, i} \triangleq \mathbf{T}_k^H \mathbf{W}_i \mathbf{T}_k$ and $\mathbf{\Psi}_k \triangleq \mathbf{T}_k^H \mathbf{R}_{\mathrm{s}} \mathbf{T}_k$. Then, based on  \eqref{Q1P1}, we can rewrite $Q_1$ as follows

\begin{equation}\label{C1-V}
\begin{aligned}
 Q_1&=\sum_{k\in \mathcal{K}} \log _2\left(\sum_{i\in\mathcal{K}} \mathbf{h}_k^H \mathbf{W}_i \mathbf{h}_k+\mathbf{h}_k^H \mathbf{R}_{\mathrm{s}} \mathbf{h}_k+\sigma_{\mathrm{b}, k}^2\right) \\
& =\sum_{k\in \mathcal{K}} \log _2\left(\begin{array}{l}
\operatorname{tr}\left(\boldsymbol{\theta}^* \boldsymbol{\theta}^T \mathbf{G}_k\left(\sum_{i\in \mathcal{K}} \mathbf{W}_i+\mathbf{R}_{\mathrm{s}}\right) \mathbf{G}_k^H\right. \\
\left.+\boldsymbol{\theta}^* \mathbf{h}_{\mathrm{d}, k}^H\left(\sum_{i\in \mathcal{K}} \mathbf{W}_i+\mathbf{R}_{\mathrm{s}}\right) \mathbf{G}_k^H\right. \\
\left.+\boldsymbol{\theta}^T \mathbf{G}_k\left(\sum_{i\in \mathcal{K}} \mathbf{W}_i+\mathbf{R}_{\mathrm{s}}\right) \mathbf{h}_{\mathrm{d}, k}\right) \\+\mathbf{h}_{\mathrm{d}, k}^H\left(\sum_{i\in \mathcal{K}} \mathbf{W}_i+\mathbf{R}_{\mathrm{s}}\right) \mathbf{h}_{\mathrm{d}, k}+\sigma_{\mathrm{b}, k}^2
\end{array}\right) \\
&  =\sum_{k\in \mathcal{K}} \log _2\left(\sum_{i\in \mathcal{K}} \operatorname{tr}\left(\boldsymbol{\Phi}_{k, i} \mathbf{V}\right)+\operatorname{tr}\left(\boldsymbol{\Psi}_k \mathbf{V}\right)+\sigma_{\mathrm{b}, k}^2\right).
\end{aligned}
\end{equation}
Similarly, $P_1$ can be equivalently transformed into
\begin{equation}
    P_1=\sum_{k \in \mathcal{K}}^K \log _2\left(\sum_{j\in \mathcal{K}_k}^K \operatorname{tr}\left(\boldsymbol{\Phi}_{k, j} \mathbf{V}\right)+\operatorname{tr}\left(\boldsymbol{\Psi}_k \mathbf{V}\right)+\sigma_{\mathrm{b}, k}^2\right).
\end{equation}

To tackle the non-concavity of the objective
function, we adopt a similar SCA method as in \eqref{P1-UPPER} to handle $P_1$. Specifically, for  any feasible $\mathbf{V}^{(m)}$, $P_1$ can be upper bounded as 
\begin{equation}
P_1 \leq P_1^{(m)}+\operatorname{tr}\left(\nabla_{\mathbf{V}}^H P_1\left(\mathbf{V}^{(m)}\right)\left(\mathbf{V}-\mathbf{V}^{(m)}\right)\right),
\end{equation}
where 
\begin{equation*}
\begin{aligned}
    &\nabla_{\mathbf{V}}^H P_1\left(\mathbf{V}^{(m)}\right)\\
    &=\frac{1}{\ln 2} \sum_{k \in \mathcal{K}} \frac{\sum_{j \in \mathcal{K}_k} \boldsymbol{\Phi}_{k, j}+\boldsymbol{\Psi}_k}{\sum_{j \in \mathcal{K}_k} \operatorname{tr}\left(\boldsymbol{\Phi}_{k, j} \mathbf{V}^{(m)}\right)+\operatorname{tr}\left(\boldsymbol{\Psi}_k \mathbf{V}^{(m)}\right)+\sigma_{\mathrm{b}, k}^2}.
\end{aligned}
\end{equation*}

\subsubsection{Covertness constraint in \texorpdfstring{$\mathrm{C2}$}{Covert-C2}}
By leveraging the following proposition, the infinite infinitely many constraints of $\mathrm{C} 2$ in \eqref{P3-1} can be transformed into an LMI form with respect to $\mathbf{V}$.
\begin{proposition}\label{pro-covert-ris}
   By inserting \eqref{channel-willie} into the covertness constraint in \eqref{P3-1}, $\mathrm{C} 2$ can be equivalently transformed into the following form
\begin{equation}\label{C2-NEW3}
\mathrm{C}2\Leftrightarrow\overline{\overline{\mathrm{C2a} }}:\overline{\mathbf{\Pi}}\triangleq\left[\begin{array}{cc}
\widehat{\mathbf{P}}+\widehat{\boldsymbol{\mu}} & \widehat{\mathbf{q}} \\
\widehat{\mathbf{q}}^H & \widehat{{q}}
\end{array}\right] \succeq \mathbf{0} \thinspace\mathrm{ and }\thinspace\overline{\overline{\mathrm{C2b} }}: \widehat{\mu}_1, \widehat{\mu}_2 \geq0, 
\end{equation}

\end{proposition}
where $\widehat{\mathbf{P}}$, $\widehat{\boldsymbol{\mu}}$, and $\widehat{q}$ are given in Appendix C, $\widehat{\mu}_1\geq0$ and $\widehat{\mu}_2\geq0$ are slack variables.

\textit{Proof}: Please refer to  Appendix C. \hfill $\blacksquare$

\subsubsection{The non-convex unit-modulus constraints in  \texorpdfstring{$\mathrm{C4}$}{unit-modulus-C4}} Based on the definition in \eqref{V-define}, the non-convex unit-modulus constraints in $\mathrm{C4}$ can be equivalently represented as
\begin{equation}
\begin{aligned}
    \mathrm{C} 4 \Leftrightarrow &\mathrm{C} 8: \operatorname{diag}(\mathbf{V})=\mathbf{1}_{N_{\mathrm{R}}+1},\\
    & \mathrm{C} 9: \mathbf{V} \succeq \mathbf{0}\thinspace \mathrm{and}\thinspace\mathrm{C} 10: \operatorname{rank}(\mathbf{V})=1,
\end{aligned}
\end{equation}
where $\mathrm{C} 8$ can guarantee the
unit modulus of $\boldsymbol{\theta}$, $\mathrm{C} 9$ and $\mathrm{C} 10$ are introduced to guarantee that $\mathbf{V}=\mathbf{v} \mathbf{v}^H$.

The non-convexity of constraint $\mathrm{C} 10$ is addressed by adopting the sequential rank-one constraint relaxation (SROCR) technique, which systematically relaxes the rank-one condition via a progressive relaxation strategy. Unlike conventional rank-one relaxation approaches \cite{cao2017sequential} that completely discard this constraint, the SROCR method achieves computationally tractable solutions while preserving essential problem structure through controlled constraint relaxation. In particular, we first equivalently rewrite the rank-one constraint as\cite{Zhangxianda} 
\begin{equation}\label{rank1-new}
\mathrm{C}10 \Leftrightarrow \overline{\mathrm{C} 10}:\lambda_{\max }(\mathbf{V})=\operatorname{tr}(\mathbf{V})=\underset{\substack{\mathbf{x} \in \mathbb{C}^{(N_{\mathrm{R}}+1)\times1},\\\|\mathbf{x}\|_2 \leq 1} }{\max}\mathbf{x}^H \mathbf{V} \mathbf{x},
\end{equation}
where $\mathbf{x}$ is a slack vector and the second equality holds if and only if $\mathbf{x}=\mathbf{u}_{\max }\left(\mathbf{V}\right)$.

According to \cite{cao2017sequential}, the basic idea of SROCR is to partially relax the rank one constraint $\overline{\mathrm{C} 10}$ in \eqref{rank1-new} as 
\begin{equation}
\label{C10-new}
\overline{\overline{\mathrm{C} 10}}: \mathbf{u}_{\max }^H(\mathbf{V}^{(m)}) \mathbf{V} \mathbf{u}_{\max }(\mathbf{V}^{(m)}) \geq \widehat{\mu}_3^{(m)} \operatorname{tr}(\mathbf{V}^{(m)}),
\end{equation}
where $\widehat{\mu}_3^{(m)}$ is a slack parameter to control the ratio of the largest eigenvalue
of $\mathbf{V}$ to the trace of $\mathbf{V}$, and the rank-one constraint is satisfied until $\widehat{\mu}_3^{(m)}$ increases to 1.

Based on the above derivations, the final optimization problem of $\boldsymbol{\theta}$ can be rewritten as
\begin{equation}\label{P3-2}
\begin{array}{ll}
\bar{\mathcal{P}}_3: & \underset{\mathbf{V}, \mu_1, \mu_2}{\max}\quad Q_1-\bar{P}_1\\
\text { s.t } & \overline{\overline{\mathrm{C2a}}}-\overline{\overline{\mathrm{C2b}}}: \eqref{C2-NEW3},\thinspace \mathrm{C} 8: \operatorname{diag}(\mathbf{V})=\mathbf{1}_{N_{\mathrm{R}}+1},\\ & \mathrm{C} 9: \mathbf{V} \succeq \mathbf{0},\thinspace \overline{\overline{\mathrm{C} 10}}:\eqref{C10-new}.
\end{array}
\end{equation}
where $\bar{P}_1=\operatorname{tr}\left(\nabla_{\mathbf{V}}^H P_2\left(\mathbf{V}^{(m)}\right)\mathbf{V}\right)$. 
It is readily observed that $\bar{\mathcal{P}}_3$ is convex and can be solved by existing convex optimization solvers. In particular, the slack parameter $\widehat{\mu}_3^{(m+1)}$ can be updated by 
\begin{equation}\label{update-socr}
\widehat{\mu}_3^{(m+1)}=\min (1, \frac{\lambda_{\max }(\mathbf{V}^{(m)})}{\operatorname{tr}(\mathbf{V}^{(m)})}+\delta^{(m)}),
\end{equation}
where $\delta^{(m)}$ is the step size that is reduced iteratively by  $\delta^{(m+1)}=\frac{\delta^{(m)}}{2}$ until $\bar{\mathcal{P}}_3$ is solvable\cite{cao2017sequential}. 

\begin{algorithm}
\caption{AO-based algorithm to solve problem $\mathcal{P}_1$}
\label{alg:1}
\begin{algorithmic}[1]
\REQUIRE
 $\varepsilon$, $\mathrm{MSE}_{\text {max }}$, $P_{\mathrm{t}}$.
\STATE { Set the outer iteration index $m_o=0$ and initialize $\mathbf{W}_k^{(0)}$, $\mathbf{R}_{\mathrm{s}}^{(0)}$, $\boldsymbol{\theta}^{(0,0)}$}.
\REPEAT
\STATE Update $\{\mathbf{R}_{\mathrm{s}}^{(m_o+1)},\mathbf{W}_k^{(m_o+1)}\}$ by solving  $\overline{\mathcal{P}2}$ with given  $\{\mathbf{R}_{\mathrm{s}}^{(m_o)},\mathbf{W}_k^{(m_o)}\}$.
\STATE Set $\mathbf{V}^{(m_o,0)}$ with given $\boldsymbol{\theta}^{(m_o,0)}$ according to \eqref{V-define}.
 \STATE Initialize the step size $\delta^{(0)} $, set the inner iteration index $m_i=0$ and the slack
parameter $\widehat{\mu}_3^{(0)}=0$.
\REPEAT
\IF{$\bar{\mathcal{P}}_3$  is feasible}
\STATE Update $\mathbf{V}^{(m_o, m_i+1)}$ by solving $\bar{\mathcal{P}}_3$ in \eqref{P3-2} with given $\mathbf{V}^{(m_o, m_i)}$.
\STATE  Update $\delta^{(m_i+1)}=\delta^{(m_i)}$.
\ELSE 
\STATE  Update $\delta^{(m_i+1)}=\frac{\delta^{(m_i)}}{2}$.
\ENDIF
\STATE Update $\widehat{\mu}_3^{(m_i+1)}$ according to \eqref{update-socr}.
\STATE $m_i=m_i+1$.
\UNTIL $(f_{\bar{\mathcal{P}}_3}^{\left(m_o, m_i+1\right)}-f_{\bar{\mathcal{P}}_3}^{\left(m_o, m_i\right)})/f_{\bar{\mathcal{P}}_3}^{(m_o, m_i)}\leq \varepsilon_{\mathrm{in,1}}$ and $\lambda_{\max }(\mathbf{V}^{\left(m_o, m_i+1\right)}) / \operatorname{tr}(\mathbf{V}^{t(m_o, m_i+1)}) \leq \varepsilon_{\mathrm{in,2}}$.

\STATE Obtain $\boldsymbol{\theta}^{(m_o+1, 0)}$ by decomposing ${\mathbf{V}^{(m_o, m_i)}}$  and $m_o=m_o+1$.
\UNTIL $\sum_{k \in \mathcal{K}}(R_k^{(m_o+1)}-R_k^{(m_o)}) / \sum_{k\in \mathcal{K}} R_k^{(m_o)} \leq \varepsilon_{\mathrm{out }}$.
\ENSURE
$\{\mathbf{W}_k, \mathbf{R}_{\mathrm{s}}, \boldsymbol{\theta}\}$.
\end{algorithmic}
\end{algorithm}

Finally, Algorithm~\ref{alg:1} summarizes all the key steps of solving the original problem in $\mathcal{P}_1$, where $f_{\bar{\mathcal{P}}_3}$ denotes the objective function value in $\bar{\mathcal{P}}_3$. After optimizing  the communication beamformers, the covariance matrix of the sensing signal, and the RIS phase shifts, Alice (assisted by RIS) can transmit communication signals covertly and meanwhile send the sensing signal to sense Willie. At the same time, based on the echo signal of the sensing signal, Alice will obtain the new measurements about Willie in time slot $n$ and then correct the estimated state information of Willie as mentioned in Section~\ref{Protocol}. 

\textit{Remark 2}:  While the proposed Algorithm~\ref{alg:1} is developed under the covertness constraint $D\left(\mathbb{P}_0 \| \mathbb{P}_1\right) \leq 2 \varepsilon^2$,  it can be readily adapted to handle the alternative $D\left(\mathbb{P}_1 \| \mathbb{P}_0\right) \leq 2 \varepsilon^2$. In this case, the resulting robust optimization problem resembles problem $\mathcal{P}_1$ in \eqref{P1}, with the primary difference being the modified covert constraint. Similar to Proposition 1,  the constraint $D\left(\mathbb{P}_1 \| \mathbb{P}_0\right) \leq 2 \varepsilon^2$  can be reformulated as $\frac{\lambda_0}{\lambda_1} \geq \eta_1$, given that $\frac{\lambda_0}{\lambda_1} \leq 1$  and the monotonicity of  $h(x)$ with respect to $x$ for $x \geq1$. Therefore, the relaxation and transformation techniques proposed in this paper remain applicable for addressing the modified optimization problem under $ D\left(\mathbb{P}_1 \| \mathbb{P}_0\right) \leq 2 \varepsilon^2$. For brevity and due to page limitation, the detailed derivations and adjustments are omitted.

\subsection{Convergence and Complexity analysis}
It is worth noting that we can obtain a lower bound on the optimal objective value of $\mathcal{P}_2$ by solving $\bar{\mathcal{P}}_2$. Then, According to \cite[Theorem 1]{cao2017sequential}, the SROCR-based method can converge to a KKT solution of $\bar{\mathcal{P}}_3$  with the original rank-one constraint $\mathrm{C10}$, which means that the lower bound of the optimal objective values of $\mathcal{P}_3$ can still be obtained by solving $\bar{\mathcal{P}}_3$. Finally, we can monotonically
tighten these lower bounds by alternatively optimizing $\left\{\mathbf{W}_k, \mathbf{R}_{\mathrm{s}}\right\}$ and $\left\{\boldsymbol{\theta}\right\}$. The objective values achieved by the sequence $\{\mathbf{W}_k^{(m_o)}, \mathbf{R}_{\mathrm{s}}^{(m_o)}, \boldsymbol{\theta}^{(m_o,0)}\}_{m_o \in \mathbb{N}}$ form a non-decreasing sequence that converges to a stationary value in polynomial time, and any limit point of the sequence $\{\mathbf{W}_k^{(m_o)}, \mathbf{R}_{\mathrm{s}}^{(m_o)}, \boldsymbol{\theta}^{(m_o,0)}\}_{m_o \in \mathbb{N}}$ is a stationary point of the original problem $\mathcal{P}_1$\cite{sun2016majorization}. 

Moreover, it is clear that the main complexity of Algorithm~\ref{alg:1}
arises from solving $\bar{\mathcal{P}}_2$ and $\bar{\mathcal{P}}_3$, and the approximated complexity of utilizing the interior point method to solve them are $\mathcal{O}((K+3) N_{\mathrm{A}}^{3.5}+(K+3)^2 N_{\mathrm{A}}^{2.5}+K^3)$ and $\mathcal{O}(l_2(3 N_{\mathrm{R}}^{3.5}+18 N_{\mathrm{R}}^{2.5}+27 N_{\mathrm{R}}^{1.5}))$, respectively, where $l_2$ denotes the iteration number required for solving $\bar{\mathcal{P}_3}$. Therefore, the total complexity of Algorithm~\ref{alg:1} is given by $\mathcal{O}(l_{\mathrm{AO}}((K+3) N_{\mathrm{A}}^{3.5}+(K+3)^2 N_{\mathrm{A}}^{2.5}+K^3))+\mathcal{O}(l_{\mathrm{AO}}l_2(3 N_{\mathrm{R}}^{3.5}+18 N_{\mathrm{R}}^{2.5}+27 N_{\mathrm{R}}^{1.5}))$, where $l_{\mathrm{AO}}$ is the number of outer iterations.

\section{Simulation Results}\label{sim-result}
\begin{figure}
 		\centering
 		\includegraphics[width=7cm,height=5.2cm]{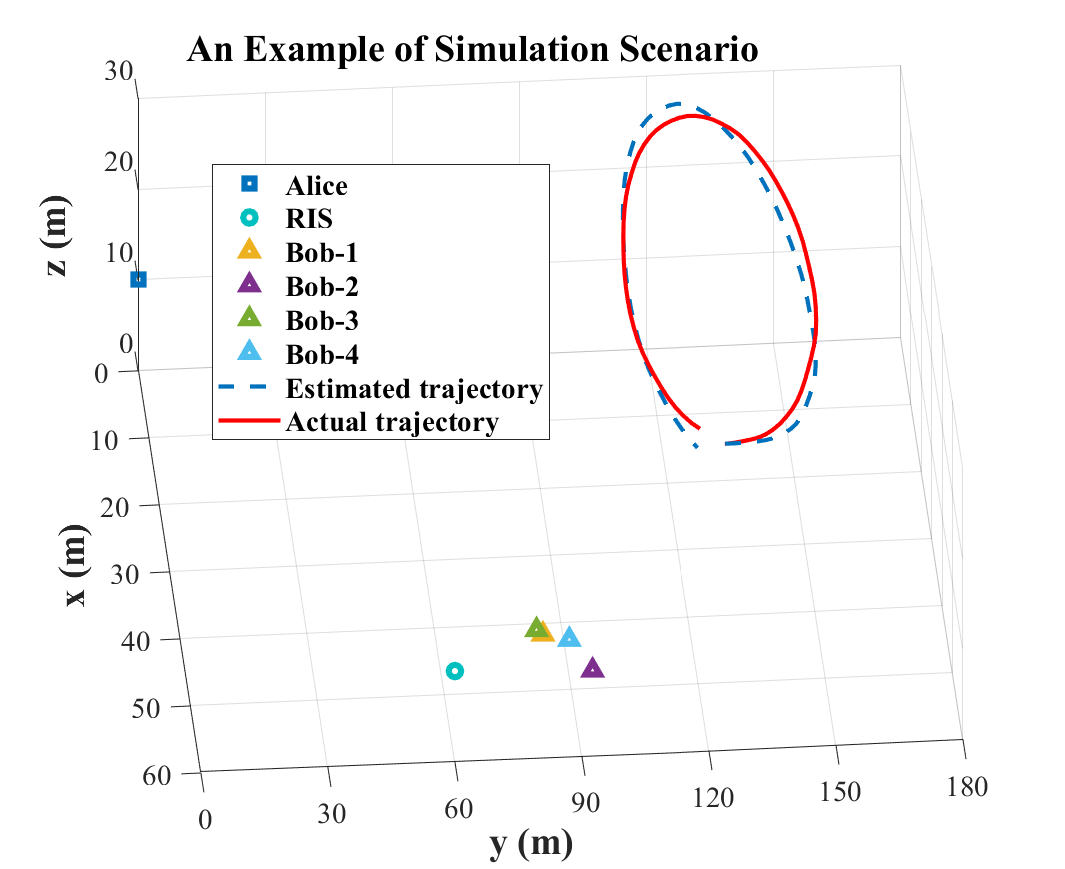}
 		\caption{An example of our simulation scenario. }
 		\label{fig_sim_sce}
        \vspace{-0.3cm} 
 \end{figure}

\begin{table*}[htbp]
\vspace{-0.7cm}
\small
\caption{System Parameters}
\label{TAB2}
\centering
\resizebox{1.98\columnwidth}{!}{
\begin{tabular}{!{\vrule width 1pt}c|c!{\vrule width 1pt}!{\vrule width 1pt}c|c!{\vrule width 1pt}}
\Xhline{1pt}
Parameter & Value & Parameter & Value \\ \Xhline{1pt}
Number of Antenna at Alice &$\left[N_H, N_V\right]=[5,2]$ &  Frame duration  & $T=7$ s  \\ \hline 
Number of Passive elements at RIS  & $N_\mathrm{R} = 8 \sim 16$ & Slot duration  & $\delta=0.1$ s \\ \hline 
RCS of Willie & $\varsigma=1$  & Number of symbols in each slot  & $L=1000$   \\ \hline 
Noise power at Bob & $\sigma_{\mathrm{b}, 1}^2=\cdots=\sigma_{\mathrm{b}, K}^2=-90$ dBm & Matched-filter gain & $\mathrm{G}_{\mathrm{MF}}=1000$\\ \hline 
Noise power at Willie & $\sigma_{\mathrm{w}}^2=-90 $ dBm & Center frequency & $f_c=3$ GHz\\ \hline 
Noise power at Alice for sensing & $\sigma_{\mathrm{r}}^2=-100 $ dBm & Modeling parameter of $\tau_{\mathrm{w}}$ & $c_{\tau_{\mathrm{w}}}=10^{-6}$ \\ \hline 
Transmit power at Alice & $P_{\mathrm{t}}=26 \sim 34$ dBm  & Modeling parameter of $v_{\mathrm{w}}$ & $c_{v_{\mathrm{w}}}=10^{5}$ \\ \hline 
Covertness level & $\varepsilon=0.05 \sim 0.25$ & Modeling parameter of $\theta_{\mathrm{aw }}$ & $c_{\theta_{\mathrm{aw}}}=1$ \\ \hline 
Sensing threshold & $\mathrm{MSE}_{\mathrm{max }}=7 \sim 12$ & Modeling parameter of $\phi_{\mathrm{aw }}$ & $c_{\phi_{\mathrm{aw}}}=1$\\ \hline  Path loss exponents of RIS-Bob link  & $\alpha_{\mathrm{RB},1}=\cdots=\alpha_{\mathrm{RB},K}=2$  & Channel power at one meter & $\rho_0=-30$ dB\\ \hline  Path loss exponents of Alice-Bob link  & $\alpha_{\mathrm{AB},1}=\cdots=\alpha_{\mathrm{AB},K}=4$   & Path loss exponents of Alice-RIS link  & $\alpha_{\mathrm{AR}}=2.2$ \\  \Xhline{1pt}
\end{tabular}}
\vspace{-0.5cm}
\end{table*}
In this section, we present some simulation results  to evaluate the effectiveness of the proposed algorithms. We consider a RIS-empowered  ISACC system as shown in Fig.~\ref{fig_sim_sce}, where Alice and RIS are located at $[0 \mathrm{~m}, 0 \mathrm{~m}, 10 \mathrm{~m}]$ and $[60 \mathrm{~m}, 60 \mathrm{~m}, 10 \mathrm{~m}]$, respectively. The legitimate communication users, i.e., Bobs, are randomly and uniformly distributed within a radius of $5 \mathrm{~m}$ in a circle centered at $[45 \mathrm{~m}, 90 \mathrm{~m}, 10 \mathrm{~m}]$, the number of Bobs is set as $K=4$, and each Alice-Bob $k$ link follows the Rayleigh flat fading model \cite{zhou2020framework}. The large-scale fading of all channels are modeled as $P L=\rho_0-10 \alpha \log _{10}(d)$ dB with the path loss exponent $\alpha$ and  the link distance $d$ in meter \cite{zhou2020framework}. We consider a mobile Willie (such as UAV) with its flight speed fixed at $20 \mathrm{~m} / \mathrm{s}$, and the variation of Willie direction is randomly distributed in $[-\pi / 18,0]$. The convergence threshold related parameters of Algorithm~\ref{alg:1} are set as $\varepsilon_{\mathrm{out}}=\varepsilon_{\mathrm{in,1}}=10^{-4}$ and $ \varepsilon_{\mathrm{in,2}}=0.999$. Other specific system parameters adopted for our simulations are listed in Table~\ref{TAB2}.

To verify the efficiency of the proposed algorithm, we also compare it with four baseline schemes as follows:
\begin{itemize}
    \item \textbf{Baseline~1}: In this case, we extend the single user scenario in \cite{wang2024sensing} to a multi-user scenario, while assuming that RIS is not optimized. Similar to \cite{Yu:2020}, we optimize the communication beamformers and sensing signal covariance matrix by setting $\boldsymbol{\Theta}=\mathbf{I}_{N_\mathrm{R}}$ and then solve $\overline{\mathcal{P}2 }$ in \eqref{P2-2}.
    \item \textbf{Baseline 2}: In this case, instead of adopting SROCR-based method, RIS is optimized based on the widely adopted SDR-based Gaussian randomization method \cite{xu2024robust,wang2023joint}, while the solutions for $\mathbf{W}_k$ and $\mathbf{R}_\mathrm{s}$ remain consistent with Algorithm \ref{alg:1}.
    \item \textbf{Baseline~3}:  In this case, we consider a non-robust design scheme that ignores the channel estimation errors in \eqref{channel-err}, i.e., the covert design in Eq.~(26) directly using $\widehat{\mathbf{h}}_{\mathrm{aw}}$ and $\widehat{\mathbf{h}}_{\mathrm{rw}}$ instead of ${\mathbf{h}}_{\mathrm{aw}}$ and ${\mathbf{h}}_{\mathrm{rw}}$\cite{shuai:2023,ma2022covert}.
    \item \textbf{Baseline~4}: In this case, the sensing signals are not used to predict or construct Willie’s CSI across time slots to assist covert communication design \cite{hu2024covert}; instead, only the initial CSI of Willie is employed for robust transmission design.
\end{itemize}

Meanwhile, we also consider an ideal case with perfect CSI at Willie (labeled as ``\textbf{Perfect CSI}'')  as an upper bound for evaluating the performance limit of the proposed scheme.
 
\textit{Remark 3}: Both Baseline 1 and Baseline 2 incorporate robust design principles, and thus serve as benchmarks for evaluating the covert sum rate. In contrast, Baseline 3 and Baseline 4 fail to achieve robust covertness, as the former neglects predicted channel errors and the latter fails to update the corresponding CSI. Therefore, they are included as comparisons to evaluate the robustness of covertness.

\begin{figure}
    		\centering
 		\includegraphics[width=8cm,height=4cm]{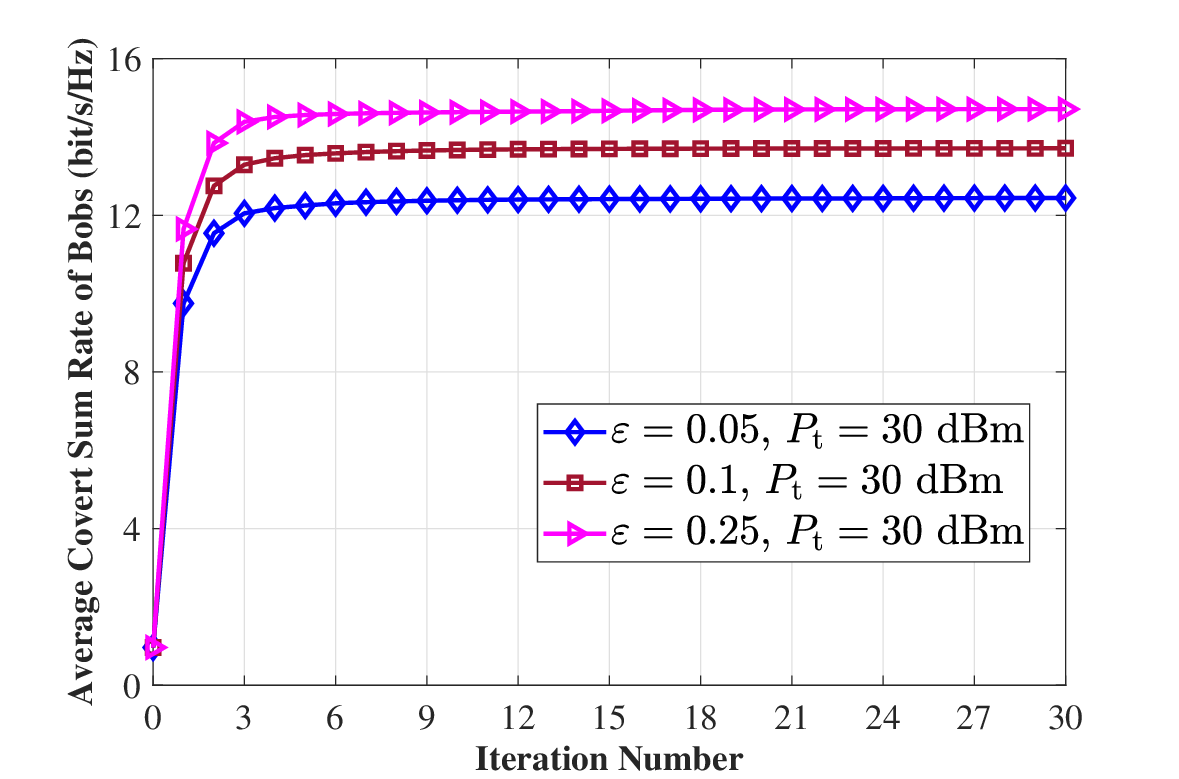}
 		\caption{Convergence properties of Algorithm~\ref{alg:1} with $N_\mathrm{A}=N_\mathrm{R}=10$ and $\mathrm{MSE}_{\max }=7$.}
 		\label{fig_converge}
                \vspace{-0.5cm} 
 \end{figure}

We first investigate the convergence properties of the proposed Algorithm~\ref{alg:1} with $N_\mathrm{A}=N_\mathrm{R}=10$ and $\mathrm{MSE}_{\max }=7$. Fig.~\ref{fig_converge} shows the relationship between the average covert sum rate of Bobs and the iteration number of algorithm under different covertness levels. It can be observed that the proposed Algorithm~\ref{alg:1} quickly converges to a steady value after about 6 iterations. Meanwhile, setting a more relaxed covertness level constraint can also improve the covert sum rate. This is because the more relaxed covertness level constraint prevents Alice from excessively reducing the energy of communication beamformers to meet the covert requirements and achieves a better covert sum rate.

\begin{figure}
\vspace{-0.5cm} 
 		\centering
 		\includegraphics[width=6.5cm,height=5.2cm]{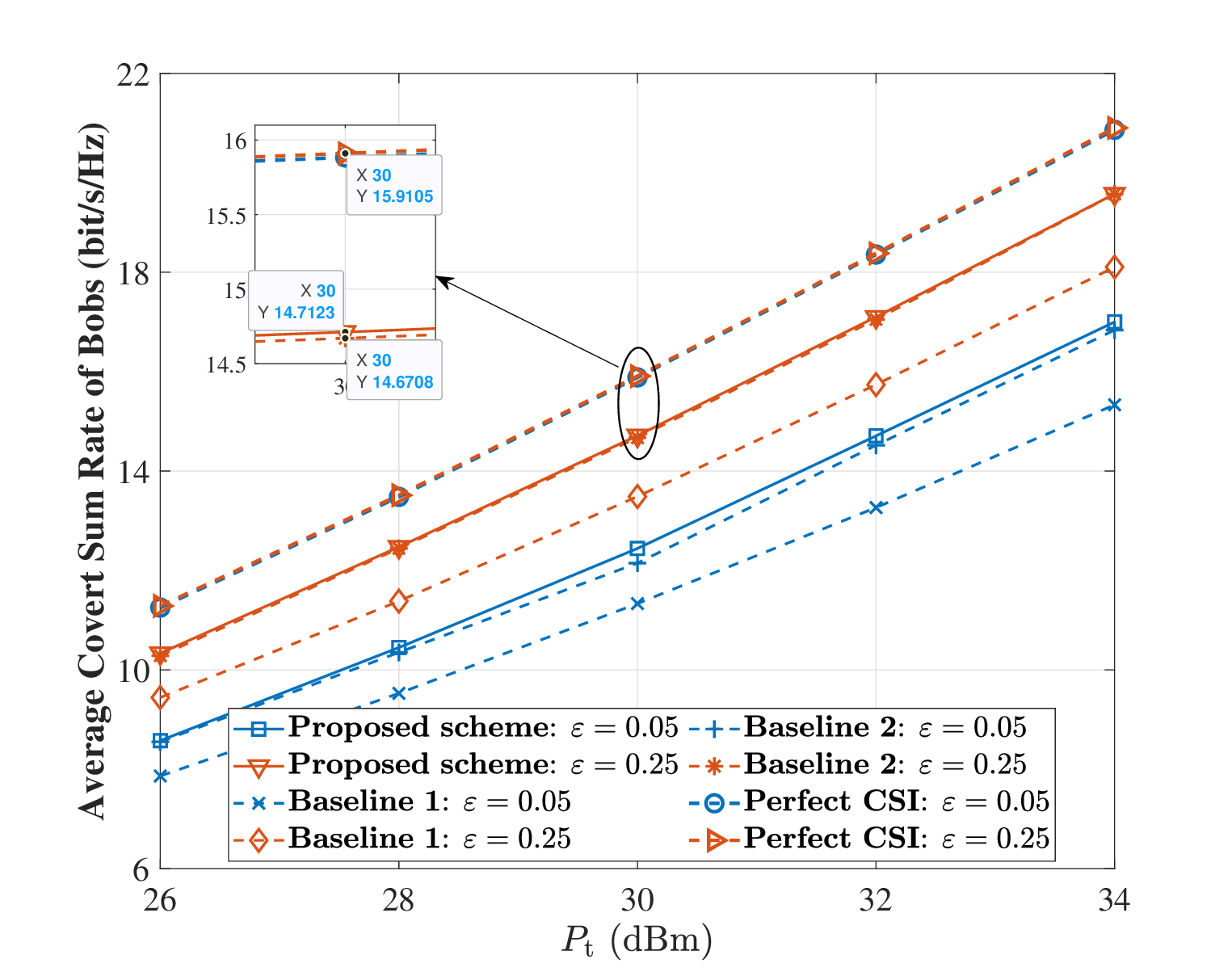}
 		\caption{ Average covert sum rate of Bobs versus the
maximum transmit power with $N_\mathrm{A}=N_\mathrm{R}=10$ and $\mathrm{MSE}_{\max }=7$.}
 		\label{fig_rate_vs_pt}
          \vspace{-0.5cm}

 \end{figure}

Fig.~\ref{fig_rate_vs_pt} illustrates the average covert sum rate of Bobs versus the maximum transmit power $P_\mathrm{t}$ at Alice. As expected, the covert sum rate increases monotonically with higher transmit power. Furthermore, the proposed algorithm and Baseline 2 consistently outperform Baseline 1, highlighting the benefit of optimizing RIS passive elements compared to scenarios without RIS. The performance gain of the proposed scheme over Baseline 2 can be attributed to the limitations of Gaussian randomization for the latter, which cannot guarantee strict convergence in alternating optimization and may fail to satisfy complex robust covertness constraints. Moreover, Gaussian randomization introduces an additional complexity of  $\mathcal{O}(LN_\mathrm{R}^3)$ for generating $L$ candidate solutions. These underscore the computational efficiency and superior solution quality of the proposed algorithm, consistent with \cite{wang2023joint}. Finally,  the proposed scheme achieves  over 90\% of the covert sum rate under perfect CSI of Willie when  $\varepsilon=0.25$, further validating its effectiveness.

 \begin{figure}
      		\centering
 		\includegraphics[width=6.5cm,height=5.2cm]{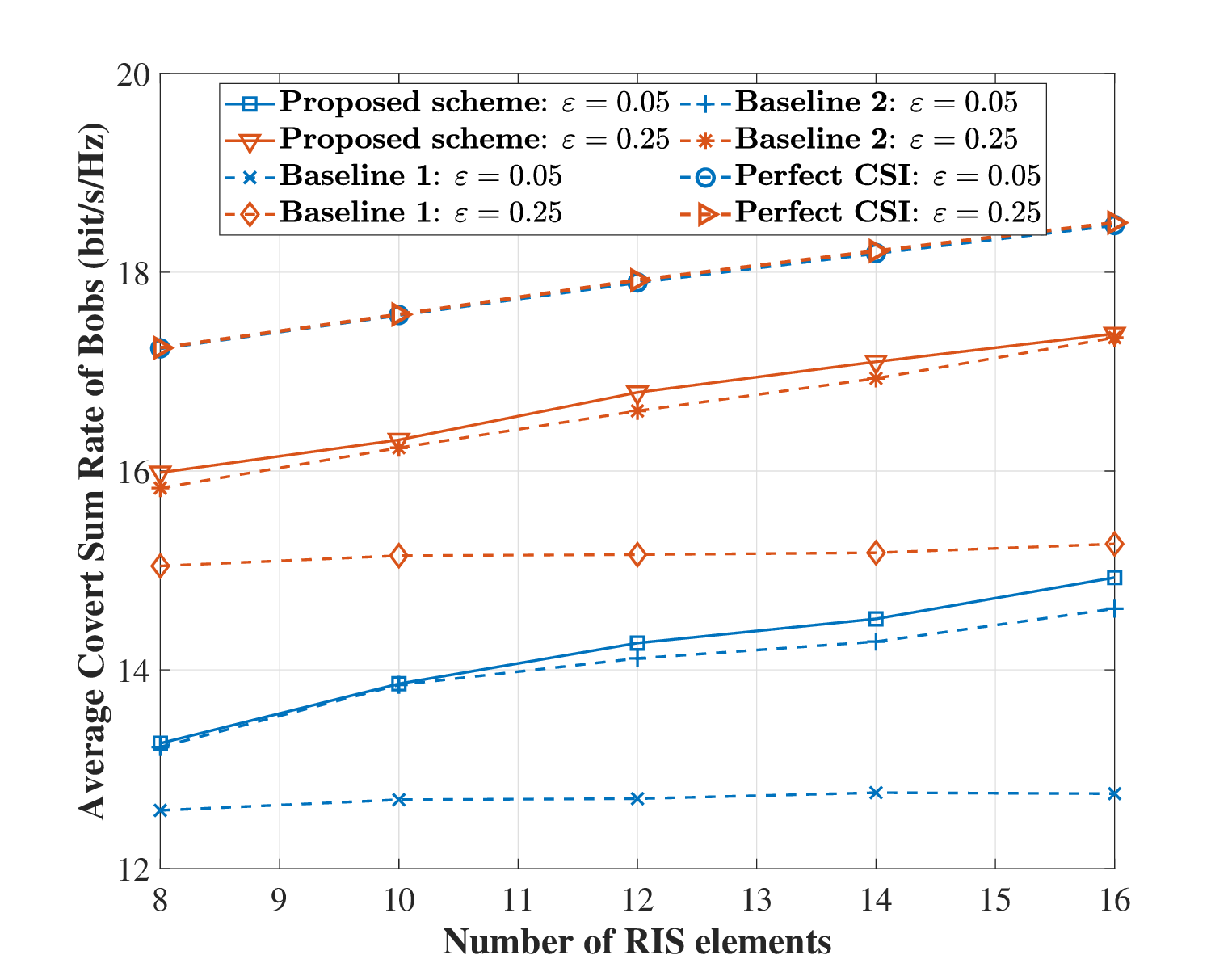}
 		\caption{Average covert sum rate of Bobs versus the number of RIS elements with $N_\mathrm{A}=10$, $P_\mathrm{t}=30$ dBm  and $\mathrm{MSE}_{\max }=7$.}
 		\label{fig_rate_riselements}
         \vspace{-0.4cm}

 \end{figure}

Fig.~\ref{fig_rate_riselements} further compares the average covert sum rate of Bobs versus the numbers of passive elements at RIS. By increasing the number of passive elements at RIS. Similar to Fig.~\ref{fig_rate_vs_pt}, the proposed scheme outperforms Baseline 1-2 across different RIS sizes and maintains more than 90\% of the performance of the perfect CSI scheme when  $\varepsilon=0.25$. Meanwhile, Fig.~\ref{fig_rate_riselements}  also reveals that enlarging the RIS size can be conducive to achieving a stricter covertness level at the same average covert sum rate of Bobs, which provides another degree of freedom to satisfy covert communication constraints rather than simply controlling the power allocation between communication and sensing.

 \begin{figure}[t]
 \vspace{-0.5cm}
      		\centering
 		\includegraphics[width=6.5cm,height=5.2cm]{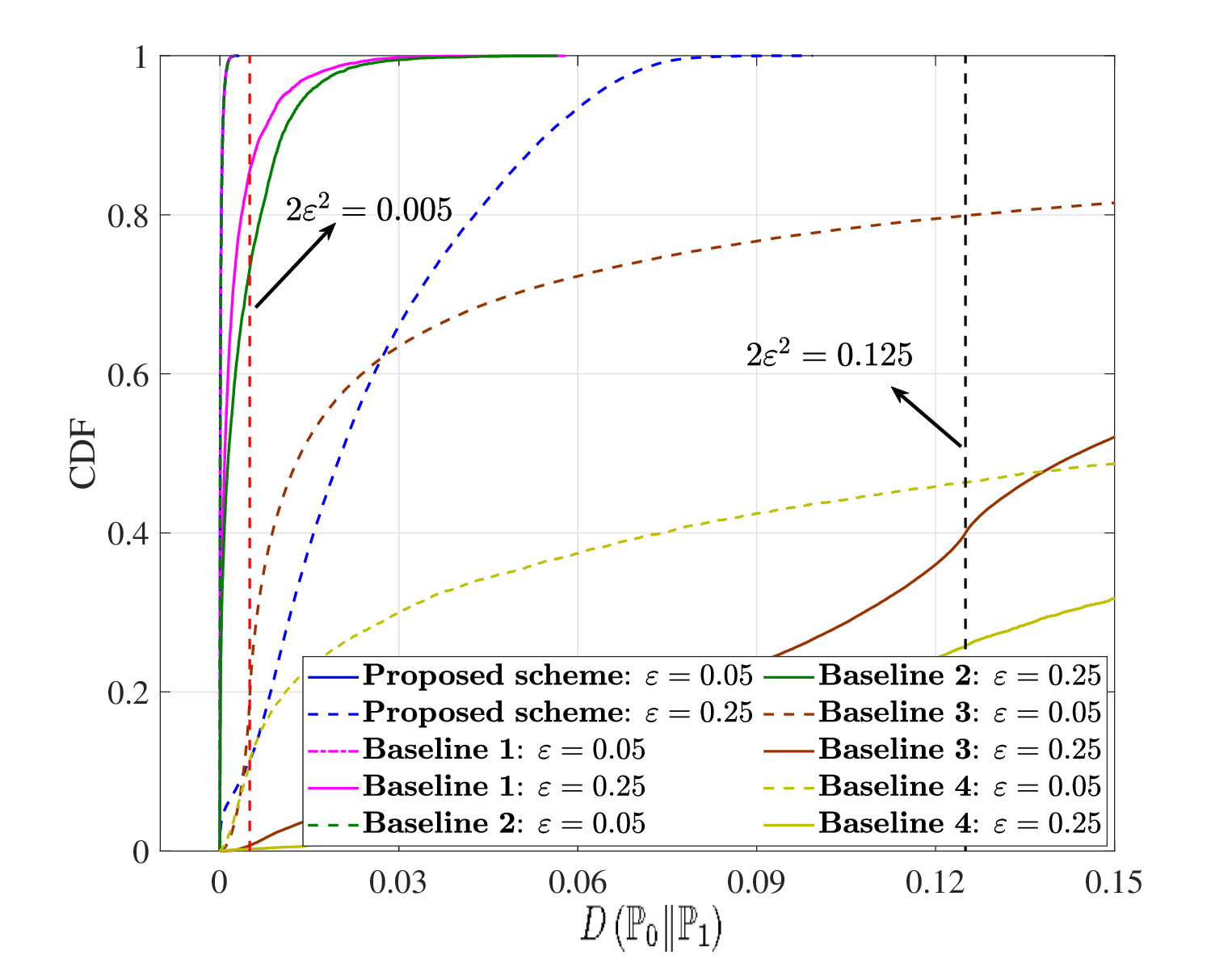}
 		\caption{The empirical CDF of the achieved $D\left(\mathbb{P}_0 \| \mathbb{P}_1\right)$ with $N_\mathrm{A}=N_\mathrm{R}=10$, $\mathrm{MSE}_\mathrm{max}=7$ and $P_\mathrm{t}=30$ dBm.}
 		\label{fig_covertCDF}
            \vspace{-0.5cm}
 \end{figure}

 \begin{figure}[!ht]
    \centering
    \begin{minipage}{0.99\linewidth} 
        \centering
        \subfloat[The missed detection probability.]{\includegraphics[width=6.5cm,height=5.2cm]{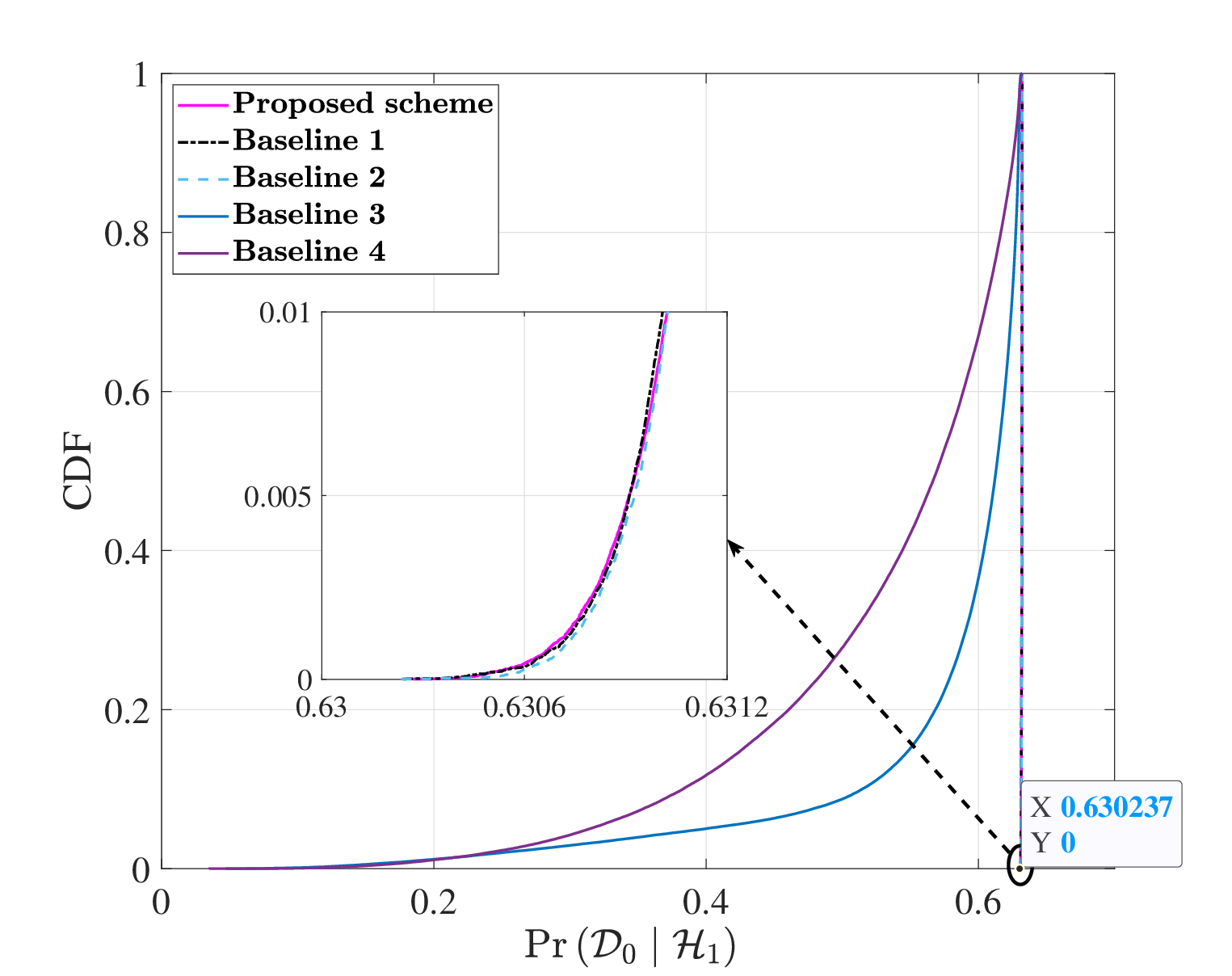}
        \label{fig_detection1}
        }
    \end{minipage} 
    \\ 
    \vspace{0cm} 
    \begin{minipage}{0.99\linewidth}
        \centering
        \subfloat[The false alarm probability.]{\includegraphics[width=6.5cm,height=5.2cm]{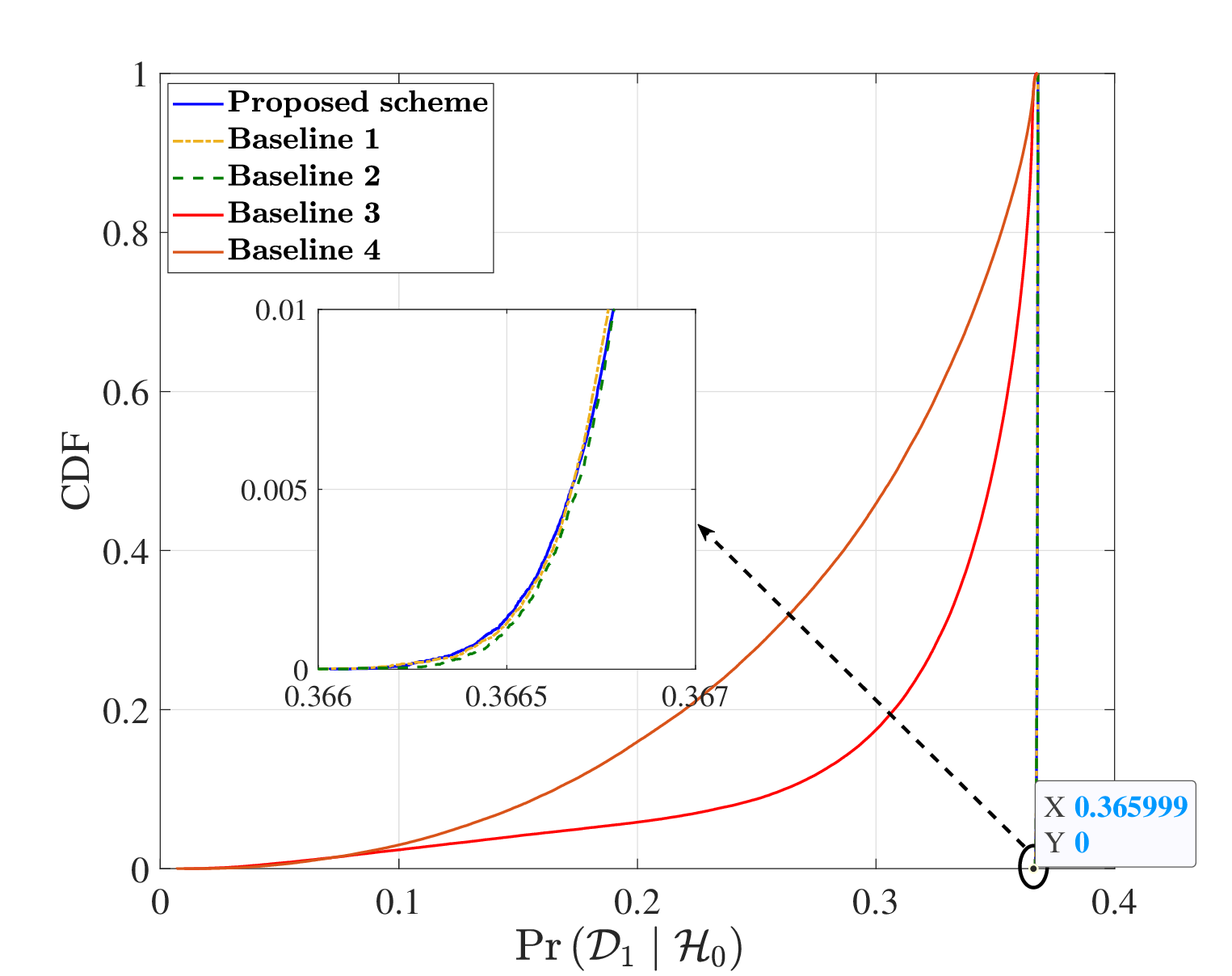}
        \label{fig_detection2}
        }
    \end{minipage}

    \caption{The empirical CDF of the false alarm probability and missed detection probability at Willie with $\varepsilon=0.05$, $N_\mathrm{A}=N_\mathrm{R}=10$, $\mathrm{MSE}_\mathrm{max}=7$ and $P_\mathrm{t}=30$ dBm.}
    \label{fig_detection}
    \vspace{-0.4cm}
\end{figure}

We also demonstrate the robustness of the proposed scheme in terms of covertness. In particular, Fig.~\ref{fig_covertCDF} illustrates the empirical cumulative density function (CDF) of the achieved $D\left(\mathbb{P}_0 \| \mathbb{P}_1\right)$ under the covertness thresholds $\varepsilon=0.05$ and $\varepsilon=0.25$, respectively. Both the proposed
scheme and Baselines 1–2 achieve $D\left(\mathbb{P}_0 \| \mathbb{P}_1\right)$ below the threshold $2\varepsilon^2$, confirming the
effectiveness of robust design.  By contrast, Baseline 3, although able to track Willie, lacks robust design and thus fails to guarantee covertness, where approximately $20\%$ of the resulting $D\left(\mathbb{P}_0 \| \mathbb{P}_1\right)$ does not exceed $2\varepsilon^2=0.005$, and about $40\%$ does not exceed $2\varepsilon^2=0.02$. In addition, Baseline 4 does not update Willie’s CSI in real time, ultimately failing to preserve covertness over the entire transmission period, where approximately $11\%$  and of the resulting $D\left(\mathbb{P}_0 \| \mathbb{P}_1\right)$ does not exceed $2\varepsilon$ at $\varepsilon=0.05$ and $\varepsilon=0.25$, respectively. These results collectively demonstrate the importance of leveraging sensing to dynamically acquire Willie’s CSI and validate the robust design of the proposed scheme. Furthermore, Fig.~\ref{fig_detection} plots the empirical CDF of the theoretical false alarm probability and missed detection probability at Willie when $\varepsilon=0.05$, calculated as per \cite{shuai:2023,yan2018delay}. It can be seen that both probabilities are higher under our proposed robust design, validating its effectiveness in enhancing covertness. Together with the results in Fig.~\ref{fig_covertCDF}, this also further validates the rationale of designing covert communication based on the relative entropy criterion $D\left(\mathbb{P}_0 \| \mathbb{P}_1\right)$, i.e., $\xi^{\star} \geq 1-\sqrt{\frac{D\left(\mathbb{P}_0 \| \mathbb{P}_1\right)}{2}}$. By contrast, Baseline 3 and Baseline 4 fail to maintain a high detection error probability because they lack robust design and real-time channel prediction.

   \begin{figure}
\vspace{-0.5cm}
 		\centering
 		\includegraphics[width=6.5cm,height=5.2cm]{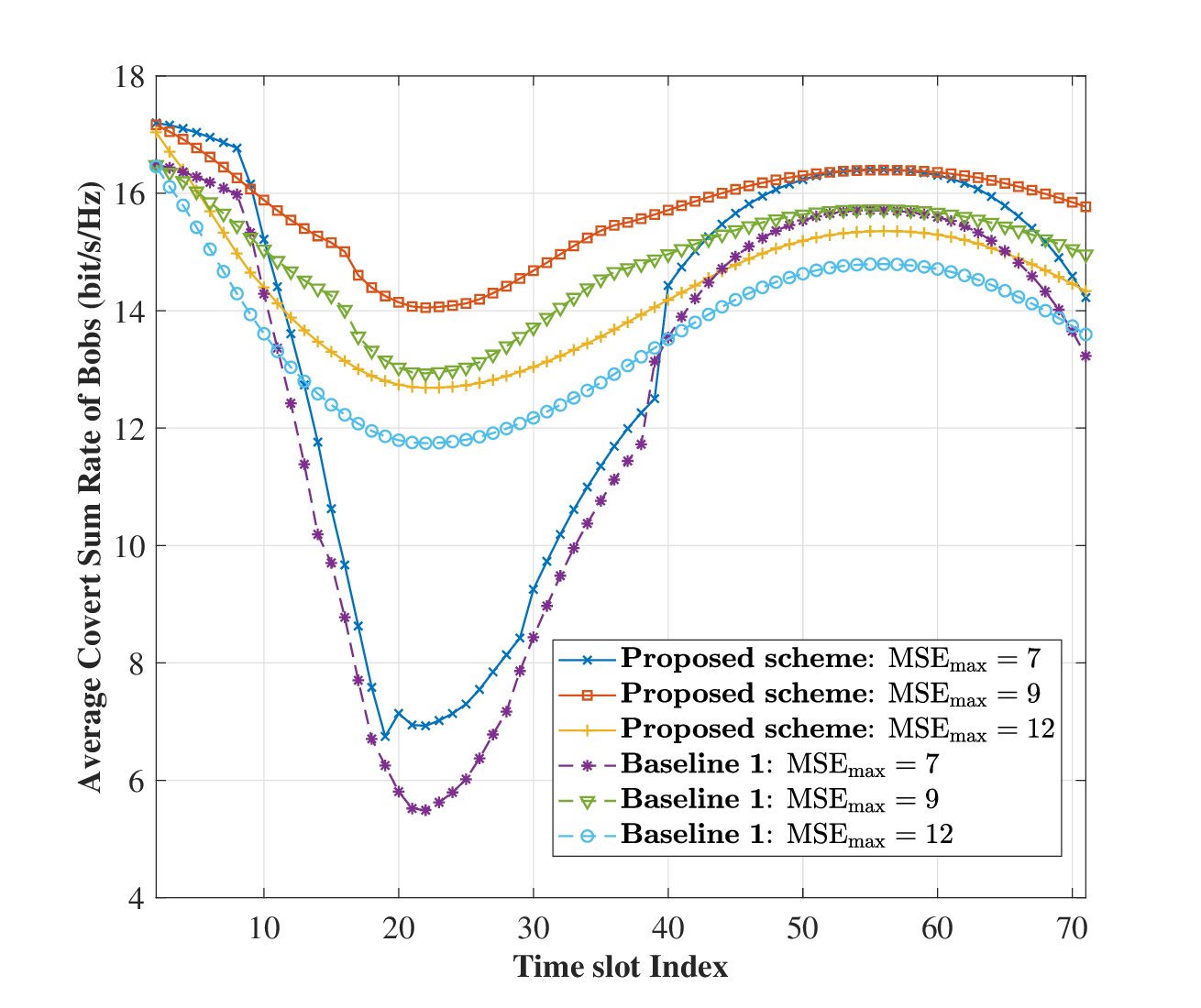}
 		\caption{Average covert sum rate of Bobs versus the time slot index with $N_\mathrm{A}=N_\mathrm{R}=10$, $P_\mathrm{t}=30$ dBm and $\varepsilon=0.25$.}
 		\label{fig_rate_vs_slot}
            \vspace{-0.5cm}
 \end{figure}
    \begin{figure}
 		\centering
 		\includegraphics[width=6.5cm,height=5.2cm]{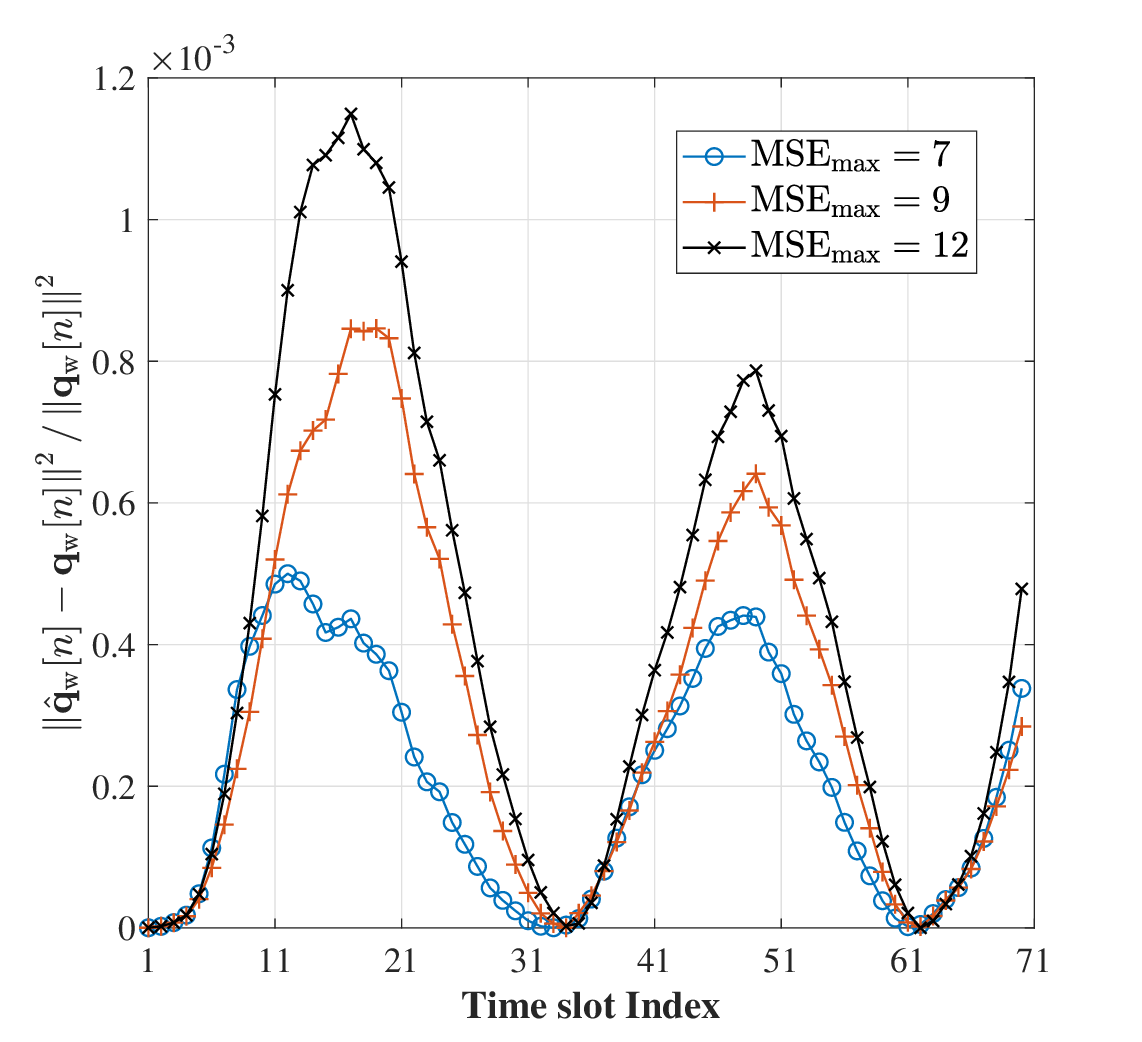}
 		\caption{The corresponding normalized posterior MSE of tracking Willie's location versus the time slot index with $N_\mathrm{A}=N_\mathrm{R}=10$ and $\varepsilon=0.25$.}
 		\label{fig_mse_vs_slot}
        \vspace{-0.4cm}
 \end{figure}

Next, to further evaluate the benefit of RIS in improving the covert communication, we examine the effective covert sum rate variation in a tracking period in Fig.~\ref{fig_rate_vs_slot}, where the proposed
scheme is compared with Baseline 1 under different $\mathrm{MSE}_\mathrm{max}$. It can be seen that the performance of our proposed scheme is always superior to that of Baseline 1 throughout the entire process simulated. However, the covert sum rate is not constant and drops in some time slots. This is because the posterior MSE for tracking the adversary target is accumulated over the time, causing Alice to allocate  more power to sensing in order to meet the sensing constraints and this situation will be improved as the EKF-based method can track Willie well to reduce the posterior MSE. This tracking MSE variation is further illustrated in Fig.~\ref{fig_mse_vs_slot}, where
the corresponding normalized posterior MSE of tracking Willie's location, i.e., $\left\|\hat{\mathbf{q}}_{\mathrm{w}}[n]-\mathbf{q}_{\mathrm{w}}[n]\right\|^2 /\left\|\mathbf{q}_{\mathrm{w}}[n]\right\|^2$, is evaluated for each time slot. Moreover, a comparison of the covert sum rate under different $\mathrm{MSE}_\mathrm{max}$ constraints reveals that stricter sensing constraint  (e.g., $\mathrm{MSE}_\mathrm{max}=7$) leads to a more significant reduction in the covert rate in the tracking process. This occurs because a greater portion of the available power is allocated to sensing for ensuring tracking accuracy.

     \begin{figure}
     \vspace{-0.5cm}
     		\centering
 		\includegraphics[width=6.5cm,height=5.2cm]{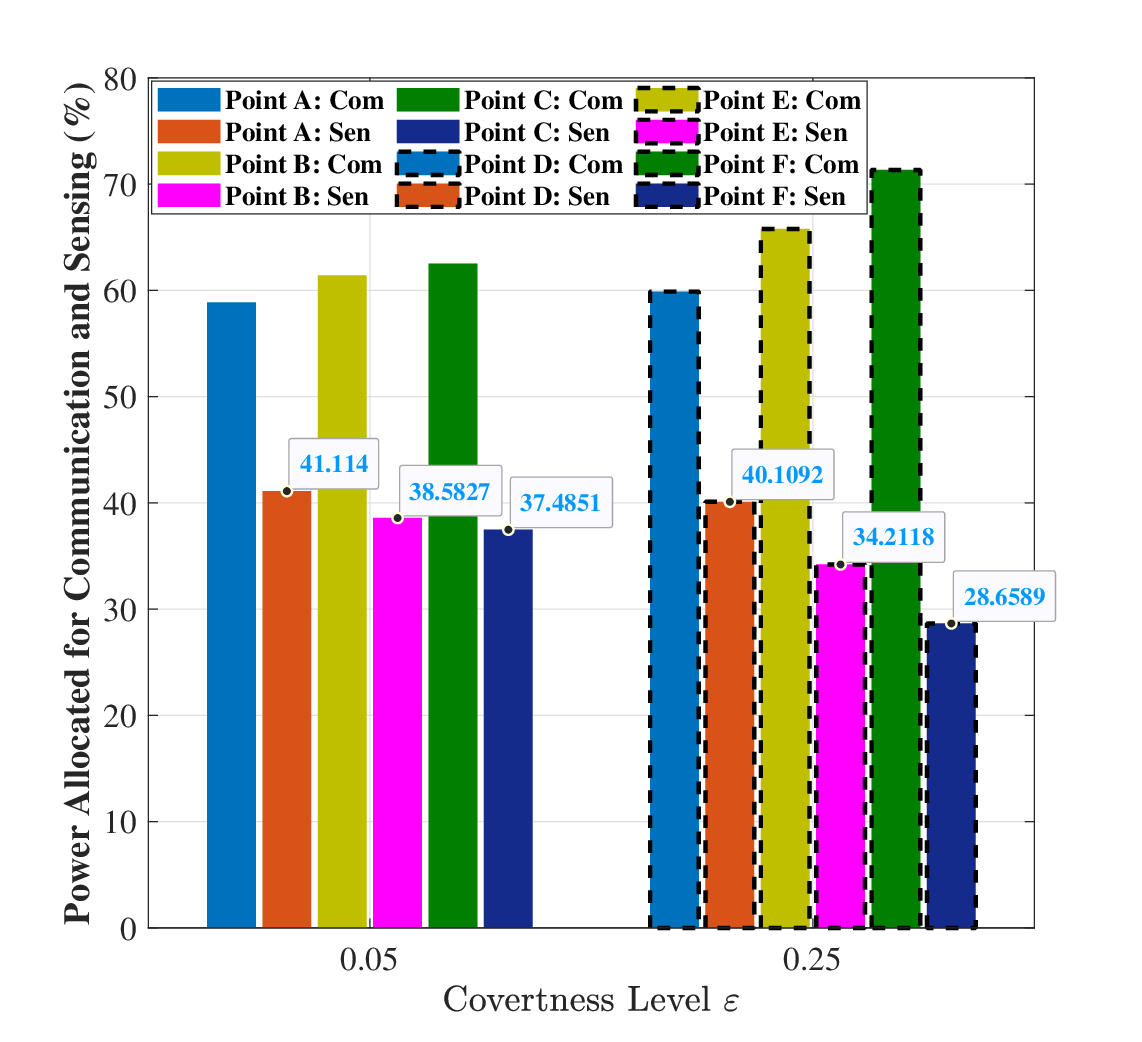}
 		\caption{Average power allocation strategy between the sensing and communication with $N_\mathrm{A}=N_\mathrm{R}=10$ and $P_\mathrm{t}=30$ dBm.}
 		\label{fig_power_allocation}
        \vspace{-0.5cm}
 \end{figure}

    \begin{figure}
        \centering
           \includegraphics[width=6.5cm,height=5.2cm]{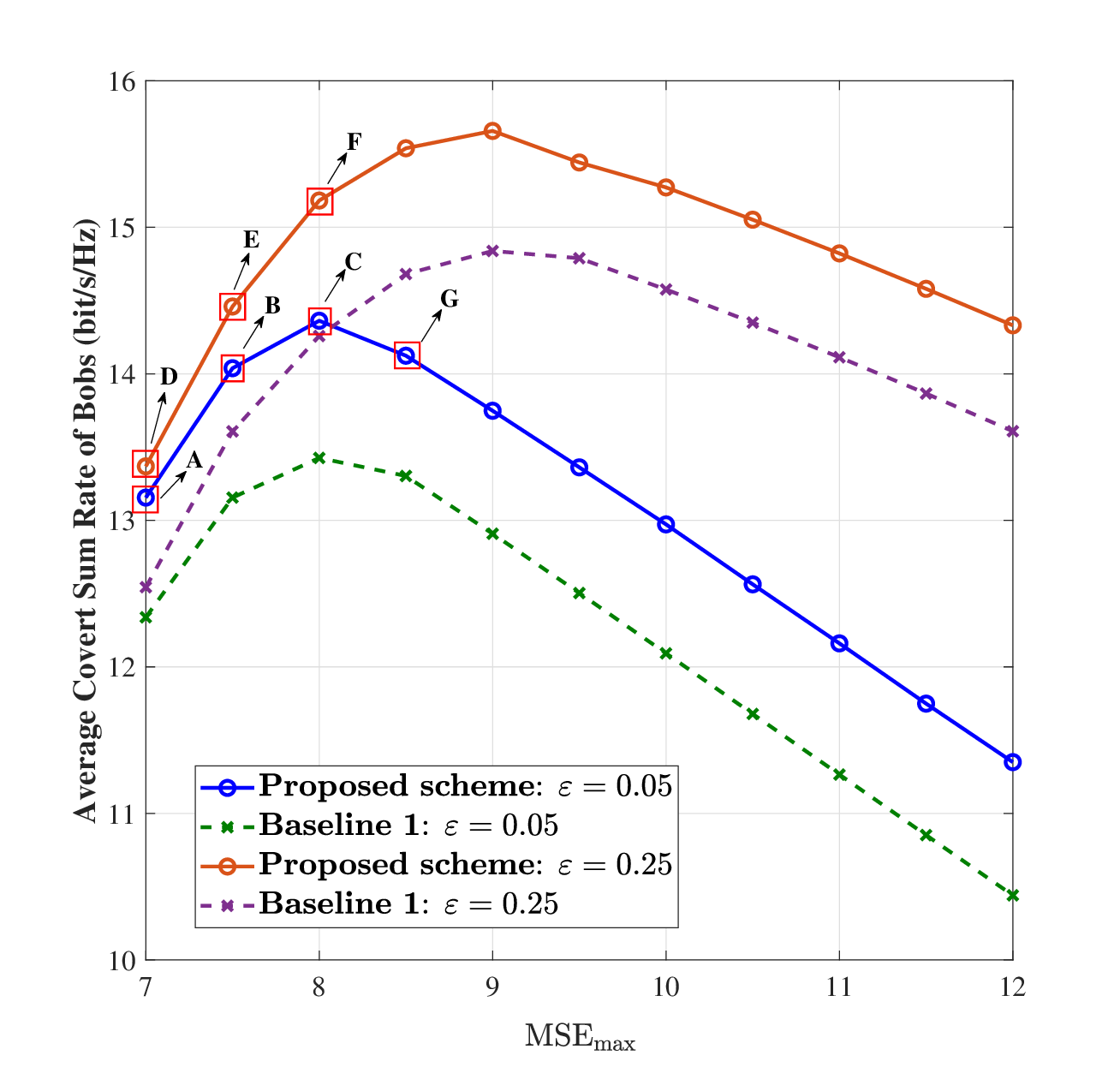}
        \caption{Tradeoff between the covert sum rate and sensing $\mathrm{MSE}_\mathrm{max}$ with $N_\mathrm{A} = N_\mathrm{R} = 10$ and $P_\mathrm{t}=30$ $\mathrm{dBm}$.}
    \label{fig_trade-off}
        \vspace{-0.5cm}
    \end{figure}

Finally, we analyze the tradeoff between covert communication and sensing performance by evaluating the covert sum rate averaged over multiple time slots in a tracking period as a function of the sensing constraint $\mathrm{MSE}_\mathrm{max}$. As shown in Fig.~\ref{fig_trade-off}, the average covert sum rate initially increases and then decreases as the sensing requirement $\mathrm{MSE}_\mathrm{max}$ is gradually relaxed. Specifically, a smaller $\mathrm{MSE}_\mathrm{max}$ demands more power to be allocated to sensing, which improves tracking accuracy and channel estimation for Willie. However, this reduces the power available for communication, leading to a lower effective covert sum rate. Conversely, as $\mathrm{MSE}_\mathrm{max}$ increases, relaxing the sensing constraint ($\mathrm{C}3$) allows more power for communication, resulting in a higher covert sum rate (e.g., from Point A to Point C with $\varepsilon = 0.05$ ). Beyond a certain point, however, further relaxation of $\mathrm{MSE}_\mathrm{max}$ introduces greater uncertainty in the predicted CSI of Willie, which degrades covert efficiency and causes the covert sum rate to decrease (e.g., from Point C to Point G with $\varepsilon=0.05$). It is also noted that the exact tradeoff curve depends on the level of covertness.

Fig.~\ref{fig_power_allocation} further presents the power allocation outcomes for different covertness levels, which align with the earlier discussions. It can be  seen that under a stricter covert requirement (e.g., $\varepsilon=0.05$), improving the tracking MSE from Point C to Point A requires  only an additional $3.6289\%$ of power allocated to sensing. In contrast, under a more relaxed covertness requirement ($\varepsilon = 0.25$), this additional power allocated to sensing increases to $11.4503\%$ from Point F to Point D. These differences arise because, under stricter covert requirements, Alice tends to allocate more power to sensing to reduce the KL divergence and meet the covertness constraint. As a result, less additional power is needed to achieve further improvements in tracking.

\section{Conclusion}\label{conclu}
This paper has proposed a sensing-then-beamforming framework for RIS-empowered ISACC systems. This framework leverages sensing capabilities to track and predict the state of a mobile warden Willie and construct the corresponding channels. Accounting for the constructed channel errors, we have formulated a robust non-convex optimization problem for the joint design of the beamformers and sensing signal covariance matrix at Alice, as well as the phase shifts at the RIS, aiming to maximize the covert sum rate of multiple legitimate Bobs. The structural properties of the problem have been analyzed and exploited to transform it into a more tractable form. An alternating-optimization-based algorithm has been developed to iteratively solve subproblems concerning Alice and RIS separately. Simulation results verified the effectiveness of our proposed scheme and demonstrated its superior performance compared to baseline designs.

Future work could extend the current framework to address scenarios involving multiple wardens. Additionally, the framework could be explored for sensing-assisted covert communication in environments with more complex scattering characteristics. Moreover, it would be intriguing to explore a fully integrated dual-functional waveform that simultaneously protects both communication and sensing functions for the ISACC system. Finally, novel hardware architectures, such as simultaneously transmitting and reflecting reconfigurable intelligent surfaces (STAR-RIS), active RIS, semi-passive RIS and movable antenna (MA) \cite{xiu2025movable,xiu2025movable2}, should be further explored to understand how they can enhance ISACC systems.

\section*{Appendix A \\ Proof of Proposition~\ref{pro-covert-w} }\label{APA}
By substituting \eqref{channel-willie} into the left term of \eqref{C2-NEW1}, we have 
\begin{equation}\label{APA-eq1}
\begin{aligned}
& \left(\mathbf{h}_{\mathrm{aw}}^H+\mathbf{h}_{\mathrm{rw}}^H \boldsymbol{\Theta} \mathbf{G}\right) \mathbf{R}\left(\mathbf{h}_{\mathrm{aw}}^H+\mathbf{h}_{\mathrm{rw}}^H \boldsymbol{\Theta} \mathbf{G}\right)^H \\
 =&\left(\boldsymbol{\Delta}_{\mathrm{rw}}^H \boldsymbol{\Theta} \mathbf{G}+\boldsymbol{\Delta}_{\mathrm{aw}}^H\right) \mathbf{R}\left(\boldsymbol{\Delta}_{\mathrm{rw}}^H \boldsymbol{\Theta}+\boldsymbol{\Delta}_{\mathrm{aw}}^H\right)^H\\&+2 \mathfrak{R}\left\{\left(\boldsymbol{\Delta} _{\mathrm{rw}}^H \boldsymbol{\Theta} \mathbf{G}+\boldsymbol{\Delta} _{\mathrm{aw}}^H\right) \mathbf{R}\left(\widehat{\mathbf{h}}_{\mathrm{aw}}^H+\widehat{\mathbf{h}}_{\mathrm{rw}}^H \boldsymbol{\Theta} \mathbf{G}\right)^H\right\}\\
&+\left(\widehat{\mathbf{h}}_{\mathrm{aw}}^H+\widehat{\mathbf{h}}_{\mathrm{rw}}^H \boldsymbol{\Theta} \mathbf{G}\right) \mathbf{R}\left(\widehat{\mathbf{h}}_{\mathrm{aw}}^H+\widehat{\mathbf{h}}_{\mathrm{rw}}^H \boldsymbol{\Theta}\right)\\=&
 \bar{\boldsymbol{\Delta} }_{\mathrm{w}}^H \bar{\mathbf{G}} \mathbf{R} \bar{\mathbf{G}}^H\bar{\boldsymbol{\Delta} }_{\mathrm{w}}+2 \mathfrak{R}\left\{\bar{\boldsymbol{\Delta} }_{\mathrm{w}}^H \bar{\mathbf{G}} \mathbf{R} \bar{\mathbf{G}}^H \overline{\mathbf{h}}_{\mathrm{w}}\right\}+\overline{\mathbf{h}}_{\mathrm{w}}^H \bar{\mathbf{G}} \mathbf{R} \bar{\mathbf{G}}^H \overline{\mathbf{h}}_{\mathrm{w}},
\end{aligned}
\end{equation}
where $\bar{\mathbf{G}} \triangleq\left[\mathbf{G}^H \boldsymbol{\Theta}^H, \mathbf{I}_{N_{\mathrm{A}}}\right]^H$, $\overline{\mathbf{h}}_{\mathrm{w}}^H\triangleq\left[\widehat{\mathbf{h}}_{\mathrm{rw}}^H, \widehat{\mathbf{h}}_{\mathrm{aw}}^H\right]$ and $  \bar{\boldsymbol{\Delta} }_{\mathrm{w}}^H\triangleq\left[\boldsymbol{\Delta} _{\mathrm{rw}}^H, \boldsymbol{\Delta} _{\mathrm{aw}}^H\right]$.

According to \eqref{channel-err_set}, we have 
\begin{equation}
\bar{\boldsymbol{\Delta}}_{\mathrm{w}}^H \mathbf{P}_1 \bar{\boldsymbol{\Delta}}_{\mathrm{w}}-\delta_{\mathrm{rw}}^2 \leq 0, \thinspace
\bar{\boldsymbol{\Delta}}_{\mathrm{w}}^H \mathbf{P}_2 \bar{\boldsymbol{\Delta}}_{\mathrm{w}}-\delta_{\mathrm{aw}}^2 \leq 0,
\end{equation}
where $\mathbf{P}_1=\operatorname{blkdig}\left(\mathbf{I}_{N_{\mathrm{R}}}, \mathbf{0}_{N_{\mathrm{A}} \times N_{\mathrm{A}}}\right)$ and $ \mathbf{P}_2=\operatorname{blkdig}\left( \mathbf{0}_{N_{\mathrm{R}} \times N_{\mathrm{R}}},\mathbf{I}_{N_{\mathrm{A}}}\right)$.
Then, by defining in \eqref{define-apxA} on the top of the next page
\begin{figure*}[!t] 
\vspace{-0.7cm}
\begin{equation}\label{define-apxA}
\mathbf{P}_0=\bar{\mathbf{G}} \mathbf{R} \bar{\mathbf{G}}^H, \mathbf{q}_0=\mathbf{P}_0 \overline{\mathbf{h}}_{\mathrm{w}}, q_0=\overline{\mathbf{h}}_{\mathrm{w}}^H \mathbf{P}_0 \overline{\mathbf{h}}_{\mathrm{w}}-(\eta_2-1) \sigma_{\mathrm{w}}^2, \mathbf{x}=\bar{\boldsymbol{\Delta}}_{\mathbf{w}}, \mathbf{q}_1=\mathbf{q}_2=\mathbf{0}_{\left(N_{\mathrm{A}}+N_{\mathrm{R}}\right) \times1}, q_1=-\delta_{\mathrm{rw}}^2, q_2=-\delta_{\mathrm{aw}}^2.
\end{equation}

\begin{equation}\label{define-apxC}
\widehat{\mathbf{P}}=-\sum_{i} \bar{\mathbf{P}}_i \mathbf{V}^* \bar{\mathbf{Q}}_i, \widehat{\mathbf{q}}=\widehat{\mathbf{P}} \overline{\mathbf{h}}_{\mathrm{w}}, \widehat{\boldsymbol{\mu}}=\operatorname{blkdiag}\left(\widehat{\mu}_1 \mathbf{I}_{N_{\mathrm{R}}}, \widehat{\mu}_2 \mathbf{I}_{N_{\mathrm{A}}}\right),\widehat{q}=\overline{\mathbf{h}}_{\mathrm{w}}^H \widehat{\mathbf{P}} \overline{\mathbf{h}}_{\mathrm{w}}+(\eta_2-1) \sigma_{\mathrm{w}}^2-\widehat{\mu}_1 \delta_{\mathrm{rw}}^2-\widehat{\mu}_2 \delta_{\mathrm{aw}}^2.
\end{equation}

\begin{equation}\label{Jacobian-matrix}
\frac{\partial \mathbf{g}}{\partial \boldsymbol{\chi}_{\mathrm{w}}}=\left[\begin{array}{cccccc}
\frac{2\left(x_{\mathrm{w}}-x_{\mathrm{a}}\right)}{c\left\|\mathbf{q}_{\mathrm{w}}-\mathbf{q}_{\mathrm{a}}\right\|} & \frac{2\left(y_{\mathrm{w}}-y_{\mathrm{a}}\right)}{c\left\|\mathbf{q}_{\mathrm{w}}-\mathbf{q}_{\mathrm{a}}\right\|} & \frac{2\left(z_{\mathrm{w}}-z_{\mathrm{a}}\right)}{c\left\|\mathbf{q}_{\mathrm{w}}-\mathbf{q}_{\mathrm{a}}\right\|} & 0 & 0 & 0 \\
\frac{\partial v_{\mathrm{w}}}{\partial x_{\mathrm{w}}} & \frac{\partial v_{\mathrm{w}}}{\partial y_{\mathrm{w}}} & \frac{\partial v_{\mathrm{w}}}{\partial z_{\mathrm{w}}} & \frac{2 f_c\left(x_{\mathrm{w}}-x_{\mathrm{a}}\right)}{c\left\|\mathbf{q}_{\mathrm{w}}-\mathbf{q}_{\mathrm{a}}\right\|} &\frac{2 f_c\left(y_{\mathrm{w}}-y_{\mathrm{a}}\right)}{c\left\|\mathbf{q}_{\mathrm{w}}-\mathbf{q}_{\mathrm{a}}\right\|} & \frac{2 f_c\left(z_{\mathrm{w}}-z_{\mathrm{a}}\right)}{c\left\|\mathbf{q}_{\mathrm{w}}-\mathbf{q}_{\mathrm{a}}\right\|} \\
\frac{-\left(y_{\mathrm{w}}-y_{\mathrm{a}}\right)\left(x_{\mathrm{w}}-x_{\mathrm{a}}\right)}{\left\|\mathbf{q}_{\mathrm{ew}}\right\|^3} & \frac{\left(x_{\mathrm{w}}-x_{\mathrm{a}}\right)^2}{\left\|\mathbf{q}_{\mathrm{ew}}\right\|^3} & 0 & 0 & 0 & 0\\
\frac{\left(y_{\mathrm{w}}-y_{\mathrm{a}}\right)^2}{\left\|\mathbf{q}_{\mathrm{ew}}\right\|^3} & -\frac{\left(x_{\mathrm{w}}-x_{\mathrm{a}}\right)\left(y_{\mathrm{w}}-y_{\mathrm{a}}\right)}{\left\|\mathbf{q}_{\mathrm{ew}}\right\|^3} & 0 & 0 & 0 & 0 \\
-\frac{\left(z_{\mathrm{w}}-z_{\mathrm{a}}\right)\left(x_{\mathrm{w}}-x_{\mathrm{a}}\right)}{\left\|\mathbf{q}_{\mathrm{w}}-\mathbf{q}_{\mathrm{a}}\right\|^3} & -\frac{\left(z_{\mathrm{w}}-z_{\mathrm{a}}\right)\left(y_{\mathrm{w}}-y_{\mathrm{a}}\right)}{\left\|\mathbf{q}_{\mathrm{w}}-\mathbf{q}_{\mathrm{a}}\right\|^3} & \frac{\left\|\mathbf{q}_{\mathrm{ew}}\right\|^2}{\left\|\mathbf{q}_{\mathrm{w}}-\mathbf{q}_{\mathrm{a}}\right\|^3} & 0 & 0 & 0
\end{array}\right].
\end{equation}

\hrulefill
\vspace{-0.5cm}
\end{figure*}

and based on Lemma 1, we can obtain the LMI as follows
\begin{equation}
 \sum_{i=1}^2 \mu_i\left[\begin{array}{cc}
\mathbf{P}_i & \mathbf{q}_i \\
\mathbf{q}_i^H & q_i
\end{array}\right] \succeq \left[\begin{array}{cc}
\mathbf{P}_0 & \mathbf{q}_0 \\
\mathbf{q}_0^H & q_0
\end{array}\right] \Rightarrow \left[\begin{array}{cc}
\mathbf{P}+\boldsymbol{\mu} & \mathbf{q} \\
\mathbf{q}^H & q
\end{array}\right] \succeq \mathbf{0},
\end{equation}
where $\mathbf{P}=-\mathbf{P}_0$, $\boldsymbol{\mu}=\operatorname{blkdiag}\left(\mu_1 \mathbf{I}_{N_{\mathrm{R}}}, \mu_2 \mathbf{I}_{N_{\mathrm{A}} }\right)$, $\mathbf{q}=-\mathbf{q}_0$, $q=-q_0-\mu_1 \delta_{\mathrm{rw}}^2-\mu_2 \delta_{\mathrm{aw}}^2$, $\mu_1\geq0$ and $\mu_2\geq0$ are slack variables.
The proof is thus completed. \hfill $\blacksquare$

\section*{Appendix B \\ Proof of Proposition~\ref{pro-rank} }\label{APB}
Problem $\bar{\mathcal{P}}2$ is jointly convex with respect to the optimization
variables and satisfies Slater’s constraint qualification. Therefore, strong duality holds, and the Lagrangian function in terms of $\mathbf{W}_k$ is given by 
\begin{equation}
\begin{aligned}
    \mathcal{L}&=Q_1-\sum_{k \in \mathcal{K}} \operatorname{tr}\left(\nabla_{\mathbf{W}_k}^H P_1\left(\mathbf{W}^{(m)}, \mathbf{R}_{\mathrm{s}}^{(m)}\right) \mathbf{W}_k\right)\\
    &\quad-\gamma \sum_{k\in \mathcal{K}} \operatorname{tr}\left(\mathbf{W}_k\right)+\sum_{k\in \mathcal{K}} \operatorname{tr}\left(\boldsymbol{\Upsilon}_k \mathbf{W}_k\right)+\operatorname{tr}(\boldsymbol{\Omega} \mathbf{\Pi})+\xi,
\end{aligned}
\end{equation}
where $\xi$  includes all the terms that do not involve $\mathbf{W}_k$.  Let $\gamma  \geq 0$, $\boldsymbol{\Upsilon}_k\in \mathbb{H}^{N_{\text {A}}}$ and $\boldsymbol{\Psi} \in \mathbb{H}^{N_{\mathrm{A}}+N_{\mathrm{R}}+1}$ denote the Lagrange multipliers associated with constraints $\overline{\mathrm{C} 1}$, $\overline{\mathrm{C} 2}$ and $\mathrm{C} 6$ in problem \eqref{P2-2}, respectively. Then, we investigate the structure of $\mathbf{W}_k$ by checking the Karush-Kuhn-Tucker (KKT) conditions of problem, which are given by
\begin{equation}\label{proof-eq2}
    \begin{aligned}
        \mathrm{K} 1: &\mathbf{\Upsilon}_k^{\star} \mathbf{W}_k^{\star}=\mathbf{0}, \quad \mathrm{K} 2: \gamma^{\star} \geq 0,  \boldsymbol{\Upsilon}_k^{\star} \succeq \mathbf{0}, \boldsymbol{\Omega}^{\star} \succeq \mathbf{0},\\\mathrm{K} 3:&\nabla_{\mathbf{W}_k} \mathcal{L}\left(\mathbf{W}_k^{\star}\right)=\mathbf{0}
    \end{aligned}
\end{equation}
where $\gamma^{\star}, \mathbf{\Upsilon}_k^{\star}$, and $\boldsymbol{\Omega}^{\star}$ are the optimal Lagrangian multipliers. Based on $\mathrm{K}3$, we have $\boldsymbol{\Upsilon}_k^*=\gamma^{\star}\mathbf{I}_{N_{\mathrm{A}}}-\boldsymbol{\Xi}_k^*$, where
\begin{equation}
\begin{aligned}
    \boldsymbol{\Xi}_k^*&=
\frac{1}{\ln 2} \sum_{j\in \mathcal{K}} \frac{\mathbf{h}_j \mathbf{h}_j^H}{\operatorname{tr}\left(\mathbf{h}_j \mathbf{h}_j^H \mathbf{R}_{\mathrm{s}}\right)+\sum_{i\in \mathcal{K}} \operatorname{tr}\left(\mathbf{h}_j \mathbf{h}_j^H \mathbf{W}_i\right)+\sigma_{\mathrm{b}, j}^2}\\
&\quad-\nabla_{\mathbf{w}_k}P_1\left(\mathbf{W}^{(m)}, \mathbf{R}_{\mathrm{s}}^{(m)}\right)
+\nabla_{\mathbf{w}_k} \operatorname{tr}(\boldsymbol{\Omega}^* \boldsymbol{\Pi}).
\end{aligned}
\end{equation}

According to $\boldsymbol{\Upsilon}_k^{\star} \succeq \mathbf{0}$ in $\mathrm{K2}$, $\gamma^{\star} \geq \lambda_{\max }(\boldsymbol{\Xi}_k^*) $ must hold. Then, we have the following conclusions:
\begin{itemize}
    \item $\gamma^{\star}>\lambda_{\max }\left(\boldsymbol{\Xi}_k^*\right)$: it  will result in a positive definite and
full-rank matrix $\boldsymbol{\Upsilon}_k^{\star} \succ \mathbf{0}$. Based on $\mathrm{K1}$, we have $\mathbf{W}_k=\mathbf{0}$, which is obviously not a feasible solution of problem $\bar{\mathcal{P}}2$.
\item $\gamma^{\star}=\lambda_{\max }\left(\boldsymbol{\Xi}_k^*\right)$: in this case, we have $ N_{\mathrm{A}}-1 \leq \operatorname{rank}\left(\mathbf{\Upsilon}_k^*\right)<N_{\mathrm{A}}$, then according to $\mathrm{K}1$ and the Sylvester inequality, we have \begin{equation*}
\begin{aligned} &N_{\mathrm{A}}-1+\operatorname{rank}\left(\mathbf{W}_k^*\right) \leq \operatorname{rank}\left(\mathbf{\Upsilon}_k^*\right)+\operatorname{rank}\left(\mathbf{W}_k^*\right) \leq N_{\mathrm{A}}\\&\Rightarrow \operatorname{rank}\left(\mathbf{W}_k^*\right) \leq 1.
\end{aligned}
\end{equation*}
\end{itemize}

Based on the above derivation,  $\mathrm{rank}\left(\mathbf{W}_k^*\right) \leq 1$ follows.\hfill $\blacksquare$

\section*{Appendix C \\ Proof of Proposition~\ref{pro-covert-ris} }\label{APC}
We first perform the singular value
decomposition (SVD) of $\mathbf{R}=\sum_{i} \mathbf{p}_i \mathbf{q}_i^H$ in Proposition~\ref{pro-covert-w} and then define that  
\begin{equation}
\begin{aligned}
    &\bar{\mathbf{P}}_i=\left[\begin{array}{cc}
\operatorname{diag}\left(\mathbf{G} \mathbf{p}_i\right) & \mathbf{0}_{N_{\mathrm{R}} \times 1} \\
\mathbf{0}_{N_{\mathrm{A}} \times N_{\mathrm{R}}} & \mathbf{p}_i
\end{array}\right],\\&\bar{\mathbf{Q}}_i=\left[\begin{array}{cc}
\operatorname{diag}\left(\mathbf{q}_i^H \mathbf{G}^H\right) & \mathbf{0}_{N_{\mathrm{R}} \times N_{\mathrm{A}}} \\
\mathbf{0}_{1 \times N_{\mathrm{R}}} & \mathbf{q}_i^H
\end{array}\right].
\end{aligned}
\end{equation}

According to \eqref{APA-eq1}, we can rewrite $\bar{\mathbf{G}} \mathbf{R} \bar{\mathbf{G}}^{H}$ as follows

\begin{equation}
\begin{aligned}
& \bar{\mathbf{G}} \mathbf{R} \bar{\mathbf{G}}^H  =\sum_{i}\left[\begin{array}{cc}
\boldsymbol{\Theta} \mathbf{G} \mathbf{p}_i \mathbf{q}_i^H \mathbf{G}^H \mathbf{\Theta}^H & \boldsymbol{\Theta} \mathbf{G} \mathbf{p}_i \mathbf{q}_i^H \\
\mathbf{p}_i \mathbf{q}_i^H \mathbf{G}^H \boldsymbol{\Theta}^H & \mathbf{p}_i \mathbf{q}_i^H
\end{array}\right] \\
& =\sum_{i}\left[\begin{array}{cc}
\operatorname{diag}\left(\mathbf{G p}_i\right) \boldsymbol{\theta} \boldsymbol{\theta}^H \operatorname{diag}\left(\mathbf{q}_i^H \mathbf{G}^H\right) & \operatorname{diag}\left(\mathbf{G} \mathbf{p}_i\right) \boldsymbol{\theta} \mathbf{q}_i^H \\
\mathbf{p}_i \boldsymbol{\theta}^H \operatorname{diag}\left(\mathbf{q}_i^H \mathbf{G}^H\right) & \mathbf{p}_i \mathbf{q}_i^H
\end{array}\right] \\
& =\sum_{i} \bar{\mathbf{P}}_i \mathbf{V}^* \bar{\mathbf{Q}}_i.
\end{aligned}
\end{equation}

Hence, similar to Proposition~\ref{pro-covert-w}, and based on the definition in \eqref{define-apxC} on the top of next page, the covertness constraint in \eqref{P3-1} can be equivalently transformed into 
\begin{equation}
\left[\begin{array}{cc}
\widehat{\mathbf{P}}+\widehat{\boldsymbol{\mu}} & \widehat{\mathbf{q}} \\
\widehat{\mathbf{q}}^H & \widehat{q}
\end{array}\right] \succeq \mathbf{0},
\end{equation}
where $\widehat{\mu}_1\geq0$ and $\widehat{\mu}_2\geq0$ are slack variables.
This completes the proof. \hfill $\blacksquare$

\section*{Appendix D \\ Derivations of Measurement Model of Willie }\label{APD}
We consider the matched-filter (MF) principle and exploit the reflected echo signal to estimate $v_{\mathrm{w}}[n], \tau_{\mathrm{w}}[n], \theta_{\mathrm{aw}}[n]$ and $\phi_{\mathrm{aw}}[n]$ in \eqref{echo}. According to \cite{1994Fundamentals,Liu:2020,Wei:2023}, the estimation error variances of these parameters can be given as
\begin{equation}\label{channel-err}
\begin{aligned}
 &  \sigma_{\tau_{\mathrm{w}}[n]}^2=\frac{c_{\tau_{\mathrm{w}}}}{\mathrm{SNR}[n]}, \sigma_{v_{\mathrm{w}}[n]}^2=\frac{c_{v_{\mathrm{w}}}}{\mathrm{SNR}[n]},\\ &  \sigma_{\theta_{\mathrm{aw}}[n]}^2=\frac{c_{\theta_{\mathrm{aw}}}}{\mathrm{SNR}[n]}, \sigma_{\phi_{\mathrm{aw}}[n]}^2=\frac{c_{\phi_{\mathrm{aw}}}}{\mathrm{SNR}[n]}, 
\end{aligned}
\end{equation}
where $c_{\tau_{\mathrm{w}}}, c_{v_{\mathrm{w}}}, c_{\theta_{\mathrm{aw}}}, c_{\phi_{\mathrm{aw}}}>0$ are determined by the adopted estimation methods,  and $\operatorname{SNR}[n]$ is the output SNR of the MF given by
\begin{equation}\label{SNR-mes}
\begin{aligned}
&\operatorname{SNR}[n] \\
&=\frac{\rho_0 \varsigma G_{\mathrm{MF}} N_{\mathrm{A}} \mathbf{a}^H\left(\theta_{\mathrm{aw}}[n], \phi_{\mathrm{aw}}[n]\right) \mathbf{R}_{\mathrm{s}}[n] \mathbf{a}\left(\theta_{\mathrm{aw}}[n], \phi_{\mathrm{aw}}[n]\right)}{\sigma_{\mathrm{r}}^2 d_{\mathrm{aw}}^4[n]},
\end{aligned}
\end{equation}
where $G_{\mathrm{MF}} $ is the MF gain related to the number of transmit symbols in a time slot. Then, based on $\boldsymbol{\chi}_{\mathrm{w}}[n]$ and $\mathbf{q}_{\mathrm{a}}$, measurement model in \eqref{mes-model} can be  further expressed as
\begin{equation}\label{mes-equ}
\left\{\begin{array}{l}
\hat{\tau}_{\mathrm{w}}[n]=\frac{2\left\|\mathbf{q}_{\mathrm{w}}[n]-\mathbf{q}_{\mathrm{a}}\right\|}{c}+n_{\tau_{\mathrm{w}}[n]}, \\
\hat{v}_{\mathrm{w}}[n]=\frac{2 \dot{\mathbf{q}}_{\mathrm{w}}^T[n]\left(\mathbf{q}_{\mathrm{w}}[n]-\mathbf{q}_{\mathrm{a}}\right) f_c}{c \left\| \mathbf{q}_{\mathrm{w}}[n]-\mathbf{q}_{\mathrm{a}} \right\|}+n_{v_{\mathrm{w}}[n]}, \\
\sin \hat{\theta}_{\mathrm{aw}}[n]=\frac{y_{\mathrm{w}}[n]-y_{\mathrm{a}}}{\sqrt{\left|x_{\mathrm{w}}[n]-x_{\mathrm{a}}\right|^2+\left|y_{\mathrm{w}}[n]-y_{\mathrm{a}}\right|^2}}+n_{\sin \theta_{\mathrm{aw}}[n]}, \\
\cos \hat{\theta}_{\mathrm{aw}}[n]=\frac{x_{\mathrm{w}}[n]-x_{\mathrm{a}}}{\sqrt{\left|x_{\mathrm{w}}[n]-x_{\mathrm{a}}\right|^2+\left|y_{\mathrm{w}}[n]-y_{\mathrm{a}}\right|^2}}+n_{\cos \theta_{\mathrm{aw}}[n]}, \\
\sin \hat{\phi}_{\mathrm{aw}}[n]=\frac{z_{\mathrm{w}}[n]-z_{\mathrm{a}}}{\left\| \mathbf{q}_{\mathrm{w}}[n]-\mathbf{q}_{\mathrm{a}} \right\|}+n_{\sin \phi_{\mathrm{aw}}[n]},
\end{array}\right.
\end{equation}
where $f_c=\frac{c}{\lambda_c}$ with $c$ and $\lambda_c$  denoting the speed of light and wavelength, respectively. The measurement noises $n_{\tau_{\mathrm{w}}[n]}$, $n_{v_{\mathrm{w}}[n]}$, $n_{\sin{\theta}_{\mathrm{w}}[n]}$, $n_{\cos{\theta}_{\mathrm{w}}[n]}$, and $n_{\sin{\phi}_w[n]}$ are assumed to follow Gaussian distribution with zero mean and variances $\sigma_{\tau_{\mathrm{w}}[n]}^2$, $\sigma_{v_{\mathrm{w}}[n]}^2$, $\sigma_{\sin \theta_{\mathrm{aw}}[n]}^2$, $\sigma_{\cos \theta_{\mathrm{aw}}[n]}^2$ and  $\sigma_{\sin \phi_{\mathrm{aw}}[n]}^2$, respectively. Similar to \cite{Wei:2023,LiuP:2023}, we adopt the following approximations: $\sigma_{\sin \theta_{\mathrm{aw}}[n]}^2 \approx \cos^2 \hat{\theta}_{\mathrm{aw}}[n] \sigma_{\theta_{\mathrm{aw}}[n]}^2$, $\sigma_{\cos \theta_{\mathrm{aw}}[n]}^2\approx \sin^2 \hat{\theta}_{\mathrm{aw}}[n] \sigma_{\theta_{\mathrm{aw}}[n]}^2$ and  $\sigma_{\sin \phi_{\mathrm{aw}}[n]}^2 \approx \cos^2 \hat{\phi}_{\mathrm{aw}}[n] \sigma_{\phi_{\mathrm{aw}}[n]}^2$ in the high SNR regime. 

Furthermore,  by defining $\tilde{c}_{\theta_{\mathrm{av}}} \triangleq c_{\theta_{\mathrm{av}}} \cos ^2 \hat{\theta}_{\mathrm{aw}}^2[n]$ and $\tilde{c}_{\phi_{\mathrm{aw}}} \triangleq c_{\phi_{\mathrm{av}}} \cos ^2 \hat{\phi}_{\mathrm{aw}}[n]$ and combining \eqref{SNR-mes}, the covariance matrix $\widehat{\mathbf{Q}}_{\mathbf{Z}_{\mathrm{W}[n]}}$ of $\mathbf{n}_{\mathbf{Z}_{\mathrm{w}}[n]}$ can be estimated as follows\footnote{Note that due to the inability to obtain the true angle information about Willie, we have to resort to the related predicted values, that is to say we can only calculate the estimated $\hat{\theta}_{\mathrm{aw}}[n]$  and $\hat{\phi}_{\mathrm{aw}}[n]$ by using the predicted state $\hat{\boldsymbol{\chi}}_{\mathrm{w}}[n|n-1]$.}
 \begin{equation}\label{mea-noise-matrix}
\begin{aligned}
& \widehat{\mathbf{Q}}_{\mathbf{Z}_{\mathrm{W}[n]}} \\
& =\operatorname{diag}\left\{\sigma_{\tau_{\mathrm{w}}[n]}^2, \sigma_{v_{\mathrm{w}}[n]}^2, \sigma_{\sin \theta_{\mathrm{aw}}[n]}^2, \sigma_{\cos \theta_{\mathrm{aw}}[n]}^2, \sigma_{\sin \phi_{\mathrm{aw}}[n]}^2\right\} \\
& =\frac{\operatorname{diag}\{\tilde{\mathbf{z}}\}}{\mathbf{a}^H\left(\hat{\theta}_{\mathrm{aw }}[n], \hat{\phi}_{\mathrm{aw }}[n]\right) \mathbf{R}_{\mathrm{s}}[n] \mathbf{a}\left(\hat{\theta}_{\mathrm{aw }}[n], \hat{\phi}_{\mathrm{aw }}[n]\right)}, 
\end{aligned}
\end{equation}
where $\tilde{\mathbf{z}}=\frac{\sigma_{\mathrm{r}}^2 d_{\mathrm{aw}}^4[n]}{\rho_0 \varepsilon G_{\mathrm{MF}} N_{\mathrm{A}}}\left[c_{\tau_{\mathrm{w}}}, c_{v_{\mathrm{w}}}, \tilde{c}_{\theta_{\mathrm{av}}}, c_{\theta_{\mathrm{aw}}}-\tilde{c}_{\theta_{\mathrm{aw}}}, \tilde{c}_{\phi_{\mathrm{aw}}}\right]$.

Finally, for the linearized measurement model in \eqref{mes-line}, $\frac{\partial \mathbf{g}}{\partial \boldsymbol{\chi}_\mathrm{w}}$ in the Jacobian matrix $\mathbf{g}$ is calculated in \eqref{Jacobian-matrix}, where
\begin{equation*}
\begin{aligned}
 \frac{\partial v_e}{\partial x_{\mathrm{w}}}&=\frac{2 f_c\left(\dot{x}_{\mathrm{w}}\left\|\mathbf{q}_{\mathrm{w}}-\mathbf{q}_{\mathrm{a}}\right\|^2-\dot{\mathbf{q}}_{\mathrm{w}}^T\left(\mathbf{q}_{\mathrm{w}}-\mathbf{q}_{\mathrm{a}}\right)\left(x_{\mathrm{w}}-x_{\mathrm{a}}\right)\right)}{c\left\|\mathbf{q}_{\mathrm{w}}-\mathbf{q}_{\mathrm{a}}\right\|^3}, \\  
 \frac{\partial v_e}{\partial y_{\mathrm{w}}}&=\frac{2 f_c\left(\dot{y}_{\mathrm{w}}\left\|\mathbf{q}_{\mathrm{w}}-\mathbf{q}_{\mathrm{a}}\right\|^2-\dot{\mathbf{q}}_{\mathrm{w}}^T\left(\mathbf{q}_{\mathrm{w}}-\mathbf{q}_{\mathrm{a}}\right)\left(y_{\mathrm{w}}-y_{\mathrm{a}}\right)\right)}{c\left\|\mathbf{q}_{\mathrm{w}}-\mathbf{q}_{\mathrm{a}}\right\|^3},\\ \frac{\partial v_e}{\partial z_{\mathrm{w}}}&=\frac{2 f_c\left(\dot{z}_{\mathrm{w}}\left\|\mathbf{q}_{\mathrm{w}}-\mathbf{q}_{\mathrm{a}}\right\|^2-\dot{\mathbf{q}}_{\mathrm{w}}^T\left(\mathbf{q}_{\mathrm{w}}-\mathbf{q}_{\mathrm{a}}\right)\left(z_{\mathrm{w}}-z_{\mathrm{a}}\right)\right)}{c\left\|\mathbf{q}_{\mathrm{w}}-\mathbf{q}_{\mathrm{a}}\right\|^3},\\ \left\|\mathbf{q}_{\mathrm{ew}}\right\|&=\sqrt{\left(x_{\mathrm{w}}-x_{\mathrm{a}}\right)^2+\left(y_{\mathrm{w}}-y_{\mathrm{a}}\right)^2}.
\end{aligned}
\end{equation*}

\bibliography{Ref/Ref.bib}
\bibliographystyle{IEEEtran}

\begin{IEEEbiography}[{\includegraphics[width=4cm,height=3.2cm,clip,keepaspectratio]{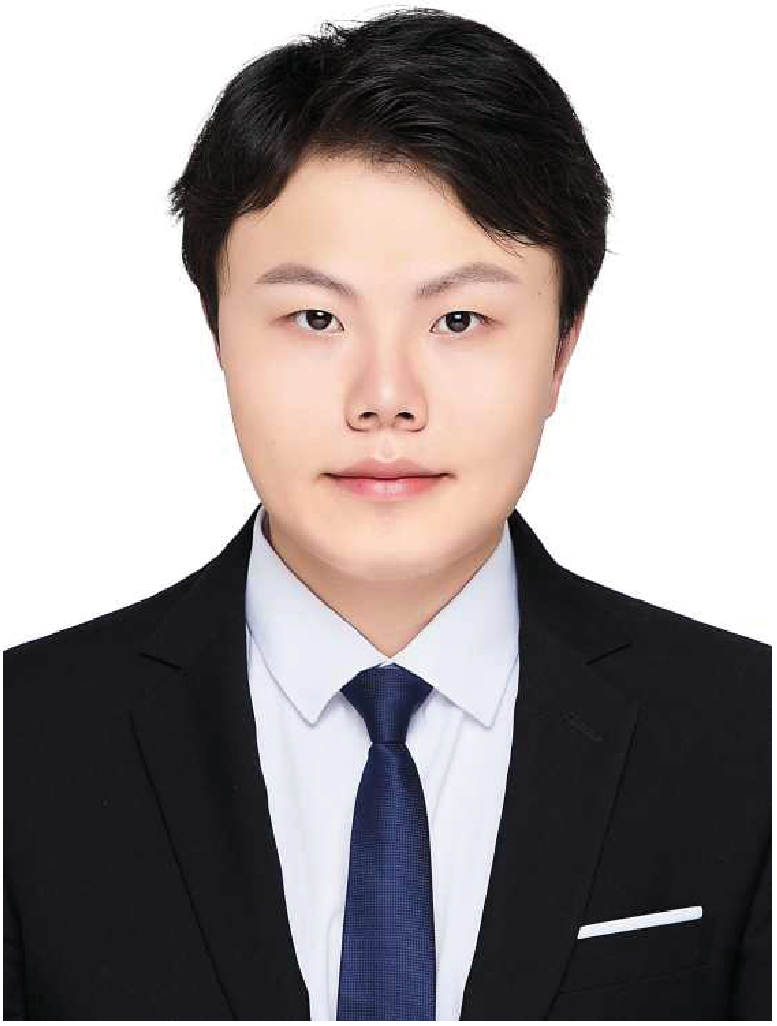}}]{Xingyu Zhao} (Graduate Student Member, IEEE) received the B.S. degree in electronic and information engineering from Northwestern Polytechnical University, Xi’an, China, in 2019, the M.S. degree in  communication and information system from Fudan University, Shanghai, China, in 2022. From 2022 to 2023,  he was a Wireless Communication Algorithm Engineer with ZTE Corporation, Shanghai, China. He is currently pursuing the Ph.D. degree with the College of Information Science and Electronic Engineering, Zhejiang University, Hangzhou, China. 

His current research interests include integrated sensing and communication systems, covert communications, and physical layer security.
\end{IEEEbiography}

\begin{IEEEbiography}[{\includegraphics[width=1in,height=1.25in,clip,keepaspectratio]{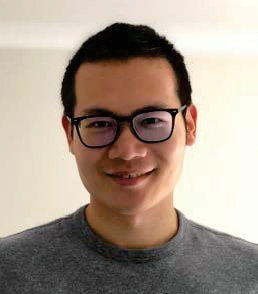}}] {Min Li} (Member, IEEE) received the B.E. degree in telecommunications engineering and the M.E. degree in information and communication engineering from Zhejiang University, Hangzhou, China, in 2006 and 2008, respectively, and the Ph.D. degree in electrical engineering from The Pennsylvania State University, USA, in 2012. He conducted postdoctoral research in Australia, holding positions at Macquarie University (2012-2016, 2018–2019) and the University of Newcastle (2016–2018). Since March 2019, he has been with the College of Information Science and Electronic Engineering, Zhejiang University, where he was appointed as a ZJU100 Young Professor and promoted to Tenured Associate Professor in July 2025. 

His research interests include network information theory, millimeter-wave cellular communications, integrated sensing and communication systems, AI-empowered communications and covert communications. He has received the Young Rising Star Award from the Information Theory Society of the Chinese Institute of Electronics in 2021. He is the Publicity Chair for the 2025 IEEE Information Theory Workshop, and co-organizes workshops at IEEE/CIC ICCC 2025 and IEEE VTC2025-Fall. He is currently serving as an Associate Editor for IEEE TRANSACTIONS ON INFORMATION FORENSICS AND SECURITY and IEEE TRANSACTIONS ON GREEN COMMUNICATIONS AND NETWORKING.
\end{IEEEbiography}

\begin{IEEEbiography}[{\includegraphics[width=3.5cm,height=3.2cm,clip,keepaspectratio]{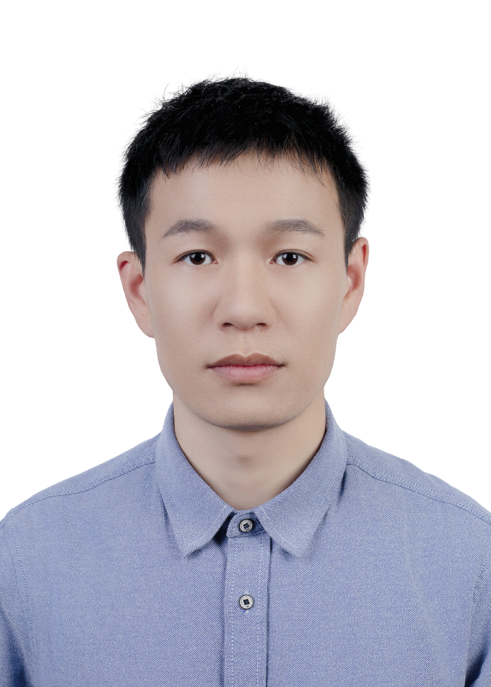}}] {Ming‑Min Zhao} (Senior Member, IEEE) received the B.Eng. and Ph.D. degrees in information and communication engineering from Zhejiang University, in 2012 and 2017, respectively. From 2015 to 2016, he was a Visiting Scholar with the Department of Electrical and Computer Engineering, Iowa State University, Ames, IA, USA. From 2017 to 2018, he worked as a Research Engineer with Huawei Technologies Co., Ltd. From 2019 to 2020, he was a Visiting Scholar with the Department of Electrical and Computer Engineering, National University of Singapore. Since 2018, he has been working with Zhejiang University, where he is currently an Associate Professor with the College of Information Science and Electronic Engineering. 

His research interests include algorithm design and analysis for advanced MIMO, signal processing for communication, channel coding, and machine learning for wireless communications. He was the recipient of the IEEE Communications Society Katherine Johnson Young Author Best Paper Award in 2024. He was the Publication and Publicity Co-Chair for the 2022 International Symposium on Wireless Communication Systems (ISWCS), and co-organized workshops at IEEE/CIC ICCC 2025 and IEEE VTC2025-Fall. He is currently serving as an Associate Editor for IEEE OPEN JOURNAL OF SIGNAL PROCESSING and IEEE WIRELESS COMMUNICATIONS LETTERS.
\end{IEEEbiography}

\begin{IEEEbiography}[{\includegraphics[width=0.95in,clip,keepaspectratio]{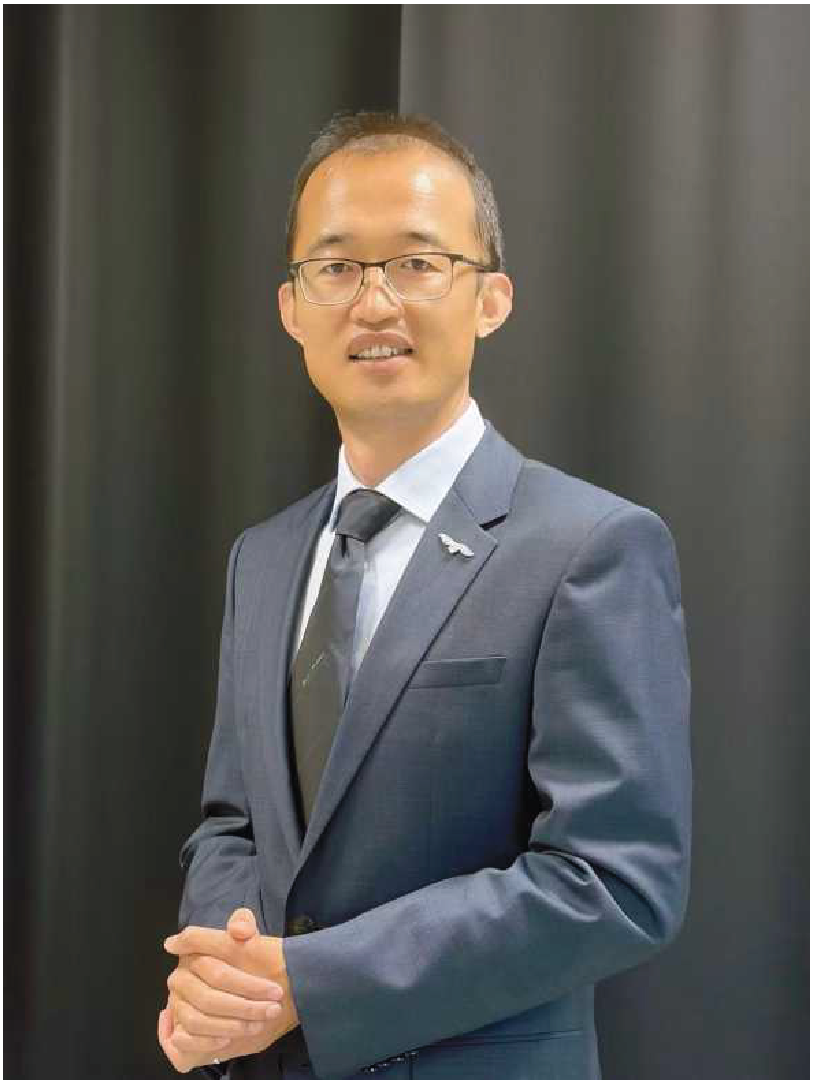}}]{Shihao Yan}
(Senior Member, IEEE) received the Ph.D. degree in Electrical Engineering from the University of New South Wales (UNSW), Sydney, Australia, in 2015. He received the B.S. in Communication Engineering and the M.S. in Communication and Information Systems from Shandong University, Jinan, China, in 2009 and 2012, respectively. He is currently an Associate Professor in the School of Science, Edith Cowan University (ECU), Perth, Australia. He is an Editorial Board Member for multiple journals, i.e., Drones, Scientific Reports and Transactions on Emerging Telecommunications Technologies, and a Committee Member of the IEEE Focus Group on Physical Layer Security. He served as the General Chair of Australian Communications Theory Workshop (AusCTW) and a Program Chair of IEEE International Workshop on Information Forensics and Security (IEEE WIFS) in 2025. He was also awarded the Humboldt Research Fellowship for experienced researchers in 2023 and Australia Endeavour Research Fellowship in 2017. 

His current research interests are in the areas of signal processing for wireless communication security and privacy, including covert communications, drone detection, satellite communications, and physical layer security. He is currently serving as an Associate Editor for IEEE TRANSACTIONS ON VEHICULAR TECHNOLOGY.
\end{IEEEbiography}

\begin{IEEEbiography}
[{\includegraphics[width=0.95in,clip,keepaspectratio]{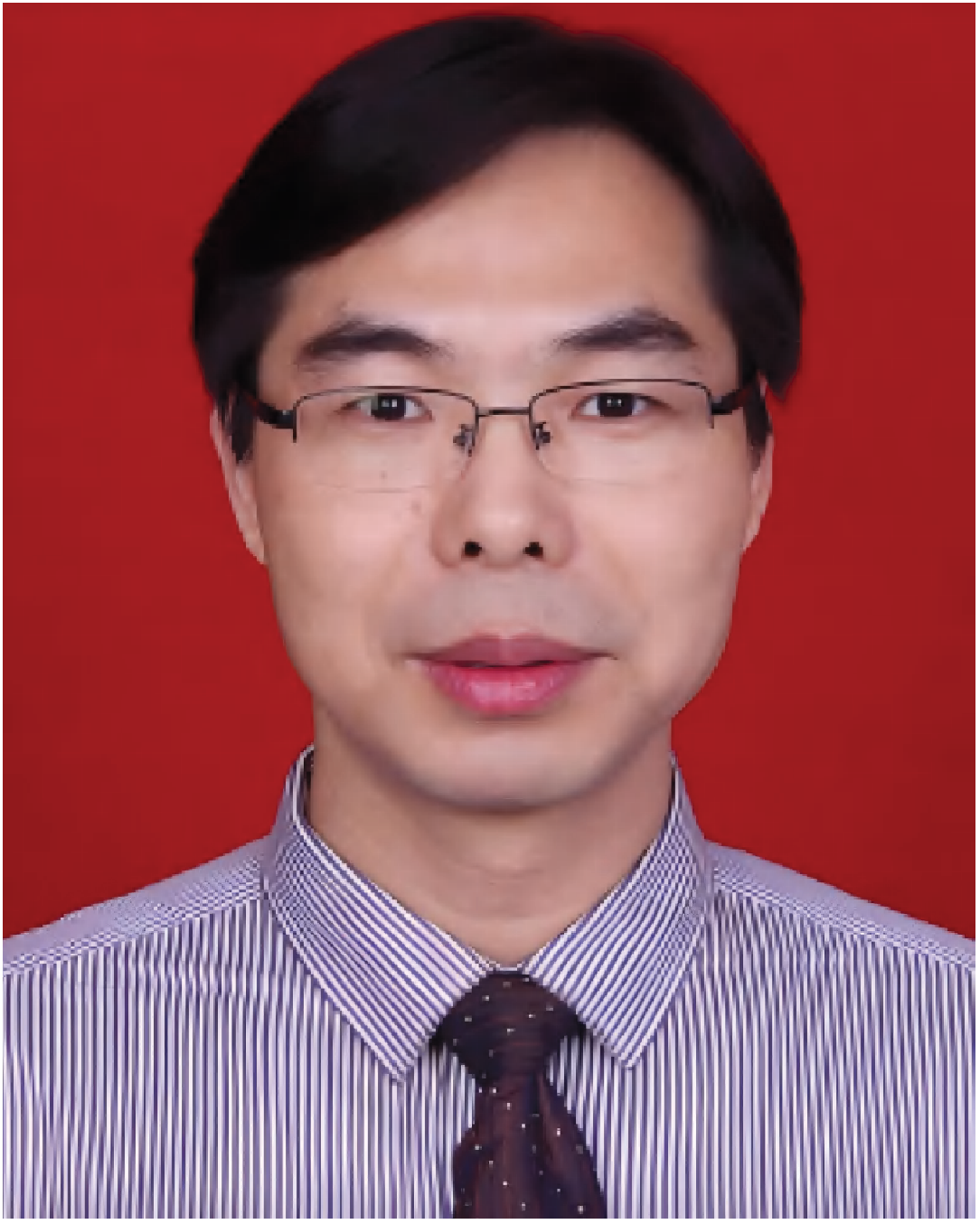}}]{Min-Jian Zhao} (Member, IEEE) received the M.Sc. and Ph.D. degrees in communication and information systems from Zhejiang University, Hangzhou, China, in 2000 and 2003, respectively. He was a Visiting Scholar with the University of York, York, U.K., in 2010. He is currently a Professor and the Deputy Director with the College of Information Science and Electronic Engineering, Zhejiang University. 

His research interests include modulation theory, channel estimation and equalization, MIMO, signal processing for wireless communications, anti-jamming technology for wireless transmission and networking, and communication SOC chip design.
\end{IEEEbiography}

\end{document}